\documentclass[twocolumn,amsmath,amssymb,showpacs,prx]{revtex4-1}
 \usepackage{graphicx,graphics,color}
\usepackage{amsmath}
\usepackage{amssymb}
\usepackage{amsfonts}
\usepackage{amsfonts}
\usepackage{epstopdf}
\usepackage{bm}
\usepackage{times,xspace}
\usepackage{color}
\usepackage[utf8]{inputenc}

\def\be{\begin{equation}}
\def\ee{\end{equation}}
\def\bea{\begin{eqnarray}}
\def\eea{\end{eqnarray}}

\def\PT{\mathcal{PT}}
\def\P{\mathcal{P}}
\def\T{\mathcal{T}}

\def\CPT{\mathcal{CPT}}

\def\PT{$\mathcal{PT}$}
\def\T{$\mathcal{T}$}
\def\P{$\mathcal{P}$}
\def\CPT{$\mathcal{CPT}$}

\begin{document}

\title{New topological invariants in non-Hermitian systems}
\author{Ananya Ghatak\footnote{gananya04@gmail.com}, and Tanmoy Das\footnote{tnmydas@gmail.com}\\
{Department of Physics, Indian Institute of Science, Bangalore-560012, India.}}

\date{\today}
\vspace{0.3cm}
\begin{abstract}
Both theoretical and experimental studies of topological phases in non-Hermitian systems have made a remarkable progress in the last few years of research. In this article, we review the key concepts pertaining to topological phases in non-Hermitian Hamiltonians with relevant examples and realistic model setups. Discussions are devoted to both the adaptations of topological invariants from Hermitian to non-Hermitian systems, as well as origins of new topological invariants in the latter setup. Unique properties such as exceptional points and complex energy landscapes lead to new topological invariants including winding number/vorticity defined solely in the complex energy plane, and half-integer winding/Chern numbers. New forms of Kramers degeneracy appear here rendering distinct topological invariants. Modifications of adiabatic theory, time-evolution operator, biorthogonal bulk-boundary correspondence lead to unique features such as topological displacement of particles, `skin-effect', and edge-selective attenuated and amplified topological polarizations without chiral symmetry. Extension and realization of topological ideas in photonic systems are mentioned. We conclude with discussions on relevant future directions, and highlight potential applications of some of these unique topological features of the non-Hermitian Hamiltonians.
\end{abstract}

\maketitle

\tableofcontents

\section{Introduction}\label{Sec:Intro}
Studies of non-Hermitian (NH) Hamiltonians date back to the early days of quantum mechanics and matrix algebra. Usage of pseudo-unitary metric in quantum theory was proposed by many legends including Dirac, Pauli, Feynman, and Sudarshan\cite{Dirac,Pauli,Feynman,Sundarshan,LeeWick} Existence of real energy ground state in NH Hamiltonians of hard-core bosons is known since 1959.\cite{TWu} Variety of NH perturbations, non-unitary singularities has been discussed in different contexts.\cite{Fisher,Cardy,NHBerry2} Over the years, there have been plenty of evidence of NH Hamiltonians which give purely real eigenvalues and ground state with unitarity and conservation laws. Scholz and others argued that a NH Hamiltonian with real eigenvalues can be casted into a Hermitian counterpart with a similarity transformation, or equivalently one can find a metric to describe the unitarity and inner product.\cite{cryptoH0,cryptoH1,cryptoH2} Such Hamiltonians are in general termed as crypto-Hermitian Hamiltonians. In recent years, studies of such systems have been substantially channelized and advanced towards the materials realization by Bender and co-workers.\cite{Bender1998,Bender2007,PToperator1,PToperator2,BenderCopp} They found that NH Hamiltonians which are invariants under the combination of parity (\P) and time-reversal (TR) $\mathcal{T}$ symmetries are guaranteed to give real eigenvalues, conserved probability, and physical observables.\cite{Bender1998,Bender2007,PToperator1,PToperator2,BenderCopp} Mostafazadeh and co-workers have also specialized the idea of crypto-Hermitian systems to pseudo-Hermitian systems with anti-linear metric operator which ensures real eigenvalues, and proper Hilbert space.\cite{PseudoHTheory,PseudoHReview} Of course, study of NH Hamiltonians has remained a constant theme of research in quantum optics, non-equilibrium systems and other open, dissipative systems. Research interests on \PT-symmetric and pseudo-Hermitian systems have received a huge boost after the experimental realization of \PT-symmetric systems in quantum optics.\cite{Ruter,guo,sch,bit,ram,kla,mir,such,uzd,kott,lon,lie,miri}

In the mean time, there have been a tremendous research progress in the field of topological insulators, mainly motivated by the proposals and materials discovery of TR invariant topological insulators in the Hermitian systems.\cite{KaneMele1,KaneMele2,ZhangQSH,ZhangQSH2,QiWuZhang,KaneRMP,ZhangRMP,DasRMP,HTSCReview,DasIISc,ShankarLN} These research activities and progresses in the Hermitian topological fields have greatly helped extending and generalizing some of these ideas to the NH Hamiltonians in the last two to three years.\cite{Levitov,NHTELee,Leykam,SSHJiang,SSHYin,SSHZhou,SSHLin,SSHPTBulkimpurity,SSHLieu,SSHYao,NHTELee,SSHElectric,SSHDiss,SSHColdAtom,NHAAHYuce,NHAAHYuce2,NHAAHTh,NHQWireTh, NHEPTI,NHRiceMeleDQ,NHKramersEsaki,NHTRKramers,NHTopUnif,NHNodaLine1,NHNodaLine2,NHNodaLine3,NHRing1,NHRing2Exp,NHRing3,NHLink1,NHHopf1,NHKnot1,NHKnot2, NHESurface,NHESurface2,WeylColdAtomTh,WeylNodallineTH,NHWeylTh,NHWeylTh2,NHRing1,NHFloquet,NHSSHFloquet,NHSSHFloquet2, NHTSC,NHKitaevWang,NHKitaevKlett,NHKitaevLi,NHKitaevYuce,NHMajoranaTh,NHChernKawabata,LFu,NHChernKunst,NHChernPhilip,NHRiceMeleDQ,NHEPSLin,NHChernDiss,NHGraphene, PhotonicTI,NHTrimer,PhotonicTI2,NHTIclass,NHTopUnif,NHDynamicalClass,NHTIMirror,NHTIRev1,NHTIRev2,NHTopCommentry,NHTopCommentry2,NHExp1,NHExp2Photon,NHExp3Floquet,NHExp4QWalk,NHExp4QWalk2,NHExp5FermiArc,NHExp6,NHExp7,NHExp8Laser,NHExp9Photon,NHExp10Photon,NHExp11,NHExp12}. At the initial stage, the existence of quantized topological invariant, topological protection of the boundary states, correspondence between Chern number and quantum Hall effect were scrutinized at length.\cite{NHHughes,NHBulkBoundary,NHTProtection,NHChernNoHall,NHBulkBoundary2} The reason was that even though \PT-symmetry protects the unitarity of the bulk states, the symmetry is lost at the boundary, and hence violates the bulk-boundary correspondence associated with topological phases. Subsequently, it was realized in various model systems that chiral symmetry, which can be simultaneously present both in the bulk and at the boundary, can rescue the bulk-boundary correspondence.\cite{NHTELee,Leykam} So far, topological phases have been investigated in most of the prototypical topological systems such as 1D Su-Schrieffer-Heeger (SSH) lattices,\cite{SSHJiang,SSHYin,SSHZhou,SSHLin,SSHPTBulkimpurity,SSHLieu,SSHYao,NHTELee,SSHElectric,SSHDiss,SSHColdAtom} Aubry-Andr\'e-Harper chain,\cite{NHAAHYuce,NHAAHYuce2,NHAAHTh} Kitaev model,\cite{NHTSC,NHKitaevWang,NHKitaevKlett,NHQWireTh,NHKitaevLi,NHKitaevYuce,NHMajoranaTh} 2D Chern insulators,\cite{NHChernKawabata,LFu,NHChernKunst,NHChernPhilip,NHRiceMeleDQ,NHEPSLin,NHChernDiss,NHGraphene}
Rice-Mele,\cite{NHRiceMeleDQ} Kane-Mele, Bernevig-Hughes-Zhang models,\cite{NHKramersEsaki,NHTRKramers,NHTopUnif} nodal line, nodal ring, Hopf link,\cite{NHNodaLine1,NHNodaLine2,NHNodaLine3,NHRing1,NHRing2Exp,NHRing3,NHLink1,NHHopf1,NHKnot1,NHKnot2} or even exceptional surface.\cite{NHESurface,NHESurface2} and Weyl semimetals in 3D,\cite{WeylColdAtomTh,WeylNodallineTH,NHWeylTh,NHWeylTh2,NHRing1} as well as in Floquet systems,\cite{NHFloquet,NHSSHFloquet,NHSSHFloquet2} and many other condensed matter settings\cite{NHChernKawabata,LFu,NHChernKunst,NHChernPhilip,NHRiceMeleDQ,NHEPSLin,NHChernDiss,NHGraphene}. Typically non-Hermiticity in these popular models is achieved by introducing NH hoppings and/or with NH gain/loss terms. More interestingly, the multifaceted non-Hermiticity lends itself its extension to photonics, optical lattice, and non-equilibrium systems.\cite{PhotonicTI,NHTrimer,PhotonicTI2} In fact, there have been a large number of experimental observations of topological phases in quantum optical lattices already placed in literature.\cite{NHExp1,NHExp2Photon,NHExp3Floquet,NHExp4QWalk,NHExp4QWalk2,NHExp5FermiArc,NHExp6,NHExp7,NHExp8Laser,NHExp9Photon,NHExp10Photon,NHExp11,NHExp12}

From conceptual perspective, the non-Hermiticity helps govern new topological invariants, which may not have a direct analog with the Hermitian Hamiltonians. In addition, some of the topological invariants of the Hermitian Hamiltonians become generalized when the Hermiticity condition is removed, or replaced with pseudo-Hermiticity or \PT-invariance. Owing to associated uniqueness in NH Hamiltonians, the new topological invariants may not be adiabatically connected to any topological phase in the corresponding Hermitian Hamiltonian if the Hermiticity is restored by a continuous tuning of a parameter without closing the energy gap between the two Hamiltonians. This should be contrasted with the adiabatic continuity theory for Hermitian topological insulators which implies that two Hamiltonians are adiabatically connected with the same topological phase if one can continuously go from one Hamiltonian to another without closing any gap between them.

Band degeneracy at discrete $k$-points is crucial for topological phases, as it serves as singular pole to yield non-trivial Berry curvature. Band degeneracy is usually considered as an apparent curse to the NH Hamiltonians, since here the corresponding eigenstates coalesce (i.e., the right and left eigenstates for a given eigenvalue become orthogonal and hence the inner product becomes singular). They are called exceptional points (EPs).\cite{EPKato,EPHeiss,EPSmilga,EPRotter} For topological phases, the EP is however a blessing, which gives complex Berry phase. Under various conditions such as with certain symmetry protection, and/or with special pseudo-Hermitian metric, the Berry phase can become purely real. Interestingly, with suitable parameter tuning, the EPs can be split in the parameter space, rendering half-integer winding number, and Chern number in non-interacting NH systems.\cite{NHBerry1,NHBerry2,NHBerryHeiss,NHBerryMailybaev,EPBerry,NHBerry3,NHTELee,Leykam,SSHJiang,SSHYin,SSHZhou,SSHLin,SSHPTBulkimpurity,SSHLieu,SSHYao,NHTELee,SSHElectric,SSHDiss,NHChernKawabata,LFu,NHChernKunst,NHChernPhilip,NHRiceMeleDQ,NHEPSLin,NHChernDiss,NHGraphene}
More uniquely, owing to complex energy spectrum in generic NH Hamiltonians, we can define a new topological winding number in the complex energy plane (we call it vorticity index to distinguish it from the typical winding number defined in the complex eigenstate case). The vorticity invariant effectively counts the number of EPs enclosed inside the adiabatic loop in a complex energy plane (note that, this does not mean that the energy spectrum needs to possess any EP, which would be a special case when the EP lies on the adiabatic loop). Furthermore, in various pseudo-Hermitian systems, both possible TR operators $\mathcal{T}^2=\pm 1$ as well as both possible charge conjugation operators $\mathfrak{C}^2=\pm 1$ can give Kramers degeneracy,\cite{NHKramersEsaki,NHTRKramers,GhatakDasNewpaper} while in the Hermitian counterparts we mainly study Kramers degeneracy for the $\mathcal{T}^2=-1$ case.\cite{KramersTheory,Bernevigbook,ShankarLN} 

Furthermore, the bulk-boundary correspondence is also modified in some of the NH settings with new topological effects. For example, the typical domain wall problem which gives a half-fermion polarization at the edge,\cite{Zak,BellRajaraman} modifies to a biorthogonal polarization with its imaginary term reflecting the loss function.\cite{NHChernKunst} NH winding number with gain/loss terms leads to topological displacement, where the displacement of the bulk states is quantized by the winding number.\cite{Levitov} Another example is the NH `skin effect\cite{SSHYao,NHEPTI2,NHChernYao,NHHopf1,NHNodaLine1,JinSong,SkinRonny} in which the bulk state can localize at the boundary, effectively loosing its atypical (Bloch-wave) extended character in a periodic system. In converse the boundary states can acquire a damped delocalization component by the chiral symmetry breaking NH term. In some NH topological systems, we find that the right and left edges selectively obtain lossy and amplified probabilities, respectively, with the total probability remaining conserved.\cite{NHTrimer} These unique advantages of the NH Hamiltonians clearly help expanding the plethora of topological phases to incorporate new topological invariants, new symmetry invariance, new Kramers pairs, new edge state characteristics, as well as encompassing the larger territory of condensed matter settings to quantum optics and non-equilibrium systems.

In this review article, we aim at discussing the core concepts of topological invariants that have been put forward in various NH systems. For readers' reference, we start with basic recapitulations of the essential ideas of topological phases in the Hermitian setting (Sec.~\ref{Sec:II}), and also a brief review on the mathematical background of NH quantum theory (Sec.~\ref{Sec:III}). We then discuss all possible topological phases in NH Hamiltonians such as Berry phase, winding number, energy vorticity, Chern number, TR topological invariants etc in various dimensions in periodic boundary condition (Secs.~\ref{Sec:IV},\ref{Sec:V},\ref{Sec:VI},\ref{Sec:VII}), as well as Zak phase, domain wall studies, `skin-effect', finite lattice simulations with open boundary conditions (Sec.~\ref{Sec:VIII}). Relevant examples are furnished for all discussions, with corresponding model NH Hamiltonians derived from popular condensed matter Hamiltonians. Discussion dedicating on photonic and cold atom systems are provided in Sec.~\ref{Sec:IX}. It is worthwhile noting that photonic systems are not entirely quantum mechanical, but the one to one correspondence between Schr\"oedinger equation and electromagnetic wave equation helps directly implementing the topological concepts of quantum systems to the photonic settings. In fact, most of the experimental observations of NH topological phases are obtained in photonic lattices and related systems.\cite{NHExp1,NHExp2Photon,NHExp3Floquet,NHExp4QWalk,NHExp4QWalk2,NHExp5FermiArc,NHExp6,NHExp7,NHExp8Laser,NHExp9Photon,NHExp10Photon,NHExp11,NHExp12} Although we direct the reader to relevant experimental observations in various appropriate contexts, we do not discuss these experimental results at length in this article. By the time of completion of this review article, we are not aware of any comprehensive review article on the topic of NH topological phases, however several research articles, and commentaries on the classifications and unifications of various NH topological phases can be referenced in this context.\cite{NHTIclass,NHTopUnif,NHDynamicalClass,NHTIMirror,NHTIRev1,NHTIRev2,NHTopCommentry,NHTopCommentry2}


\section{Brief review on topological invariants in Hermitian Hamiltonians}\label{Sec:II}
We begin with a basic review on some of the relevant topological properties that arise in Hermitian Hamiltonians. This recapitulation is meant to serve as an inspiration to formulate NH topological invariants as well as to compare and contrast between the two paradigms. A building block of the topological phases of matter is the Berry phase $-$ a geometric phase acquired by the eigenstates in an adiabatic cycle in the parameter space such as time, position, or momentum.\cite{Pancharatnam,HBerry} One way to grasp an intuitive understanding of the Berry phase is to compare it with the Peierls phase ($p$) acquired by a charged particle ($e$) under the application of a vector potential ${\bf A}$ as $p=\frac{e}{\hbar}\int {\bf A}({\bf r})\cdot d{\bf r}$.\cite{Sakuraibook} The Peierls phase becomes quantized in a periodic boundary condition, which is called the Aharonov-Bohm phase.\cite{ABPhase} Similarly, the Berry phase acquired by the adiabatic evolution can be casted into a geometrical `vector potential' ${\bf \mathcal{A}}({\bf k})$, called the Berry connection. For a non-interacting Hermitian Hamiltonian $H({\bf k})$ with eigenvalues $E_n({\bf k})$ and eigenfunctions $|\psi_n({\bf k})\rangle$, the Berry phase is defined as
\bea
\label{HBerry}
\gamma_n &=& \oint_{\mathcal{C}} {\bf \mathcal{A}}_n({\bf k})\cdot d{\bf k},\\
{\rm where}\quad{\bf \mathcal{A}}_n({\bf k})&=&-i\left\langle \psi_{n}({\bf k})|{\bf \nabla}_{\bf k} \psi_{n}({\bf k})\right\rangle.
\label{HBerryConn}
\eea
${\bf \mathcal{A}}_n({\bf k})$ is the Berry connection and the corresponding Berry curvature (analogous to the magnetic field) is ${\bf \mathcal{B}}_n={\bf \nabla}_{\bf k}\times {\bf \mathcal{A}}_n({\bf k})$. In 2D systems, the line integral in Eq.~\eqref{HBerry} can be converted into a surface integral using the Stokes' theorem to obtain $\frac{1}{2\pi}\oint_{\mathcal{S}} {\bf \mathcal{B}}_n({\bf k})\cdot d^2{\bf k}=C_n$. $C_n$ is called the Chern number, which measures the amount of Berry flux passing through the area $\mathcal{S}$ (BZ) enclosed by the loop. As in the case of Aharonov-Bohm phase, the Berry phase and Chern number are quantized in a periodic system.\cite{Bernevigbook,Shinbook,KaneRMP,ZhangRMP,DasRMP,DasIISc,ShankarLN,KaneChapteFranzabook}  

The source of the Berry phase can be traced back to the presence of discrete singularity in ${\bf \mathcal{A}}_n({\bf k})$ $-$ Berry monopole. Berry monopole arises at discrete band degenerate points, i.e., where all gap terms in the Hamiltonian vanish. Hence the valence and conduction bands are assumed to be inverted across these degenerate points. If all gap terms vanish at the {\it same} ${\bf k}$-point, we obtain discrete degenerate points in the energy spectrum across which the bands are inverted. They are called Dirac or Weyl points. In this case the Berry monopole lies on the contour. Otherwise, if different gap terms vanish inside the BZ but at {\it different} ${\bf k}$-points, we obtain a topological insulator, where the Berry monopole moves inside the contour and gives a finite Berry phase. On the other hand, if one or more gap terms do not vanish inside the BZ, we obtain a trivial band insulator with zero Berry phase. The Berry connection essentially constitutes the basis for different topological invariants such as winding number, Chern number, and various TR symmetric topological invariants as they appear in different dimensions and with different symmetry combinations. In a word, {\it finite Berry phase, i.e., non-trivial topological phase, is intimately connected to the constraint that all gap terms must change sign inside the BZ, and the adiabatic loop must enclose such band degenerate points}. 

Above we discussed Berry phase obtained in a periodic boundary condition. As one solves Eqs.~\eqref{HBerry},\eqref{HBerryConn} with an open boundary condition, we find that the $\pi$ Berry phase is nothing but a polarization term with {\it half-integer} fermion number at the boundary. This is called the Zak phase in 1D,\cite{Zak} and is analogous to the Jackiw-Rebbi's solution\cite{JaciwRebbi,BellRajaraman} where a Dirac fermion is localized (soliton) at the domain wall where the potential changes sign. This establishes the intriguing connection to the topological systems where Dirac/Weyl cone emerges near discrete degenerate points, and the sign reversal of the gap terms (domain wall) conspire to give a Berry/Zak phase in periodic/open boundary conditions, respectively. This is the essential bulk-boundary correspondence associated with the topological phases. In higher dimensions, the same bulk-boundary mechanism is also at play, giving rise to Thouless' charge pumping in a Chern insulator (quantum Hall insulator without a magnetic field)\cite{Thoulesspump,Thoulesspump2} or Kane-Mele's proposal of TR polarization (polarization of Kramers pair on the edges) for the case of TR symmetric topological insulators.\cite{KaneMele1,KaneMele2} With an eye on the contradictory evidence of bulk-boundary correspondence in the NH cases, it is important to recognize that in some Hermitian topological insulators, the bulk-boundary correspondence can also be lost,\cite{JMooreTAI,GGuptaTCC,VanderbiltAxion} or modified\cite{HOTI1,HOTI2}.

\begin{figure}[t]
\hspace{-.20in} \includegraphics[width=0.99\columnwidth]{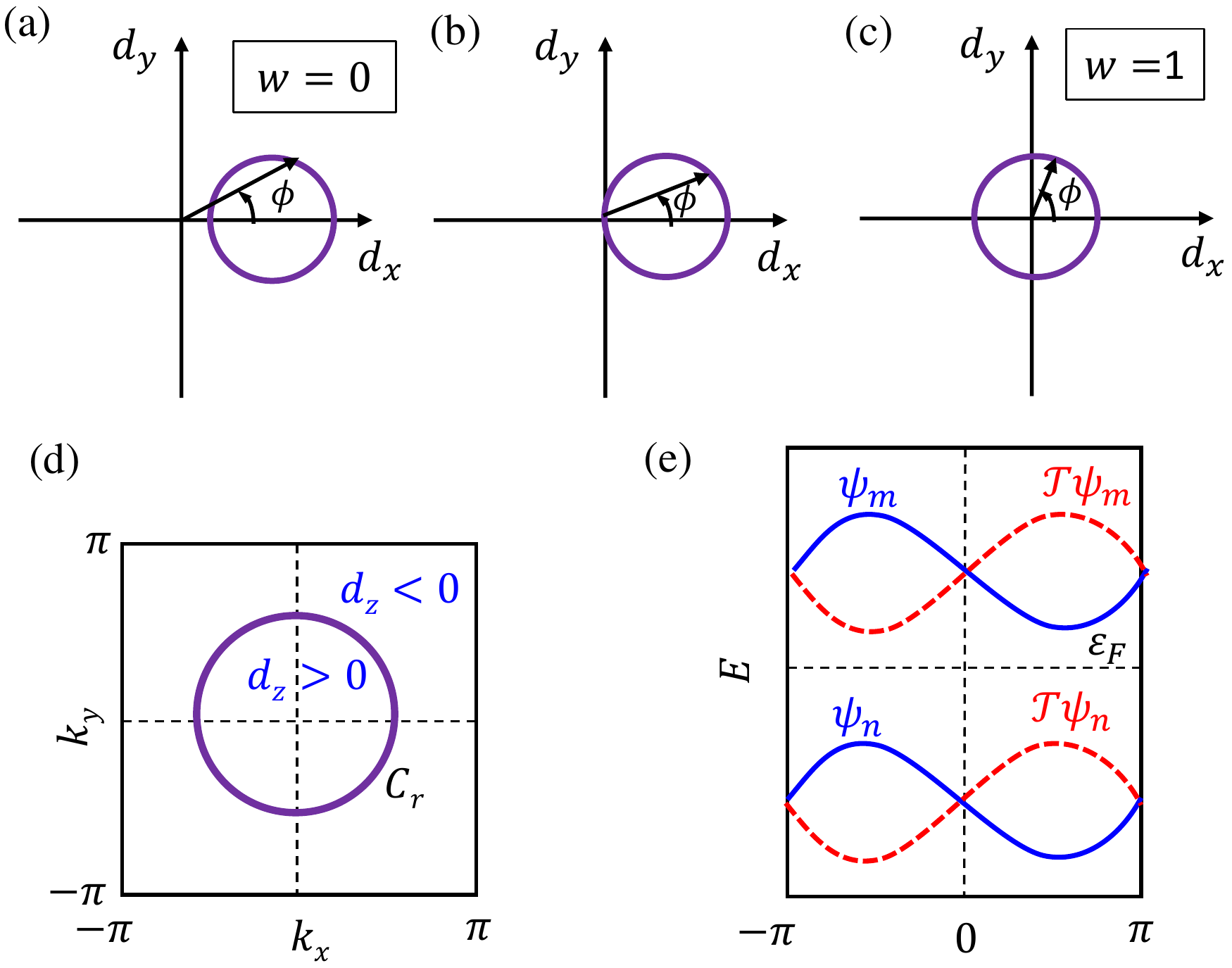} 
\caption{Schematic description of a topological phase transition determined by the winding of the argument of the $d_x$, $d_y$ gap terms in Eq.~\ref{GenNHH}. (a) A topologically trivial phase arises when any of the gap terms ($d_x$ in this particular example) does not go to zero inside the BZ (the contour is defined by the path going around the BZ). (b) A topological critical point is obtained when $d_x=d_y=0$ at the same $k$-point (Dirac point) and the Berry monopole lies on the contour. (c) Topological insulator phase is obtained when all gap terms vanish inside the BZ but at different $k$-points such that the contour encloses the ${\bf d}=0$ point. In this case, according to Eq.~\eqref{Hwinding}, the winding number $w=\pm 1$ for the filled/empty states, respectively. (d) As $d_z$ term is turned on in Eq.~\ref{GenNHH}, assuming it is even under parity, it vanishes on a closed contour (solid line). In this case, the net Berry flux from the two regions with $d_z<0$, and $d_z>0$ gives the Chern number. (e) Schematic band dispersion is shown for two Kramers pair for a TR symmetric system with $\mathcal{T}^2=-1$.}
\label{Fig:Hermitian}
\end{figure}

\subsection{A generic $2\times 2$ Hermitian Hamiltonian}
It is worthwhile discussing an example of a Hermitian topological system which can eventually be made NH, pseudo-Hermitian, and \PT-invariant Hamiltonian with tunable parameters. We can delineate few generic properties of the winding/Chern number with a general $2\times 2$ Hamiltonian:
\be
H({\bf k})  = \xi({\bf k})I_{2\times 2} + {\bf d}({\bf k})\cdot {\bm\sigma}.
\label{GenNHH}
\ee
where $\xi({\bf k})$ is real onsite term, and ${\bf d}({\bf k})$ is the real gap vector. $\sigma_{\mu}$ with $\mu=x,y,z$ are the $2\times 2$ Pauli matrices and $I_{2\times 2}$ is a unit matrix. The onsite $\xi({\bf k})$ term does not eventually appear in the Berry phase formula and thus we set $\xi({\bf k})=0$ without loosing generality. In other words, the following discussions are restricted to systems with particle-hole (PH) symmetric eigenvalues given by $E_{\pm}({\bf k})=\pm d({\bf k})$, where $d=\sqrt{\sum_{\mu=1}^3 d_{\mu}^2}$ is real here. The  eigenvectors are (${\bf k}$-dependence is suppressed for simplicity)
\be
|\psi_{n}\rangle =\frac{1}{\sqrt{2E_{n}(E_{n}-d_z)}}
\begin{pmatrix}
    d_x-i d_y    \\
    E_{n}-d_z \\
\end{pmatrix},
\label{genHES}
\ee
where $n=\pm$ is the band index. Using Eq.~\eqref{HBerryConn}, we find the Berry connection and curvature to be\cite{Bernevigbook} 
\bea
\label{HBerryConn2}
\mathcal{A}_{n}^{\mu} &=& \frac{1}{2E_{n}(E_n-d_z)} \epsilon_{ab}d_a \partial_{\mu} d_b,\\
\mathcal{B}_{n}^{\mu\nu} &=& \frac{1}{2E_n^3} \epsilon_{abc}d_a \partial_{\mu} d_b\partial_{\nu} d_c,
\label{HBerryCurv}
\eea
where $\mu$,$\nu$=$x$,$y$,$z$, and $\epsilon_{ab}$, $\epsilon_{abc}$ are the rank-2/rank-3 Levi-Civita tensors, respectively, where $a,b,c$ run over the three component of ${\bf d}$-vector.

Clearly, the Berry curvature acquires singularity at $E_n=0$, where all three gap terms ${\bf d}=0$ and change sign. Across the degenerate point, two bands $E_{\pm}$ are inverted. If all components $d_{a}$ vanishes at the {\it same} ${\bf k}$-point, we obtain a degenerate Dirac/Weyl point, which marks the topological critical point. If all $d_{a}$ terms vanish but at different ${\bf k}$-points inside the BZ, we obtain a topological insulator, while when one or more $d_{a}$ term(s) remain finite, we obtain a trivial insulator. We illustrate this criterion separately for (a) $d_z=0$, and (b) $d_z \ne 0$ cases. 

(a) For $d_z=0$, the Hamiltonian becomes off-diagonal. In such cases, the Berry connection (Eq.~\eqref{HBerryConn2}) simply depends on the argument of the ${\bf d}$-vector, i.e., $\phi_n({\bf k})=\tan^{-1}(d_y/d_x)$ as $\mathcal{A}_n({\bf k})=\partial_{\bf k}\phi_n({\bf k})$. The corresponding Berry phase in a 1D periodic lattice is called the winding number:
\be
w_n = \frac{1}{2\pi}\int_{-\pi}^{\pi} \partial_{k}\phi_n( k)\cdot dk.
\label{Hwinding}
\ee
In Fig.~\ref{Fig:Hermitian}, we discuss three cases where trivial to non-trivial winding number evolution is obtained. As we move in the complex $d_x+id_y$-plane in a periodic loop from $k=-\pi$ to $\pi$, we enclose a contour $\mathcal{C}$. ${\bf d}=0$ in the degenerate point. In Fig.~\ref{Fig:Hermitian}(a), $d_x$ does not change sign in the BZ, and hence the contour does not include the degenerate point, and we get $w_n=0$. In Fig.~\ref{Fig:Hermitian}(b), $d_x=d_y=0$ at the same ${\bf k}$-point, and hence the degenerate point lies on the contour, giving a Dirac point, where the winding number is undefined. In Fig.~\ref{Fig:Hermitian}(c), both $d_x$ and $d_y$ change sign inside the BZ but at different ${\bf k}$-points, and hence we obtain a topological insulator with $w_-=1$ for the filled band $E_-$.

(b) Next we consider a system where all three gap terms are present. Without loosing generality, let us assume $d_z$ is an even parity term. Therefore, $d_z=0$ is a closed contour in 2D as shown in Fig.~\ref{Fig:Hermitian}(d). We split the BZ into two regions with $d_z>0$, and $d_z<0$. Lets assume $\mathcal{C}_r$ is a closed contour where $d_z=0$. Then the Chern number is defined as\cite{Bernevigbook,SSHJiang}
\bea
C_n &=&\frac{1}{2\pi} \left (\oint_{\mathcal{C}_r}\frac{\epsilon_{ab}d_a \partial_{k} d_b}{2E_{n}(E_n-d_z)}dk +  \oint_{-\mathcal{C}_r}\frac{ \epsilon_{ab}d_a \partial_{k} d_b}{2E_{n}(E_n+d_z)} dk \right ),\nonumber\\
&=&\frac{1}{2\pi} \oint_{\mathcal{C}_r}\frac{ \epsilon_{ab}d_a \partial_{k} d_b}{d_x^2 + d_y^2} dk,\nonumber\\
&=&\frac{1}{2\pi}\oint_{\mathcal{C}_r} \partial_k\phi_n(k) dk.
\label{HChern}
\eea 
(In the first equation, we assume that the BZ is a torus with no boundary, and hence the radius of the contour is immaterial). Here $\phi_n(k)$ is the same argument of the $d_x,d_y$-plane defined above. Therefore, once the sign reversal of the $d_z$ is included, we find that the Chern number is determined by the winding number of the remaining two gap terms ($d_x,d_y$) as defined in Eq.~\eqref{Hwinding}. Of course, the underlying assumption is that both $d_z<0$, and $d_z>0$ regions include degenerate points $d_x=d_y=0$.

\subsection{Time-reversal symmetric topological invariant}\label{Sec:IIA}
Finally, we briefly discuss the fate of topological invariants under the TR symmetry and the concepts of a new $\mathbb{Z}_2$ symmetric topological invariant.\cite{KaneMele1,KaneMele2,FuKanePRL,FuKanePRB} Due to TR symmetry, the total Berry phase is zero. However, Kane and Mele proposed a novel idea of TR polarization for the specific TR symmetry with TR operator $\mathcal{T}^2=-1$.\cite{KaneMele1,KaneMele2} The Kramers theory ensures that for every state $|\psi_{n}\rangle$, its TR partner $|\mathcal{T}\psi_{n}\rangle$ is degenerate, and they are linearly independent to each other.\cite{KramersTheory} If the inversion symmetry is absent, such a degeneracy occurs at the high-symmetric ${\bf k}$-points (also called TR invariant ${\bf k}$-points) as shown in Fig.~\ref{Fig:Hermitian}(e), while with inversion symmetry, all $k$-points are degenerate. For such cases, we can compute the Berry phase from Eq.~\eqref{HBerry} separately for each Kramers partner. Their sum is zero, as demanded by TR invariance, but their difference is finite. This is called the TR polarization as defined by\cite{KaneMele1,KaneMele2}
\be
P^{\mathcal{T}}_n = \frac{1}{2\pi}\oint \left[{\bf \mathcal{A}}^{I}_n -  {\bf \mathcal{A}}^{II}_n\right]\cdot d{\bf k},
\label{HTRpolarization}
\ee
where ${\bf \mathcal{A}}^{\mu}_n({\bf k})=-i\left\langle \psi^{\mu}_{n}({\bf k})|{\bf \nabla}_{\bf k} \psi^{\mu}_{n}({\bf k})\right\rangle$, with $\mu=I,II$ denoting the two Kramers partners. In fact owing to the periodic boundary condition, $P^{\mathcal{T}}_n$ can be shown to be quantized, taking values either 0 or 1 for topologically trivial and non-trivial phases, respectively. Thus, such a system has the $\mathbb{Z}_2$-symmetry. 

In analogy with the Zak phase (half-integer charge polarization) in 1D or charge polarization in the TR breaking Chern insulators, Kane-Mele dubbed the TR invariant topological index in Eq.~\eqref{HTRpolarization} as TR polarization.\cite{KaneMele1,KaneMele2} However, the TR polarization is not an observable quantity unless each Kramers partner represents a physical quantity such as half-integer spin. In fact, for spinfull cases in 2D when spin is a good quantum number, each term $I$ and $II$ in Eq.~\eqref{HTRpolarization} corresponds to different spins, and is individually quantized. Each term then corresponds to a Chern number for each spin with the constraint $C_{\uparrow}=-C_{\downarrow}$ (TR invariance). The difference of the two spin-resolved Chern numbers, as called the spin Chern number, gives a new spin-resolved Hall effect which is called the quantum spin-Hall (QSH) effect.\cite{ZhangQSH} In 3D, even for spinfull case, each term in Eq.~\eqref{HTRpolarization} is not necessarily quantized, while their difference is quantized.\cite{FuKanePRL,FuKanePRB}



\section {Non-Hermitian quantum theories: General introduction}\label{Sec:III}

NH Hamiltonians are often studied in three different approaches, namely, generic NH Hamiltonians with complex eigenvalues,\cite{cryptoH0,cryptoH1,cryptoH2} and pseudo-Hermitian\cite{PseudoHTheory,PseudoHReview} and \PT symmetric Hamiltonians\cite{Bender1998,BenderCopp, PToperator1,PToperator2,Bender2007} with real eigenvalues. There have already been a considerable amount of works to study topological phases in all three types of systems. 

For a generic NH Hamiltonian $H$, we obtain pairs of eigenstates, namely right and left eigenstates, $|\psi_n^{\rm R/L}\rangle$ ($n$ is the eigenvalue index) as:
\begin{subequations}
\bea
\label{Righteigen}
H|\psi_n^{\rm R} \rangle  &=& E_n |\psi_n^{\rm R}\rangle,\\ 
H^{\dag}|\psi_n^{\rm L} \rangle &=& E^*_n |\psi_n^{\rm L}\rangle,
\label{Lefteigen}
\eea
\end{subequations}
where $E_n$, $E_n^*$ are the corresponding eigenvalues. The eigenstates are linearly independent (except at the EPs) and thus can be bi-orthogonalized as $|\tilde{\psi}_n^{\rm R}\rangle=|\psi_n^{\rm R}\rangle/\sqrt{\langle \psi_n^{\rm L}|\psi_n^{\rm R} \rangle}$, and $|\tilde{\psi}_n^{\rm L}\rangle=|\psi_n^{\rm L}\rangle/\sqrt{\langle \psi_n^{\rm L}|\psi_n^{\rm R} \rangle}$, which then gives $\langle \tilde{\psi}_n^{\rm L}|\tilde{\psi}_m^{\rm R}\rangle=\delta_{nm}$, and $\sum_n |\tilde{\psi}_n^{\rm R}\rangle \langle \tilde{\psi}_n^{\rm L}|=1$. Henceforth we drop the `tilde' symbol for simplicity, and assume all such eigenstates are bi-orthogonal.

As we have learned in the Hermitian case, band degeneracy play an important role to topological phases. For NH systems, the EP plays the similar important role. EPs are defined by the complex energy $E_n({\bf k})=0$, where the bi-orthogonal condition breaks down, and $|\psi_n^{\rm R}\rangle$, $|\psi_n^{\rm L}\rangle$ states coalesce, i.e., they become orthogonal $\langle \psi_n^{\rm L}|\psi_n^{\rm R}\rangle=0$, and the Hamiltonian becomes non-diagonalizable.\cite{EPKato,EPHeiss,EPSmilga,EPRotter} The situation is equivalent at the degenerate points $E_n(k_0)=E_m(k_0)$ for $m\ne n$, where not only two eigenvalues become equal, but also their eigenfunctions become parallel. When the degeneracy occurs on a continuous contour, such contours are called Exceptional Line (EL),  Exceptional Hopf Link (EHL), or Exceptional Ring (ER), classified according to the characteristics of the degeneracy.\cite{NHNodaLine1,NHNodaLine2,NHNodaLine3,NHNodaLine4,NHRing1,NHRing2Exp,NHRing3,NHLink1,NHHopf1,NHKnot1,NHKnot2} We will learn below that as the adiabatic contour encloses such an EP, one obtains a complex Berry phase.\cite{NHBerry1,NHBerry2,NHBerryHeiss,NHBerryMailybaev,NHBerry3,EPBerry,PTBerry,NHBerry4,NHBerrySong}


\subsection{Pseudo-Hermitian Hamiltonians}\label{Sec:IIIA}

For specific NH (namely, crypto-Hermitian) Hamiltonians with real eigenvalues, it is long known that there may exists a similarity matrix which can transform the NH to a Hermitian one. \cite{cryptoH0,cryptoH1,cryptoH2} In a more general term, if there exists a similarity transformation which can transform between the $H$ and $H^{\dag}$, the right and left eigenfunctions can be connected by the same similarity transformation. Such special NH Hamiltonians are called pseudo-Hermitian or pseudo-anti-Hermitian Hamiltonians.\cite{PseudoHTheory,PseudoHReview} We denote the corresponding similarity operator as $\eta_{\pm}$ (for pseudo-, and pseudo-anti-Hermitian cases, respectively) under which the Hamiltonian transforms as 
\be
\eta_{\pm}^{-1} H^{\dag}\eta_{\pm}=\pm H.
\label{PseudoH}
\ee
Since $\eta_{\pm}$ appears as the result of the inner product in general, and thus it is also called the metric operator. With further generalization, $\eta_{\pm}$ operators can be either linear or anti-linear. We distinguish an anti-linear metric as $\eta_{\pm}=\bar{\eta}_{\pm}\mathcal{K}$, where $\mathcal{K}$ gives the complex conjugation operator, and $\bar{\eta}_{\pm}$ is linear under which the Hamiltonian transforms as $\bar{\eta}_{\pm} H^{\dag}\bar{\eta}_{\pm}^{-1}=\pm H^*$. Such Hamiltonians can be distinguished as conjugated pseudo-Hermitian or {\it anti-linear} pseudo-Hermitian Hamiltonians. The canonical form of the metric is $\eta_+=\sum_n|\psi_n^{\rm L}\rangle\langle \psi_n^{\rm L}|$, and its inverse is $\eta_+^{-1}=\sum_n|\psi_n^{\rm R}\rangle\langle \psi_n^{\rm R}|$.\cite{PseudoHReview}.

Let us consider the case of an {\it anti-linear} pseudo-Hermitian Hamiltonian $\eta_+$. Using Eqs.~\eqref{Lefteigen}, \eqref{PseudoH} and the anti-linear property of $\eta_+$, we obtain 
\be
H(\eta_+^{-1}|\psi_n^{\rm L}\rangle)=E_n(\eta_+^{-1}|\psi_n^{\rm L}\rangle).
\label{PseudoH2}
\ee
Comparing Eq.~\eqref{PseudoH2} with \eqref{Righteigen}, we can conclude that under this metric $\eta_+^{-1}|\psi_n^{\rm L}\rangle$ and $|\psi_n^{\rm R}\rangle$ are now degenerate if they are linearly independent, or linearly dependent with the same energy. In the latter case, we can expand $\eta_+^{-1}|\psi_n^{\rm L}\rangle$ in the basis of $|\psi_n^{\rm R}\rangle$.\cite{PseudoHTheory,PseudoHReview,NHKramersEsaki,NHTRKramers} We take even a special case of the second condition in which these two states are parallel to each other: 
\be
\eta_+^{-1}|\psi_n^{\rm L}\rangle = c_n|\psi_n^{\rm R}\rangle,
\label{PseudoHcond2}
\ee
where $c_n$ is, in general, a complex number. When Eq.~\eqref{PseudoH} and \eqref{PseudoHcond2} are satisfied together, we obtain {\it real eigenvalues $E_n$}. In what follows, the inner product of the eigenstates modifies to $\langle \psi_n^{\alpha}|\eta_+|\psi_n^{\alpha}\rangle=1$ for $\alpha={\rm L/R}$ since $\eta$ is Hermitian. For further details of the pseudo-Hermitian systems, we encourage readers to consult these relevant review articles \cite{PseudoHTheory,PseudoHReview,TDPseudoH,PTBerry,pseudoHBerry2}.

\subsection{\PT - symmetric Hamiltonians}\label{Sec:IIIB}
Next we discuss a \PT - symmetric NH Hamiltonian which obeys
\be
\mathcal{PT}H(\mathcal{PT})^{-1}=H,
\label{PToperation}
\ee
where \P, ~\T ~ are the parity, and TR operators, respectively. \T~is an anti-linear operator, \P~is linear and Hermitian, and hence \PT~is also anti-linear (as the $\eta_+$ operator above, except here $H$ and \PT commute). Using Eqs.~\eqref{Righteigen}, \eqref{PToperation} and the anti-linear property of \PT, we obtain 
\be
H\mathcal{PT}|\psi_n^{\rm R}\rangle=E^*_n\mathcal{PT}|\psi_n^{\rm R}\rangle.
\label{PTHam}
\ee
(\PT)$^2=\pm 1$. For (\PT)$^2= -1$, the $|\psi_n^{\rm R}\rangle$, and $\mathcal{PT}|\psi_n^{\rm R}\rangle$ are linearly independent, and does not necessarily guarantee real eigenvalues (but if the energy is real, it gives Kramers degeneracy). On the other hand, for (\PT)$^2= +1$,  $|\psi_n^{\rm R}\rangle$ is also an eigenstate of the \PT~operator 
\be
\mathcal{PT}|\psi_n^{\rm R}\rangle = \mathfrak{p}_n|\psi_n^{\rm R}\rangle,
\label{PTeigenvalue}
\ee
where $\mathfrak{p}_n$ are the corresponding eigenvalues. Substituting Eq.~\eqref{PTeigenvalue} in Eq.~\eqref{PTHam}, we find  $H |\psi_n^{\rm R}\rangle=E^*_n|\psi_n^{\rm R}\rangle$. Combining Eq.~\eqref{PTeigenvalue} with \eqref{Righteigen}, we conclude that under (\PT)$^2= +1$ symmetry, NH Hamiltonian is guaranteed to give {\it real eigenvalues}.\cite{Bender1998,Bender2007,PToperator1,PToperator2,BenderCopp} In this case, the left and right eigenfunctions become \PT conjugate to each other, i.e., $|\psi_n^{\rm L}\rangle$, $\mathcal{PT}|\psi_n^{\rm R}\rangle$ become proportional to each other, and the inner product $\langle\psi_n^{\rm R}|\mathcal{PT}|\psi_m^{\rm R}\rangle$ is orthogonal, but {\it not yet positive, definite}.

To overcome this problem, an intrinsic, linear symmetry, say $\mathcal{C}$, is introduced such that  $\langle \psi_n^{\rm R}|\mathcal{CPT}|\psi_m^{\rm R}\rangle=\delta_{nm}$, which is called the $\mathcal{CPT}$ inner product. A canonical form of the $\mathcal{C}$-operator is $\mathcal{C}=\sum_n|\psi_n^{\rm R}\rangle \langle\mathcal{PT}\psi_n^{\rm R}|$.\cite{BenderCopp} Much like the $\eta$-operator above, $\mathcal{C}$-operator may not have any physical significance and is Hamiltonian dependent with the property that $[H,\mathcal{C}]=0$, $[\mathcal{PT},\mathcal{C}]=0$, and hence $[H,\mathcal{CPT}]=0$. Most importantly, {\it $\mathcal{C}$-operator is parameter dependent}, and helps in achieving real Berry phase. 

\section{Topological invariant with complex energy}\label{Sec:IV}

\begin{figure*}[t]
\includegraphics[width=1.9\columnwidth]{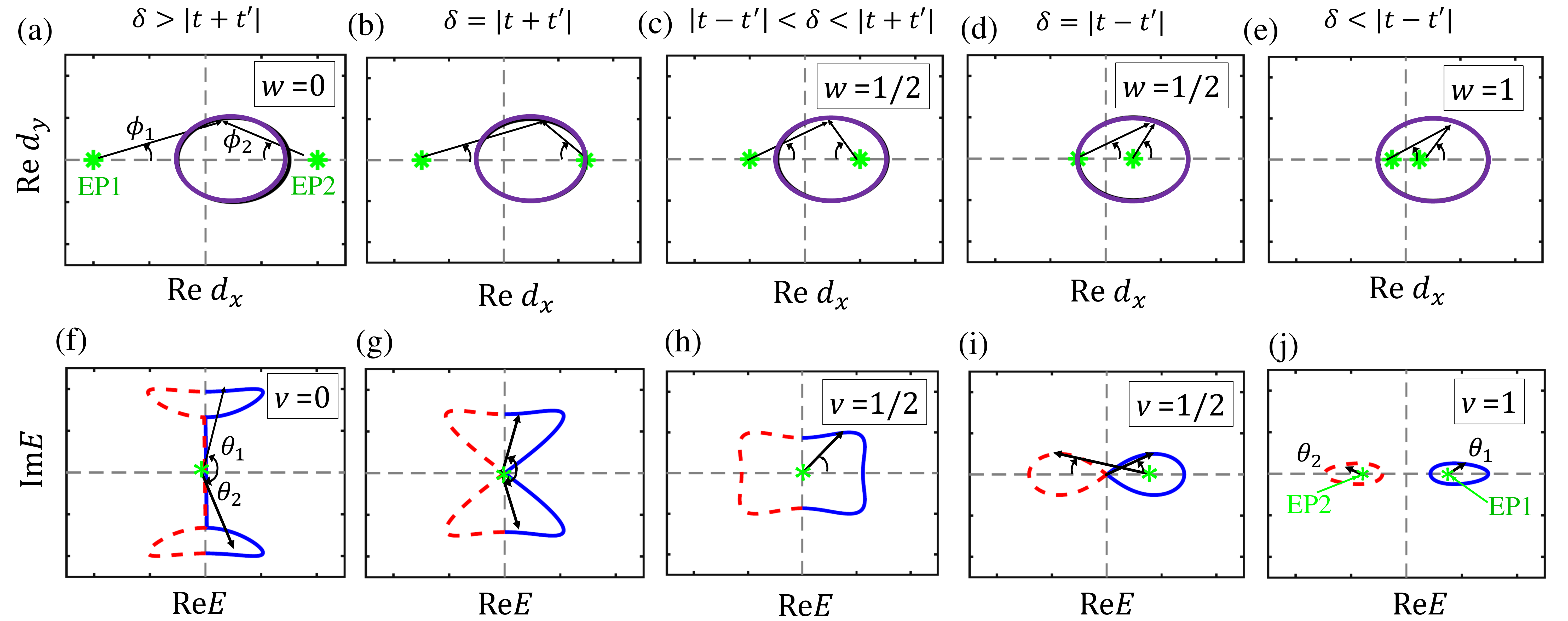} 
\caption{Evolution of eigenstate winding number $w_n$ and eigenvalue vorticity $v_n$ for a 1D 2$\times$2 NH Hamiltonian in Eq.~\eqref{SSH1}. In analogy with the corresponding Hermitian system (Fig.~\ref{Fig:Hermitian}), the winding number can be defined by a contour integral in the $({\rm Re}d_x,{\rm Re}d_y)$-plane as shown in the upper panel. Star symbols denote the locations of the EPs (Eq.~\eqref{EPs}), and circle gives the contour encircled in this plane as one completes a periodic path in the ${\bf k}$-space. We kept $|t'/t|\sim 0.5$ which gives a topological phase for the Hermitian case ($\delta=0$) in Fig.~\ref{Fig:Hermitian}. For different values of NH term $\delta$, we obtain different topological phases, determined by how many EPs lie inside the contour. (a) $\delta>|t+t'|$ gives a trivial topological insulator. (b) Topological critical point at $\delta=|t+t'|$. (c) Topological insulator with $w=1/2$ for $|t'-t|<\delta<|t+t'|$, enclosing one of the EPs. (d) The second EP enters into the contour at $\delta=|t'-t|$. (e) $w=1$ when both EPs lie inside the contour for $\delta<|t'-t|$. (f-j) For the same parameter range, the values of the energy vorticity $v$ defined with respect to the contour in the complex energy plane (Eq.~\eqref{Windingmultiband}). Solid and dashed contours are for the two PH energy states $\pm E({\bf k})$.
}
\label{Fig:Multiband}
\end{figure*}

\subsection{Complex Berry phase}\label{Sec:IVA}
The complex geometric or Berry phase for NH Hamiltonians is evaluated earlier \cite{NHBerry1,NHBerry2,NHBerryHeiss,NHBerryMailybaev,EPBerry,PTBerry,NHBerry3,Dattoli1} Following Ref.~\cite{LFu}, we can define four Berry phases for this case as
\be
\gamma^{\alpha\beta}_{n} =i\oint_{\mathcal{C}} \left\langle {\psi}_{n}^{\alpha}({\bf k}) |{\bf \nabla}_{\bf k} {\psi}_{n}^{\beta}({\bf k})\right\rangle \cdot d{\bf k},
\label{Eq:NHBerry}
\ee
where $\alpha,\beta={\rm L/R}$. ${\bf k}$ is the relevant parameter space. In general, all four components are complex. The origin of the complex Berry phase remains equivalent to those of the Hermitian one, i.e., at each EP, $\langle \psi_n^{\rm L}|\psi_n^{\rm R}\rangle=0$, and hence Eq.~\eqref{Eq:NHBerry} obtains singularity. As the contour $\mathcal{C}$ encloses an EP, one obtains a finite winding number. The total winding number is determined by the number of EPs inside the contour. The real part of $\gamma^{\alpha\beta}_n$ gives the usual geometric phase acquired in each cycle, and is a topological invariant (does not depend on the specific details of the parameters as long as number of EPs inside the contour remains the same). The corresponding imaginary part gives the decay part of the probability, and is not necessarily a topological invariant. 

It is easy to show that (see Appendix~\ref{AppendixA}) $\gamma_n^{\rm LR}=(\gamma_n^{\rm RL})^*$, and hence the total Berry phase $\gamma_{+}=\frac{1}{2}(\gamma_n^{\rm LR}+\gamma_n^{\rm RL})$ is a real number (topological invariant). Furthermore, for the diagonal Berry phase $\gamma^{\alpha\alpha}_{n}$, we can show that (see Appendix~\ref{AppendixA})  its imaginary part ${\rm Im}\left[{\gamma}_n^{\alpha\alpha}\right] = \pm \frac{1}{2}\langle  \psi_n^{\alpha}|\delta H|\psi_n^{\alpha}\rangle$, where $\delta H=H^{\dag}-H$ is the NH part of the Hamiltonian, and $\pm$ signs are for $\alpha={\rm L/R}$ states. Therefore, the imaginary part of the Berry phase comes solely from the NH part ($\delta H$) of the Hamiltonian, and as $\delta H\rightarrow 0$, the Berry phase formula coincide with that of the Hermitian one (Eq.~\eqref{HBerry}) and becomes purely real. 

Above 1D, the closed contour integral in Eq.~\eqref{Eq:NHBerry} can be converted into a surface integral, and the corresponding topological invariant is called the Chern number. Following Ref.~\cite{LFu,NHChernKawabata,NHChernKawabata,NHChernChen,NHChernYao}, we can define four Chern numbers
\begin{equation}
\mathcal{C}_{n}^{\alpha\beta}=\frac{1}{2\pi}\int_{\rm BZ}\epsilon_{\mu\nu\rho}{\bf \mathcal{B}}^{\alpha\beta}_{n,\mu\nu}  dS_{\rho}, 
\label{Chern} 
\end{equation}
where the Berry curvature and the corresponding Berry connections are defined as (suppressing the $k$-dependence for simplicity) ${\bf \mathcal{B}}^{\alpha\beta}_n = \nabla\times{\bf \mathcal{A}}^{\alpha\beta}_n$, and ${\bf \mathcal{A}}^{\alpha\beta}_{n} =  i\left\langle {\psi}_{n}^{\alpha}|\nabla {\psi}_{n}^{\beta}\right\rangle$. $dS_{\rho}$ gives the area section through which the Berry flux is being considered. Based on this definition, all four Chern numbers $\mathcal{C}^{\alpha\beta}_n$ are complex and the same.\cite{LFu}

It is easy to grasp that each EP corresponds to a (complex) pole in the Berry curvature, and thus it represents a complex monopole in the momentum space. In the case of Hermitian Hamiltonian, \PT~invariance of the system ensures that the Berry curvature at each ${\bf k}$-point vanishes. And as \PT~symmetry is broken, each Dirac cone splits in pairs, as called Weyl cones with opposite Chern numbers.\cite{WeylRMP,DasRMP,WeylRao,WeylDas} For the NH case, the \PT symmetry ensures real eigenvalues. In this case, the Berry phase and Chern number become purely real. For a pseudo-Hermitian case, when the energy eigenvalues are real, we also obtain the Berry phase and Chern number to be real. 

For symmetric Hamiltonians, i.e., $H=H^{T}$, the left and right eigenvectors are complex conjugate to each other.\cite{NHBerryMailybaev} Now since $\partial_t \langle \phi_n^{\rm R}|\phi_n0^{\rm L}\rangle ={\rm Re}[\langle \phi_n^{\rm R}|\partial_t\phi_n^{\rm L}\rangle]=0$, and hence the Berry phase, Chern number are purely {\it real}.

\subsubsection{A generic 2$\times$2 Non-Hermitian Hamiltonian}\label{Sec:IVA1}
We take the same generic 2$\times$2 Hamiltonian as in Eq.~\eqref{GenNHH}, but with all three gap terms $d_i$ are generally complex now. This NH Hamiltonian is discussed in Refs.~\cite{SSHJiang,SSHYin}. The complex eigenvalues are $E_{\pm}({\bf k})=\pm d({\bf k})$, where $d=\sqrt{\sum_{\mu=1}^3 d_{\mu}^2}$. The right eigenstate $|\psi_{n}^{\rm R}\rangle$ is same as Eq.~\eqref{genHES} while the left eigenstate is 
\be
\langle \psi_{n}^{\rm L}| =\frac{1}{\sqrt{2E_{n}(E_{n}-d_z)}}
\begin{pmatrix}
    d_x+i d_y & E_{n}-d_z \\
\end{pmatrix}.
\label{genNHLES}
\ee
The expression for the Berry connection ${\bf \mathcal{A}}^{\rm LR}_n$, and Berry curvature ${\bf \mathcal{B}}^{\rm LR}_n$ are also the same as that of the Hermitian case in Eq.~\eqref{HBerryCurv}.\cite{SSHJiang}

Clearly, $\gamma_n^{\rm LR}$ is complex and not quantized for each band. However, Liang and Huang\cite{NHBerry3} showed that the total Berry phase of the two PH pair, $\gamma^{\rm LR}_{\rm tot}=\gamma^{\rm LR}_{+}+\gamma^{\rm LR}_{-}$, is integer, and is a topological invariant. In fact, the total Berry phase/Chern number of the PH pair assumes the same form as Eq.~\eqref{HChern} [see Ref.~\cite{SSHJiang}] which was obtained for the Hermitian counterpart by including both the $d_z<0$ and $d_z>0$ regions. Therefore, $\gamma^{\rm LR}_{\rm tot}$ does not depend on the $d_z$ term, while the role of $d_z$ is to relocate the EPs across the contour to drive topological phase transition. Jiang {\it et al.}\cite{SSHJiang} considered a special unitary operator $U$ under which the Hamiltonian transforms as $UH({\bf k})U^{-1} = H^{\dag}(-{\bf k})$. This is the pseudo-Hermitian criterion discussed in Eq.~\eqref{PseudoH}, in addition to inversion ${\bf k}\rightarrow -{\bf k}$, and hence they name it pseudo-inversion symmetry. For this case, the energy eigenvalues should be either $E_{n}({\bf k})=E_n^*(-{\bf k})$  or $E_{n}({\bf k})=E_{m\ne n}^*(-{\bf k})$. In the first case, it is found that $\gamma^{\rm LR}_{n}=\big(\gamma^{\rm LR}_{m\ne n}\big)^*$, hence the net Berry phase is real, and quantized. However, for the second case of $m\ne n$, $\gamma^{\rm LR}_{n}=\big(\gamma^{\rm LR}_{n}\big)^*$, i.e., the Berry phase for each band is real but not necessarily quantized.

\subsubsection{Winding number}

We first consider the $d_z=0$ case of the Hamiltonian in Eq.~\eqref{GenNHH} with complex $d_x$ and $d_y$ terms. Here the bi-orthogonalized eigenstates are $|{\psi}_{\pm}^{\rm R}\rangle= \frac{1}{\sqrt{2}}(e^{-i\phi(k)}, \pm 1)^{T}$, and $\langle {\psi}_{\pm}^{\rm L}|= \frac{1}{\sqrt{2}}(e^{i\phi(k)}, \pm 1)$, where $\phi(k) = {\rm tan}^{-1}(d_y/d_x)$. The winding number for this case is the same as Eq.~\eqref{Hwinding} except $w_n$ is now complex, since $\phi(k)$ is complex. In the same spirit as Fig.~\ref{Fig:Hermitian}, we study the phase winding in the $({\rm Re}d_x,{\rm Re}d_y)$-space, as shown in Fig.~\ref{Fig:Multiband}. The EPs are located at the root of $d_x^2+d_y^2=0$ equation, giving two EPs (EP1, EP2) at\cite{SSHYin} 
\bea 
({\rm Re}d_x,{\rm Re}d_y)=(\mp {\rm Im}d_y,\pm {\rm Im}d_x),
\label{EPs}
\eea
In the Hermitian case, the angle $\phi (k)$ is defined with respect to the degenerate point which was at $(d_x,d_y)=(0,0)$. In the NH case, we can redefine the angle $\phi(k)$ with respect to EP1 and EP2, respectively, as
\bea 
\phi_{1,2}(k)&=& \tan^{-1}\left(\frac{{\rm Re}d_y \pm {\rm Im}d_x}{{\rm Re}d_x \mp {\rm Im}d_y}\right).
\label{phis}
\eea
Clearly, $\phi_{1,2} (k)$ are real and related to the complex $\phi(k)$ as $\tan{(2{\rm Re}\phi)} = \tan{\phi_1+\phi_2}$, leading to ${\rm Re}\phi = (\phi_1+\phi_2)/2 + n\pi$, where $n$ is an integer. Yin {\it et al.}\cite{SSHYin} showed that owing to the PH symmetry of the energy level, the imaginary part of the winding number vanishes, and the real part becomes
\bea
\label{totalwinding}
w_{\rm tot}&=&\frac{1}{2}(w_1+w_2),\\
\label{EPwinding}
{\rm where}~~~w_i &=& \frac{1}{2\pi}\oint_{\mathcal{C}}\partial_k \phi_i(k) dk,
\eea
where $i=1,2$ for the two EPs. $w_{i}$ are real and individually contribute half-integer winding number if the corresponding EP is enclosed inside the contour. In Fig.~\ref{Fig:Multiband}, we illustrate three cases where the total winding number is 0, $1/2$, and 1, as the contour encloses 0, 1, and 2 EPs. When EPs lie on the contour, the winding number becomes ill-defined, which marks the topological critical points. 

To illustrate this situation, we take a SSH model but with two different hoppings along right and left-hand dimensions:
\begin{eqnarray}
H({\bf k})=\left(
\begin{array}{cc} 
  0 & t-\delta +t'e^{-i k}\\ 
  t+\delta+t'e^{i k} & 0
\end{array} \right), 
\label{SSH1}
\end{eqnarray}
 where $t$, $t'$, and $\delta$ are real. This NH Hamiltonian is studied extensively in the literature.\cite{SSHJiang,SSHYin,SSHZhou,SSHLin,SSHPTBulkimpurity,SSHLieu,SSHYao,NHTELee,SSHElectric,SSHDiss,SSHColdAtom} The eigenvalues are PH symmetric. The two gap terms read $d_x=t+t'\cos k$ (real), and $d_y=t'\sin k-i\delta$ (complex). The two EPs are located at $({\rm Re}d_x,{\rm Re}d_y)=(\pm \delta,0)$. The first condition that $d_x$ vanishes inside the BZ requires that $|t/t'|<1$, which is also the case when the corresponding Hermitian Hamiltonian ($\delta=0$) gives a topological phase. The  boundary of the contour on the $({\rm Re}d_x,{\rm Re}d_y)$ plane is restricted between $(|t\pm t'|,0)$. This naturally restricts the value of $\delta$ to be $|t-t'|<\delta<|t+t'|$ for which the EPs will lie inside the contour. Keeping $t/t'=0.5$ fixed, we explore a representative range of $\delta$ values with different winding numbers in Fig.~\ref{Fig:Multiband}. A full phase diagram of this model is discussed in Refs.~\cite{SSHJiang,SSHYin}.

Next we introduce a gain and loss term $d_z=i\lambda$ ($\lambda$ is real) on two different lattice sites. The corresponding Hamiltonian is 
\be
H_{2}({\bf k}) = H({\bf k}) + i\lambda\sigma_z,
\label{SSHwithoutChiral}
\ee
where $H({\bf k})$ is the same as Eq.~\eqref{SSH1}. As discussed before, the winding number formula (Eq.~\eqref{EPwinding}) remains the same, while the EPs are only relocated to $({\rm Re}d_x,{\rm Re}d_y)=(\pm \sqrt{\delta^2+\lambda^2},0)$. In this case the topological phase boundary modifies to $|t-t'|<\sqrt{\delta^2+\lambda^2}<|t+t'|$. It is interesting to note that one can transform between $H_2$ and $H$ by simply tuning $\lambda$ continuously. And, if the band gap does not close in between the two Hamiltonians, and none of the EPs moves out the contour, both the topological phases are adiabatically connected, despite the loss of chiral symmetry in $H_2$. This conclusion holds for other external tuning parameters and interaction as well.\cite{NoTICYuce}

An important implication of the presence of the complex $d_z$ term here is that the Hamiltonian $H_2({\bf k})$ breaks chiral symmetry, while  $H({\bf k})$ is chiral symmetric.  The chiral operator is $\mathcal{S}=\sigma_z$ giving $\sigma_zH({\bf k})\sigma_z=-H({\bf k})$. We evaluate the edge state solution of this Hamiltonian in Sec.~\ref{Sec:VIII}, where we find that in the topologically non-trivial phase for $H({\bf k})$ and $H_2({\bf k})$, edge states exist without and with an imaginary component in the energy term, respectively. Therefore, the NH topological systems with chiral symmetry host robust edge state without any dissipation, while the chiral symmetry breaking costs dissipation of the gapless edge states $-$ a conclusion that also holds in higher dimensions.\cite{NHTELee, Leykam,LFu,NHKramersEsaki,NHChernKawabata,SSHJiang,NHDomainWall,SSHYao,CommentEdge} In fact, for such a balanced gain and loss terms ($\pm i\lambda$), one of the edge state (say, right hand edge) becomes lossy (probability/polarization attenuates) while the other edge state (say, left hand edge) becomes amplified.   

\subsection{Vorticity in the energy spectrum $-$ a new topological invariant}\label{Sec:IVB}

\begin{figure}[h]
\includegraphics[width=0.8\columnwidth]{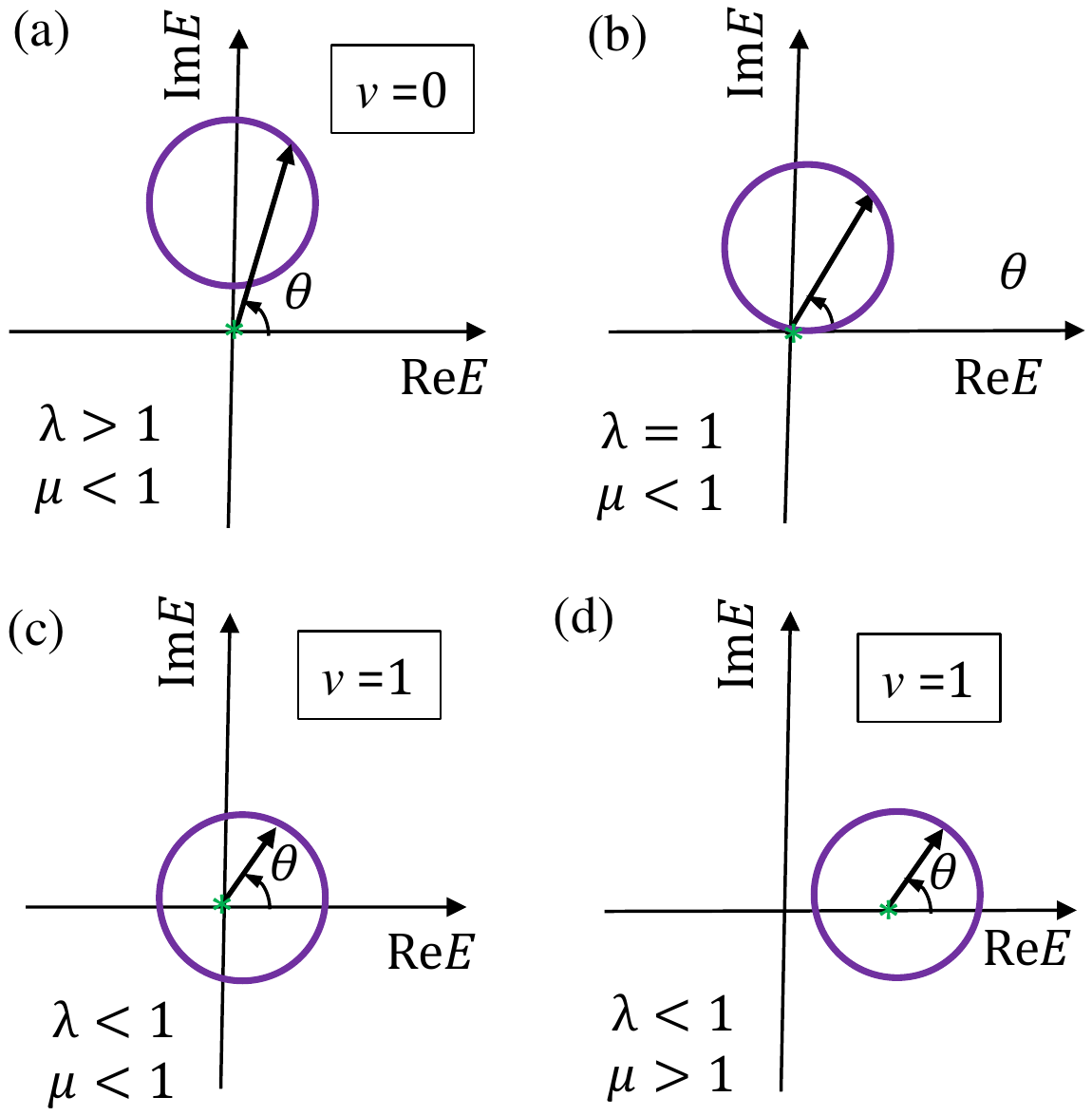} 
\caption{Contour in a single band complex energy plane (Eq.~\eqref{Eq:Singleband}) is shown for different parameter regimes as one traverses between $k=-\pi$ and $k=\pi$. We set $t=1$ for all cases. (a) For $|\lambda|>1$, the contour does not cross the ${\rm Im}E=0$ point, and hence the vorticity $v=0$, according to Eq.~\eqref{Eq:Singlevorticity}. (b) Critical point for the topological phase transition at $|\lambda|=1$. (c) For $|\lambda|<1$, the contour encloses one EP, and hence we get $v=1$. (d) As $|\mu|>1$, the contour shifts along the Re $E$ axis. In this case, we can shift the base energy to any point inside the contour on the real energy axis and obtain the same vorticity $v=1$. In this definition, $\mu$ does not cause a topological phase transition for the energy vorticity index.}
\label{Fig:Singleband}
\end{figure}

In the previous section, we mainly focused on the winding number associated with the Berry phase acquired by the eigenstates. For the case of NH Hamiltonians, the complex eigenvalues have a   unique inversion property across the EP. This can be demonstrated even for a single complex band $E(k)=|E(k)|e^{i\theta(k)}$, where $\theta=\tan^{-1}({\rm Im}E/{\rm Re}E)$. We observe that in a periodic cycle, $\theta(k)$ can evolve as $\theta(k)\rightarrow \theta(k)+2v\pi$ ($v$ being integer) without essentially violating the periodicity. $v=0$ if the energy spectrum does not enclose an EP in the real axis, see Fig.~\eqref{Fig:Singleband}(a) but is quantized as long as the contour includes an EP (Fig.~\eqref{Fig:Singleband}(c-d)). $v$ is a topological invariant. Because $v$ does not depend on any specific details of the energy, but only depends on how many EPs are present inside the contour, and one cannot change the value of $v$ unless an EP is removed from the  loop. 

In this spirit the winding number or vorticity in the complex energy plane can be defined as \cite{Leykam,NHTIclass,NHChernKawabata}
\begin{equation}
v_n=\frac{1}{2\pi}\oint_{\mathcal{C}} {\bf \nabla}_{\bf k}{\rm arg}[E_n({\bf k})]\cdot d{\bf k},
\label{Eq:Singlevorticity}
\end{equation}
$n$ is the band index. This vorticity counts the number of EPs inside the contour. Unlike the eigenstate winding number ($w_n$), the energy vorticity $v_n$ is always {\it real}, and is quantized in a periodic system.

We take an example of a single band dispersion 
\be
E(k)=te^{ik}-i\lambda-\mu,
\label{Eq:Singleband}
\ee
where $t$, $\lambda$, and $\mu$ are real. Various topological phases for this example is discussed in Fig.~\ref{Fig:Singleband}. The energy is periodic from $k=-\pi$ to $\pi$. ${\rm Im}E(k)= t\sin(k)-\lambda$ changes sign inside the BZ for $|t/\lambda|<1$, giving vorticity $v=1$. $t/\lambda=1$ is the critical point while the EP lies on the contour. Interestingly, ${\rm Re}E(k)=t\cos(k)-\mu$ technically shifts the contour in the real axis away from the $E(k)=0$ point for $|t/\mu|>1$. However, one can shift the base energy along the real axis, and redefine the angle $\theta(k)$ with respect to the new base energy (here $\mu$) to compute the vorticity, as shown in Fig.~\eqref{Fig:Singleband}(d). We note that the shift of the base energy (Dirac cone) in the real energy plane causes a band gap closing in the Hermitian topological phase and separate the two topological phases by a change in topological invariant. For energy vorticity invariant, Gong {\it et al.}\cite{NHTIclass} argues that these two cases are topologically connected. In this sense, while the parameter $\lambda$ drives a topological phase transition, topological invariants for different values of $\mu$ (keeping all other parameters unchanged) are topologically connected.\cite{NHTIclass}

\begin{figure}[t]
\includegraphics[width=0.99\columnwidth]{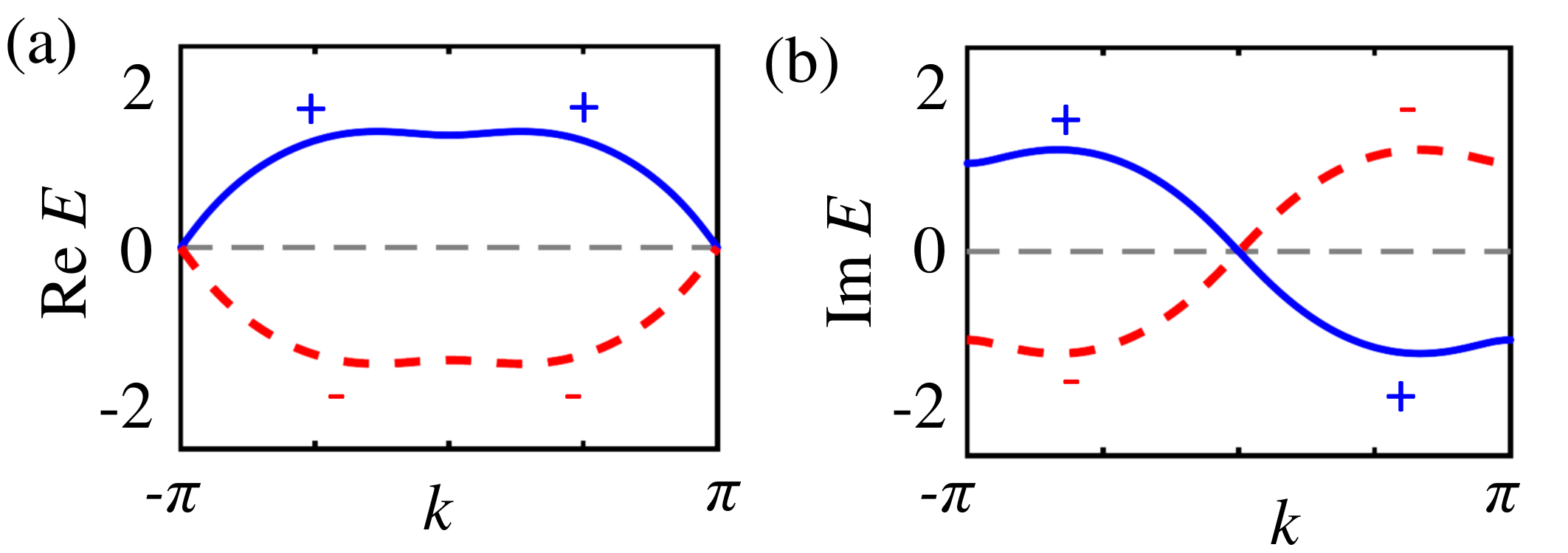} 
\caption{Band inversion property of complex energy spectrum, corresponding to Eq.~\eqref{SSH1}. Solid and dashed lines correspond to two different bands $E_{\pm}(k)=\pm d(k)$. We notice that in the topological region, (a) while the real part maintains the same sign, (b) the corresponding imaginary parts are interchanged across the EP at $k=0$.}
\label{Fig:NHBandinversion}
\end{figure}

For a complex multi-band system, when there is an EP at ${\bf k}_0$ as $E_m({\bf k}_0)=E_n({\bf k}_0)$ between two different bands, $m\ne n$, the adiabatic theory does not work any longer (adiabatic theory hold for degenerate point with real energies). In this case, as one wraps around the EP, we end up in a different band. To illustrate this situation, we consider the SSH Hamiltonian given in Eq.~\eqref{SSH1}. The real and imaginary parts of the two eigenvalues are plotted in Fig.~\ref{Fig:NHBandinversion} by solid and dashed lines from $k=-\pi$ to $\pi$ in the topological phase. The eigenstates $E_{\pm}({\bf k})=\pm d({\bf k})$ are PH symmetric, however, we observe an interesting inversion of the imaginary part of the band in the occupied state as shown in Fig.~\ref{Fig:NHBandinversion}(b). If we stick to the real part of the occupied band, its imaginary part switches band from ${\rm Im}E_{+}(k=-\pi)$ to ${\rm Im}E_{-}(k=\pi)$. This fact inspires us to define an inter-band winding number as \cite{Leykam,LFu}
\begin{eqnarray}
v_{mn} = \frac{1}{2\pi}\oint_{\mathcal{C}}{\bf \nabla}_{\bf k}{\rm arg}[E_m({\bf k})-E_n({\bf k})]\cdot d{\bf k}.
\label{Windingmultiband}
\end{eqnarray}
$v_{mn}$ is real. This formula is general to all NH multiband Hamiltonians with and without PH symmetry. When the energy spectrum is PH symmetric, energy states are paired as $E_n=-E_{-n}$ (we set $m=-n$ for the PH partner of $n$). In this case the inter-band vorticity reduces to the simple intra-band one $v_{n} = \frac{1}{2\pi}\oint_{\mathcal{C}}{\bf \nabla}_{\bf k}{\rm arg}[E_n({\bf k})]\cdot d{\bf k}$. Hence, we get $v_{n}=v_{-n}$. For such cases, ${\rm arg}[{E_n}]={\rm arg}[{E_{-n}}]=\frac{1}{2}{\rm arg}[{E_n^2}]$, and therefore the winding number of $E_{\pm n}$ becomes half of the winding number of $E^2_{\pm n}$.\cite{SSHJiang} 

In general there is no direct connection between the eigenstate winding number $w_n$ and energy vorticity $v_n$, except in few specific examples.\cite{Leykam,SSHYin} The example we presented in Eq.~\eqref{SSH1} has this property that $w_n\equiv v_n$. In Fig.~\ref{Fig:Multiband} we plot the contour in the complex energy plane for the two energies in solid and dashed lines in the corresponding lower panel. We find that for the trivial phase ($w_n=0$), the two energy contours are separated along the Im$E$ axis, and do not enclose any EP, and hence according to Eq.~\eqref{Windingmultiband} we get $v_n=0$. At the critical point of $\delta=|t-t'|$, both EPs lie on the contour. For $|t'-t|<\delta<|t+t'|$, both bands create a single contour, enclosing only one EP and hence $v_n=w_n=1/2$. Finally, for $\delta>|t+t'|$, the two contours are detached on the Re$E$ axis, and hence one can pass the EP with a constant energy shift inside the contour, and obtain a total of $v_n=1$.
 
The key reason behind the topological invariance of the energy vorticity is the same as the winding number. As long as the EPs remain inside the contour in the complex energy plane, we obtain a topological invariant. When ${\rm Re}E_n$, and ${\rm Im}E_n$ do not vanish at the same point, we obtain an `insulator' (sometimes they are referred as separable bands\cite{LFu}). All `insulators' that possess complex energy with its imaginary part changing sign within the BZ belong to the same topological class. We cannot change this topological invariant unless closing the gap through a nodal point where both ${\rm Re}E_n$, and ${\rm Im}E_n$ vanish simultaneously, i.e., unless we remove an EP outside the contour. 

\subsection{Chern number}\label{Sec:IVC}

 \begin{figure}[th]
\includegraphics[width=1.\columnwidth]{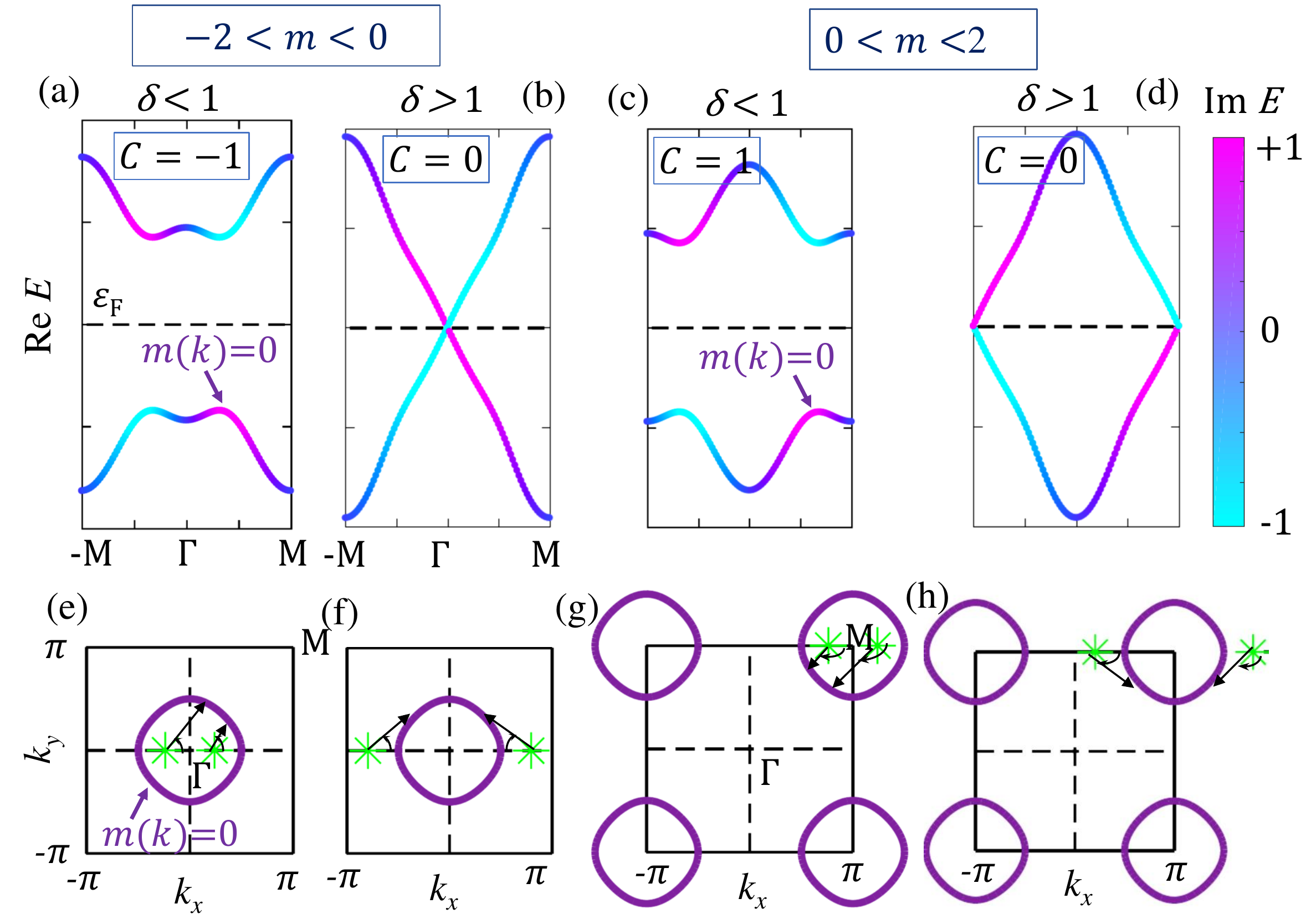} 
\caption{NH Chern insulator for the Hamiltonian in Eq.~\eqref{RiceMele1}. Top panel shows band diagram in the two regions of the topologically non-trivial phases. Cyan to magenta color gradient paints the corresponding value of Im$E$ at the same $k$-point. For $\delta<1$, EPs lie inside the nodal line of $m({\bf k})=0$ giving a non-trivial topological insulator.}
\label{Fig:Chern}
\end{figure}

Now we focus on a NH Chern insulator in 2D. Here we take a popular model for Chern insulator, namely Rice-Mele model\cite{RiceMele}. However the conclusions drawn here are general for other 2D Chern insulators, quantum spin-Hall insulators, and graphene\cite{Haldane,ZhangQSH,KaneMele1,KaneMele2,QiWuZhang,DasNC}. Yao {\it et al.}\cite{NHChernYao} considered a general case with complex potentials in all three ${\bf d}$ vector components. To ease the discussion, here we use a simplified Hamiltonian\cite{NHChernKawabata,LFu,NHChernKunst,NHChernPhilip,NHRiceMeleDQ,NHEPSLin,NHChernDiss,NHGraphene}
with NH off-diagram terms as: 
\bea
H({\bm k})=\left(
\begin{array}{cc} 
  m({\bf k}) & \sin k_x - \delta - i \sin k_y \\ 
  \sin k_x + \delta + i \sin k_y &  -m({\bf k}) 
\end{array} \right), 
\label{RiceMele1}
\eea
where $m({\bf k}) = m + \cos k_x + \cos k_y$. The components of the ${\bf d}$-vector are $d_x = \sin k_x$, $d_y=\sin k_y - i\delta$, and $d_z=m({\bf k})$, with $\delta$ and $m$ being real. The essential criterion for topological phase transition remains the same, i.e., all three gap terms must vanish inside the BZ. As in Fig.~\ref{Fig:Hermitian}(d), $d_z$ vanishes on a contour of radius $k_0$ where $k_0=\cos^{-1}(1-m)$. The EPs on the $({\rm Re}d_x, {\rm Re}d_y)$- plane must be enclosed inside the nodal contour of $d_z$ to obtain finite Chern number, while as an EP crosses outside the $d_z$ nodal contour, the Chern number becomes complex and ill-defined.  

To demonstrate the emergence of topological phase in this case, it is easier to take the long wavelength limit of Eq.~\eqref{RiceMele1}. The $d_z=0$ nodal ring encircles either the $\Gamma=(0,0)$-point for $-2<m<0$, or M = $(\pi,\pi)$-point for $0<m<2$. Hence we can expand the Hamiltonian across these two points: Around the $\Gamma$-point, keeping the linear-in-${\bf k}$ terms, we have $d_z\sim m+2$, $d_x\sim k_x$, $d_y\sim k_y-i\delta$. Similarly, around the M-point, we have $d_z\sim m-2$, $d_x\sim -k_x$, $d_y\sim -k_y-i\delta$. As a byproduct, the $({\rm Re}d_x, {\rm Re}d_y)$ plane of Fig.~2(a-e) now simply becomes the $(k_x,k_y)$-plane. Therefore, according to Eq.~\eqref{EPs}, the EPs are now placed on the ${\bf k}$-space at $(\pm \delta,0)$, and $(\pi\pm \delta,\pi)$ in the two topological regions of $|m|$, as shown in Fig.~\ref{Fig:Chern}. This provides the restriction on the topological phase transition: the EPs must lie inside the nodal contour, i.e. $\delta<\cos^{-1}(1-m)$. Therefore, following Eqs.~\ref{HChern}, and ~\eqref{EPwinding}, we can now define the Chern number in terms of four winding numbers as
\bea
C &&= \frac{{\rm sgn}[m(\Gamma)]}{2}\big(w_1(\Gamma) + w_2(\Gamma)\big) \nonumber\\
&& - \frac{{\rm sgn}[m(M)]}{2}\big(w_1(M) + w_2(M)\big),
\eea
where ${\rm sgn}[m({\bf k}_i)]$ are the sign of the Dirac mass $m({\bf k})$ at $i=\Gamma$, M, while $w_j({\bf k}_i)$ are winding numbers (Eq.~\eqref{EPwinding}) for the nodal contour with respect to the ${\bf k}_{i=\Gamma,{\rm M}}$ points. $j$ index stands for the two EPs at $(\pm \delta,0)$ for ${\bf k}_1=\Gamma$, and $(\pi\pm \delta,\pi)$ for ${\bf k}_2=$M. ${\rm sgn}[m({\bf k}_{\Gamma/{\rm M}})]={\rm sgn}[m\pm 2]$. Therefore, $d_z$ changes sign between $\Gamma$ and M-points only when $-2<m<2$, defining the non-trivial topological region. To obtain finite winding number $w_j$, the corresponding EPs must be encircled inside the nodal contour which occurs for $\delta<\cos^{-1}(1-m)$, marking another boundary for topological phase transition. In addition, for $-2<m<0$, we have ${\rm sgn}[m\pm 2]=\mp$, while in the region $0<m<2$, we have ${\rm sgn}[m\pm 2]=\pm$, with Chern number $C=\mp 1$, assuming $\delta<\cos^{-1}(1-m)$. Based on these conditions, we draw the phase diagram as shown in Fig.~\ref{Fig:Chern}. Further details of such NH Chern insulators can be found in Refs.~\cite{NHChernKawabata}. It is worthwhile mentioning that a more recent paper argues that the Chern number in NH systems may not always correspond to a Hall effect.\cite{NHChernNoHall}

\section{Symmetry protected topological invariants}\label{Sec:V}

So far, we have considered the topological invariants based on Berry phases without considering any associated symmetry properties. In fact, the above discussions are free from any symmetry properties. However, symmetry can help classify the topological phases.

In Hermitian systems, Atland and Zirnbauer (AZ) found a way to classify all non-interacting topological phases in terms of TR, charge conjugation, and chiral symmetries.\cite{AZ} The antiunitary TR symmetry is of two types with $\mathcal{T}^2=\pm 1$, if present, or 0 when absent. Similarly, the charge conjugation ($\mathfrak{C}$) operator is also an anti-unitary operator taking three values of $\pm 1$ (present) and 0 (absent). Finally the chiral symmetry (which mostly comes from the sublattice symmetry of the lattice) $\mathcal{S}=\mathcal{T}\mathfrak{C}$ is unitary, and takes two values of 1 (present) and 0 (absent). Based on these symmetries, AZ proposed that there can be ten distinct classes of topological phases belonging to either $\mathbb{Z}$ or $\mathbb{Z}_2$ homotopy groups in different dimensions. This classification scheme is often termed as {\it ten-fold way}.\cite{TenfoldKitaev,TenfoldRyu,TenfoldKane,TenfoldRMP} 

Gong {\it et al} \cite{NHTIclass} have recently classified the non-interacting NH Hamiltonians in a slightly different, and perhaps with weaker classification scheme than the AZ classification scheme employed in Hermitian Hamiltonians.\cite{AZ} This can be exemplified by comparing Fig.~\ref{Fig:Singleband}(c) and \ref{Fig:Singleband}(d). According to AZ classification scheme, these two topological phases are distinct since they are separated by a band gap closing. However, according to Gong {\it et al}\cite{NHTIclass} these two topological phases are connected. According to the latter classification scheme, there exists no non-trivial topological phase in 2D for NH systems with and without any symmetry. Finite Chern number is however obtained in 2D in many model NH Hamiltonians as discussed in Sec.~\ref{Sec:IVC}.\cite{NHChernKawabata,NHChernChen,NHChernYao}  

Dynamical classification of topological phases under quench is also presented recently.\cite{NHDynamicalClass} Furthermore, NH topological phases can also be classified with Mirror symmetry.\cite{NHTIMirror} Symmetry invariants topological phases with particular examples are also studied by various groups.\cite{NHTopUnif,NHSPTBoson,NHTIclass2,NHTIclass3} 

A unique feature of the NH Hamiltonian is the presence of Kramers degeneracy for both the TR symmetries with $\mathcal{T}^2=\pm 1$,\cite{NHKramersEsaki,NHTRKramers} and also with chiral symmetry $\mathfrak{C}^2=\pm 1$.\cite{GhatakDasNewpaper} Note that in Hermitian systems, we are aware of Kramers degeneracy only for $\mathcal{T}^2=-1$.\cite{Bernevigbook} Below we study the fate of various topological phases with TR and/or PH symmetry and/or with pseudo-Hermiticity properties.

\subsection{Topological invariants with particle-hole symmetry}\label{Sec:VA}

Note that by PH symmetry, we refer here to the existence of $\pm E_n({\bf k})$ pair in the eigenvalues. Such cases arise when Hamiltonian either has chiral  or charge conjugate symmetry or is 
pseudo-anti-Hermitian (Eq.~\eqref{PseudoH}). Below we discuss all three cases. 

\subsubsection{Chiral and charge conjugation symmetries}\label{Sec:VA1}

Chiral ($\mathcal{S}$) and charge conjugate ($\mathfrak{C}$) symmetries anticommute with the Hamiltonian, with the former being an linear operator and  the latter is an anti-linear operator. Multiplying $\mathcal{S}$ from left in Eqs.~\eqref{Righteigen}, Eqs.~\eqref{Lefteigen} and using anti-commutation relation $\{H,\mathcal{S}\}=0$, we have
\begin{subequations}
\bea
H(\mathcal{S}|\psi_n^{\rm R}\rangle)&=&-E_n\mathcal{S}|\psi_n^{\rm R}\rangle),\\
\label{ChiralH2}
H^{\dag}(\mathcal{S}|\psi_n^{\rm L}\rangle)&=&-E^*_n\mathcal{S}|\psi_n^{\rm L}\rangle).
\label{ChiralH3}
\eea
\end{subequations}
Therefore, under chiral symmetry,  the eigenstates $|\psi_n^{\rm R/L}\rangle$ states with eigenvalues $E_n$ ($E^*_n$)  come in pair with states $\mathcal{S}|\psi_n^{\rm R/L}\rangle$ with eigenvalues $-E_n$ ($-E_n^*$), respectively. 
Similarly, for charge conjugation operator $\mathfrak{C}$ with $\{H,\mathfrak{C}\}=0$, the eigenstates $|\psi_n^{\rm R/L}\rangle$ states come in pair with states $\mathfrak{C}|\psi_n^{\rm R/L}\rangle$ with eigenvalues $-E^*_n$ ($-E_n$), respectively. The following discussion remains the same for both chiral and charge conjugation symmetries. For the cases where the energy spectrum is PH symmetric, one can always find a suitable basis in which the symmetry operator ($\mathcal{S}$ or $\mathfrak{C}$) is diagonal, and then one can prove that in this new basis the Hamiltonian reduces to a block off-diagonal form as
\be
\mathcal{H}=
\begin{pmatrix}
    0            &    h_1\\
   h_2 &    0 \\
\end{pmatrix}.
\label{blockdiagonalH}
\ee
Specific relation between $h_1$, and $h_2$ depends on all the symmetries present in the system.\cite{NHTIclass} For Hermitian case $h_2=h_1^*$. In general, we can define two different winding numbers for the two reduced Hamiltonians $h_{\mu}$ with $\mu=1,2$.  In analogy with the Berry connection in Eq.~\eqref{HBerryConn2} above (with $d_z=0$), the winding number in 1D is\cite{TenfoldRyu,NHTIclass}
\be
w^{\mu}=\frac{1}{2\pi i}\oint_{C}{\rm Tr}\left[h_{\mu}^{-1}\partial_k h_{\mu}\right]dk.
\label{ChiralWinding}
\ee
$w^{\mu}$ in Eq.~\eqref{ChiralWinding} counts the number of times we wrap around $h_{\mu}(k)=0$ point in the BZ. Let us specialize to an 1D BZ where $h(k)$ is periodic in $k$ from $-\pi$ to $\pi$. As in Eq.~\eqref{Hwinding}, using polar coordinate $h_{\mu}(k)=|h_{\mu}(k)|e^{i\theta_{\mu}(k)}$, Eq.~\eqref{ChiralWinding} becomes\cite{NHKramersEsaki}
\bea
w^{\mu}_{\rm 1D}&=&\frac{1}{2\pi}\int_{-\pi}^{\pi}\frac{\partial\theta_{\mu}}{\partial k}dk +  \frac{1}{2\pi i}\int_{-\pi}^{\pi}\frac{\partial [{\rm ln}|h_{\mu}|]}{\partial k} dk\nonumber\\
&=&\frac{1}{2\pi}[\theta_{\mu}(k=\pi)-\theta_{\mu}(k=-\pi)].
\label{ChiralWinding2}
\eea
The second term disappears for the cases where $|h_{\mu}|\ne 0$, i.e., when the BZ does not contain any EP. For general NH case, $\theta_{\mu}$ is complex and hence $\omega_{\rm 1D}^{\mu}$ is also complex. Only under additional symmetry (such as anti-linear pseudo-Hermitian $\eta_+$ or \PT), one obtains purely real winding number. Owing to different symmetries, one may obtain either $\omega^{1}= \pm\omega^{2}$ or $\omega^{1}= \pm(\omega^{2})^*$. Under such circumstances, we may define purely real topological invariants such as $(\omega^{1}\pm \omega^{2})/2$ or  $(\omega^{1}\pm \omega^{2*})/2$ as appropriate. When such purely real net winding number is obtained, topological boundary states exist.\cite{NHTIclass} The calculation can be easily generalized to higher dimensions, see Sec.~\ref{Sec:VI} below.

The SSH Hamiltonian in Eq.~\eqref{SSH1} indeed has the chiral symmetry with $\mathcal{S}=\sigma_z$ which gives: $\sigma_zH(k)\sigma_z=-H(k)$. For this Hamiltonian, $h_{1,2}$-terms can be easily identified, implying that the two winding numbers, defined in Eq.~\eqref{EPwinding}, can also be obtained easily by studying the winding of the two off-block-diagonal $h_{1,2}$ Hamiltonians. Note that here $E_{n}^2=h_1h_2$, hence we obtain, ${\rm arg}[E_{n}^2]={\rm arg}[h_1h_2]$ (see the discussion of such a case below Eq.~\eqref{Windingmultiband}). This implies that for all chiral Hamiltonians, the eigenstate winding number and energy vorticity are the same. 

\subsubsection{Pseudo-anti-Hermiticity}\label{Sec:VA2}

While chiral and charge conjugation represent physical symmetries, there may exist hidden operators under which the right and left eigenvectors become related to each other. Here we consider a pseudo-anti-Hermitian case with $\eta_-$. Using Eqs.~\eqref{Lefteigen}, \eqref{PseudoH} and assuming $\eta_-$ to be a linear operator with $\eta_-^2=+1$, we obtain 
\begin{subequations}
\bea
H(\eta_-|\psi_n^{\rm L}\rangle)=-E^*_n(\eta_-|\psi_n^{\rm L}\rangle),\\
\label{PseudoantiH2}
H^{\dag}(\eta_-|\psi_n^{\rm R}\rangle)=-E_n(\eta_-|\psi_n^{\rm R}\rangle).
\label{PseudoantiH3}
\eea
\end{subequations}
Comparing these equations with Eqs.~\eqref{Righteigen}, and Eq.~\eqref{Lefteigen}, we find that $|\psi_n^{\rm R/L}\rangle$ states with eigenvalues $E_n$ ($E^*_n$)  come in pair with eigenstates $\eta|\psi_n^{\rm L/R}\rangle$ with eigenvalues $-E_n^*$ ($-E_n$). 

Unlike the case of chiral and charge conjugation operators, $\eta_-$ does not anticommute with $H$. Therefore, Eq.~\eqref{blockdiagonalH} cannot be directly applied here. We can follow the strategy of adiabatic continuity theory used for Hermitian cases in Refs.~\cite{TenfoldRyu,TenfoldRMP}, and for a NH case in Ref.~\cite{NHKramersEsaki}. Let us assume there is no EP on the contour, and that the adiabatic theorem is applicable. In what follows, the topological properties remain unchanged if we adiabatically deform all the eigenvalues and reduce them to simply $E_{\pm n}\rightarrow \pm 1$ for unoccupied and occupied states. In other words, we may project our Hamiltonian $H$ to a simplified form $\mathcal{Q}$ whose eigenvalues are $E_{\pm n} = \pm 1$ (this is called `spectral flattening'). Therefore, $H$, and $\mathcal{Q}$ have the same set of eigenfunctions $|\psi_{n}\rangle$, and share the same {\it sign} for all eigenvalues. Therefore, by definition, $H$ and $\mathcal{Q}$ are topologically connected and possess the same topological invariant. 

Following Refs.~\cite{TenfoldRyu,TenfoldRMP,NHKramersEsaki,NHTIclass} we can construct the `$\mathcal{Q}$-Hamiltonian' as
\bea
&&\mathcal{Q}=I_{n\times n}-{P}_{+}-{P}_{-},\\
&&{P}_{+} =\sum_{n>0} \big|\psi_{n}^{\rm L}\big\rangle\big\langle \psi_{n}^{\rm R}\big|, \quad {P}_{-} =\sum_{n<0} \big|\psi_{n}^{\rm R}\big\rangle\big\langle \psi_{n}^{\rm L}\big|.
\eea
${P}_{\pm}$ are called the spectral projectors for the unoccupied and occupied states, respectively, with the property that ${P}^2_{\pm}={P}_{\pm}$. $\mathcal{Q}$ is Hermitian (real eigenvalues $\pm1$), and $\mathcal{Q}^2=1$. $\mathcal{Q}$ also follows Eq.~\eqref{PseudoH}, and thus unlike $H$, it follows $\{\mathcal{Q},\eta_-\}=0$. Hence  as in Eq.~\eqref{blockdiagonalH}, $\mathcal{Q}$-matrix is block off-diagonalizable to the form
\be
\mathcal{Q}=
\begin{pmatrix}
    0            &    q\\
   q^{\dag} &    0 \\
\end{pmatrix},
\label{blockdiagonalQ}
\ee
where $q$ is a $n\times n$ Unitary matrix. Now we can use Eq.~\eqref{ChiralWinding} to define the winding number in terms of $q$-matrix. Restricting to an 1D BZ where $q(k)$ is periodic in $k$, we define
\bea
\label{PseudoAntiHWinding1}
w_{\rm 1D}&=&\frac{1}{2\pi i}\int_{-\pi}^{\pi} {\rm Tr}\left[q^{-1}\partial_k q\right] dk,\\
&=&\frac{1}{2\pi}\int_{-\pi}^{\pi} \partial_k \theta (k) dk=N.
\label{PseudoAntiHWinding2}
\eea
In the second equation, we have used $q(k)=|q(k)|e^{i\theta (k)}$. $N\in \mathbb{Z}$. Since $q$ is Hermitian, $\theta$ is real here and thus one obtains purely real (single) winding number. The calculation can be easily generalized to higher dimensions (see Sec.~\ref{Sec:VI} below). 

Let us consider an example for this case.\cite{NHKramersEsaki} We take a different SSH model with gain and loss terms $\pm i\lambda$ as 
\begin{eqnarray}
H({\bm k})=\left(
\begin{array}{cc} 
  i\lambda & t+t'e^{-i k}\\ 
  t+t'e^{i k} & -i\lambda
\end{array} \right), 
\label{su11}
\end{eqnarray}
where $t$, $t'$ and $\lambda$ are real. $\pm i\lambda$ term makes the Hamiltonian looses chiral symmetry, but it is pseudo-anti-Hermitian with {\it anti-linear} metric $\eta_-=\sigma_z\mathcal{K}$. Here $d_x=t+t'\cos k $, $d_y=t'\sin k$ are real, while $d_z=i\lambda$ is purely imaginary. The PH symmetric eigenvalues are $E_{\pm}(k)=\pm d(k)$, where $d=\pm\sqrt{d_x^2+d_y^2 -\lambda^2}$. Here $q({\bm k})=(d_x+id_y)/d$, with its argument $\theta(k)=\tan^{-1}{d_y/d_x}$. Now following Eq.~\eqref{PseudoAntiHWinding2}, we obtain the same topological phase diagram as the Hermitian SSH model gives, i.e. $w_{\rm 1D}=1$ for $|t/t'|<1$, and zero otherwise.

\subsection{Topological invariants with Kramers degeneracy}\label{Sec:VB}

For an antilinear operator, even when it commutes with the Hamiltonian, both the Hamiltonian and the operator may or may not share the same eigenfunction. TR symmetry for spinless fermions with $\mathcal{T}^2=+1$ shares the same eigenfunction with the Hamiltonian. On the other hand, for half-integer spin fermions, TR operator gives $\mathcal{T}^2=-1$. In this case, the Kramers theory implies that all eigenstates $|\psi_n\rangle$, and its TR conjugate $\mathcal{T}|\psi_n\rangle$ are degenerate, and orthogonal, as known by Kramers degeneracy.\cite{KramersTheory} For the antiunitary charge conjugation operator $\mathfrak{C}$ anticommutes with the Hamiltonian, giving PH symmetric energy spectrum, the Kramers degeneracy is not generally applicable here, unless at zero energy. 

This Kramers theory is not generally applicable to EPs, since here the eigenfunctions coalesce. However, especially for the case of pseudo-Hermitian and pseudo-anti-Hermitian Hamiltonians, new forms of Kramers degeneracy can commence for the TR symmetry with both $\mathcal{T}^2=\pm 1,  $\cite{NHKramersEsaki,NHTRKramers} as well as for the charge conjugation symmetry {\bf $\mathfrak{C}^2=\pm 1$}.\cite{GhatakDasNewpaper} In analogy with the quantum spin-Hall insulator in Hermitian Hamiltonians,\cite{ZhangQSH,ZhangQSH2,KaneMele1,KaneMele2}  here also new TR polarization, and/or spin Chern number can be defined for such Kramers degenerate systems.

\subsubsection{Kramers degeneracy under time-reversal symmetry}\label{Sec:VB1} 
Let us consider a pseudo-Hermitian Hamiltonian as discussed in Eq.~\eqref{PseudoH} under a linear or anti-linear metric $\eta_+$. We also assume the case where the pseudo-Hermitian Hamiltonian is TR symmetric with $[H,\mathcal{T}]=0$, and in addition, also has one of the following properties:
\begin{subequations}
\bea
\label{PseudoHCond1}
{\rm either}~~&&\{\eta_+,\mathcal{T}\}=0,~{\rm and}~\mathcal{T}^2=+1,\\
{\rm or}~~&&[\eta_+,\mathcal{T}]=0,~{\rm and}~\mathcal{T}^2=-1.
\label{PseudoHCond2}
\eea
\end{subequations}
Applying $\mathcal{T}$ from left on Eq.~\eqref{PseudoH2} and using $[H,\mathcal{T}]=0$ we arrive at
\be
H(\mathcal{T}\eta_+^{-1}|\psi_n^{\rm L}\rangle)=E_n(\mathcal{T}\eta_+^{-1}|\psi_n^{\rm L}\rangle).
\label{TRKramersH2}
\ee
Comparing Eq.~\eqref{TRKramersH2} with Eq.~\eqref{Righteigen}, we find that $\mathcal{T}\eta_+^{-1}|\psi_n^{\rm L}\rangle$, and $|\psi_n^{\rm R}\rangle$ have the same eigenvalue $E_n$. With either Eq.~\eqref{PseudoHCond1} or Eq.~\eqref{PseudoHCond2}, we can easily show that $\mathcal{T}\eta^{-1}|\psi_n^{\rm L}\rangle$, and $|\psi_n^{\rm R}\rangle$ are {\it linearly independent} (see Eq.~\eqref{AntiLinPseudoHKramers} for a similar proof). Hence under these conditions, the Hamiltonian $H$ has {\it two-fold degeneracy}.\cite{NHKramersEsaki,NHTRKramers} 

\subsubsection{Kramers degeneracy under charge conjugation symmetry}\label{Sec:VB2}
Let us now consider another unique case of an {\it anti-linear} pseudo-anti-Hermitian Hamiltonian Eq.~\eqref{PseudoH} with {\it anti-linear} metric $\eta_-$. Using Eqs.~\eqref{Lefteigen}, \eqref{PseudoH} and the anti-linear property of $\eta_-$, we obtain 
\be
H(\eta_-^{-1}|\psi_n^{\rm L}\rangle)=-E_n(\eta_-^{-1}|\psi_n^{\rm L}\rangle).
\label{PseudoantiH2}
\ee
Next we specialize to the Hamiltonians which are invariant under the charge conjugation symmetry $\mathfrak{C}$ with $\{H,\mathfrak{C}\}=0$, and has one of the following properties
\begin{subequations}
\bea
\label{PseudoAntiHCond1}
{\rm either},~~&&\{\eta_-,\mathfrak{C}\}=0,~{\rm and}~\mathfrak{C}^2=+1,\\
{\rm or},~~&&[\eta_-,\mathfrak{C}]=0,~{\rm and}~\mathfrak{C}^2=-1.
\label{PseudoAntiHCond2}
\eea
\end{subequations}
Applying $\mathfrak{C}$ from left in Eq.~\eqref{PseudoantiH2} and using the anticommutation relation with $H$ we get
\be
H(\mathfrak{C}\eta_-^{-1}|\psi_n^{\rm L}\rangle)=E_n(\mathfrak{C}\eta_-^{-1}|\psi_n^{\rm L}\rangle).
\label{PseudoantiH3}
\ee
Comparing Eqs.~\eqref{PseudoantiH2}, and \eqref{Righteigen}, we find that $\mathcal{C}\eta_-^{-1}|\psi_n^{\rm L}\rangle$, and $|\psi_n^{\rm R}\rangle$ have the same eigenvalue $E_n$. In addition, we have
\bea
\label{AntiLinPseudoHKramers}
\langle \psi_n^{\alpha} |\mathfrak{C}\eta_-^{-1}|\psi_n^{\alpha}\rangle&=&\langle \mathfrak{C}^2\eta_-^{-1}\psi_n^{\alpha} |\mathfrak{C}\psi_n^{\alpha}\rangle \\
&=& \pm \langle\psi_n^{\alpha} |\eta^{-1}\mathfrak{C}|\psi_n^{\alpha}\rangle =-\langle \psi_n^{\alpha}|\mathfrak{C}\eta^{-1}|\psi_n^{\alpha}\rangle,\nonumber
\eea
for $\alpha={\rm R/L}$. In the first equation, we have used the antiunitary property of $\mathfrak{C}$ that $\langle \phi|\psi\rangle =\langle \mathfrak{C}\psi|\mathfrak{C}\phi\rangle$ for any two states $|\phi\rangle$ and $|\psi\rangle$. We consider $\eta_-$ to be Hermitian. In the second equation $\pm$ signs refer to $\mathfrak{C}^2=\pm 1$, which with either Eq.~\eqref{PseudoAntiHCond1} or Eq.~\eqref{PseudoAntiHCond2}, respectively, yields the third equation. Therefore,  $\langle \psi_n^{\alpha} |\mathfrak{C}\eta_-^{-1}|\psi_n^{\alpha}\rangle=0$, i.e.,  $\mathfrak{C}\eta_-^{-1}|\psi_n^{\alpha}\rangle$, and $|\psi_n^{\alpha}\rangle$ are {\it linearly independent}.  Hence under the conditions of either Eq.~\eqref{PseudoAntiHCond1} or \eqref{PseudoAntiHCond2}, the Hamiltonian $H$ has {\it two-fold degeneracy}.\cite{GhatakDasNewpaper} 

\subsubsection{Split-quaternion}\label{Sec:VB3}
If we interchange the dual conditions between Eqs.~\eqref{PseudoAntiHCond1} and \eqref{PseudoAntiHCond2}, we obtain a PH energy spectrum $\pm E_n({\bf k})$ with linearly independent eigenspectrum $-$ split quaternion, rather than a degeneracy.\cite{NHTRKramers,GhatakDasNewpaper} If we consider a linear pseudo-anti-Hermitian Hamiltonian with metric $\eta_-$, which is TR symmetric, and follows either Eq.~\eqref{PseudoHCond1} or \eqref{PseudoHCond2}, then $\mathcal{T}\eta_-^{-1}|\psi_n^{\rm L}\rangle$, and $|\psi_n^{\rm R}\rangle$ are {\it linearly independent} and {\it possess eigenvalues $\pm E_n$}, respectively. Similarly, if we consider a pseudo-Hermitian Hamiltonian with metric $\eta_+$, which is invariant under $\mathfrak{C}$, and follows either Eq.~\eqref{PseudoAntiHCond1} or \eqref{PseudoAntiHCond2}, then $\mathfrak{C}\eta_+^{-1}|\psi_n^{\rm L}\rangle$, and $|\psi_n^{\rm R}\rangle$ also become split quaternion.

\subsubsection{$\mathbb{Z}_2$ topological invariants} \label{Sec:VB4}

From the Hermitian counterparts,\cite{KaneMele1,KaneMele2,FuKanePRL,FuKanePRB,ZhangQSH,ZhangQSH2,AZ,TenfoldRMP} we have learned that Kramers degeneracy plays an important role in modifying the $\mathbb{Z}$-symmetry of the topological invariants (such as integer winding number, integer Chern number) to $\mathbb{Z}_2$ symmetry where the topological invariant only takes 0 (trivial) and 1 (non-trivial) value. For NH systems with Kramers degeneracy, such a $\mathbb{Z}_2$ symmetry can also emerge.\cite{NHKramersEsaki,NHTRKramers,NHTIclass,NHSPTBoson} 

Let us first focus on an {\it anti-linear} pseudo-Hermitian Hamiltonian (Eq.~\eqref{PseudoH2}) with $\eta_+$ metric, which possess real energy eigenvalues. Owing to real eigenvalues, Eq.~\eqref{PseudoHcond2} is also satisfied with $c_n=\pm 1$ for each state $n$. [For a linear pseudo-Hermitian Hamiltonian where the eigenvalues are not real, one can still proceed by shifting the eigenstates of $H$ to the basis where they follow Eq.~\eqref{PseudoHcond2} with $c_n=\pm 1$.\cite{NHKramersEsaki,NHTRKramers}] We denote these two states as $|\psi_n^{\alpha,\pm}\rangle$ for $\alpha={\rm L/R}$, where $\pm$ corresponds to $c_n=\pm 1$. We also assume that the system is TR symmetric with either Eqs.~\eqref{PseudoHCond1}, or Eq.~\eqref{PseudoHCond2}, such that $\mathcal{T}\eta_+^{-1}|\psi_n^{\rm L,\pm}\rangle$, and $|\psi_n^{\rm R,\pm}\rangle$ are Kramers partners. They are related to each other as follows:\cite{NHKramersEsaki}
\be
|\psi_n^{\alpha,-}({\bf k})\rangle=e^{i\theta_n({\bf k})}\mathcal{T}\eta_+^{-1}|\psi_n^{\beta,+}(-{\bf k})\rangle,
\label{Kramerspair}
\ee
for $\alpha \ne \beta={\rm L/R}$. $\theta_n$ is complex in general. Hence similar to the Hermitian case in Eq.~\eqref{HTRpolarization}, we can define the Berry connection separately for the two Kramers pairs: ${\bf \mathcal{A}}^{\mu}_n({\bf k})=-i\left\langle \psi^{\mu}_{n}({\bf k})|{\bf \nabla}_{\bf k} \psi^{\mu}_{n}({\bf k})\right\rangle$, with $\mu=\pm$ denoting the two Kramers partners. We can then similarly define the TR polarization as 
\be
P^{\mathcal{T}}_n = \frac{1}{2\pi}\oint \left[{\bf \mathcal{A}}^{+}_n -  {\bf \mathcal{A}}^{-}_n\right]\cdot d{\bf k},
\label{NHTRpolarization}
\ee
Unless in special cases (such as quantum spin-Hall insulators), the polarization of each component of the Kramers pair is not quantized, but their difference is quantized with values either 0 or 1 for topologically trivial and non-trivial phases, respectively. Thus such systems have the $\mathbb{Z}_2$-symmetry. 

In 2D, where the global inversion symmetry is absent, each term on the right hand side of Eq.~\eqref{NHTRpolarization} corresponds to a Chern number. Hence the TR polarization is called the TR invariant Chern number $C^{\mathcal{T}}=(C^+ - C^-)/2$, where $C^{\pm}$ corresponds to first and second terms on the right hand side of Eq.~\eqref{NHTRpolarization}. Due to TR symmetry, $C^{\pm}\rightarrow -C^{\mp}$ and hence the charge Chern number $(C^+ + C^-)$ vanishes. 

For Hermitian Hamiltonians, Kane-Mele found an elegant method to compute $P^{\mathcal{T}}_n$ by generalizing Eq.~\eqref{NHTRpolarization} in terms of Pfaffian of the matrix with components made of the inner products of two Kramers partners.\cite{KaneMele1,KaneMele2} In fact for systems with both TR and parity (both symmetries are individually present), they found that the total TR polarization can be evaluated by the product of the parity eigenvalues at the TR invariants ${\bf k}$-points.\cite{FuKanePRL,FuKanePRB} Kawabata {\it et al.}\cite{NHTopUnif} argued that such a framework also works for NH Hamiltonians with Kramers degeneracy. In such cases, the $\mathbb{Z}_2$ invariant is defined by
\be
(-1)^{P^{\mathcal{T}}} = \prod_{i=1}^I \prod_{E_n<E_F} p_n({\bf k}_i),
\label{NHTRparity}
\ee
where $p({\bf k}_i)$ is the parity eigenvalue at the TR ${\bf k}$-points ${\bf k}_i$ for the filled bands, and $I$ is the total number of TR invariant ${\bf k}$-points. Therefore, the system possesses a non-trivial topological invariant if its filled bands possess odd number of parity inversions. 

Some of the famous examples of the TR invariant $\mathbb{Z}_2$ symmetric 2D topological insulators for Hermitian Hamiltonians include Kane-Mele model of graphene with spin-orbit coupling,\cite{KaneMele1,KaneMele2} and Bernevig-Hughes-Zhang's  model for HgTe/CdTe quantum well states.\cite{ZhangQSH} These models have been generalized in the literature with complex spin-orbit coupling and/or hopping terms to study the robustness of the corresponding TR invariant topological index in the NH counterparts.\cite{NHKramersEsaki,NHTRKramers,NHTopUnif}

\subsubsection{Examples}\label{Sec:VC}

\begin{figure}[t]
\includegraphics[width=1.\columnwidth]{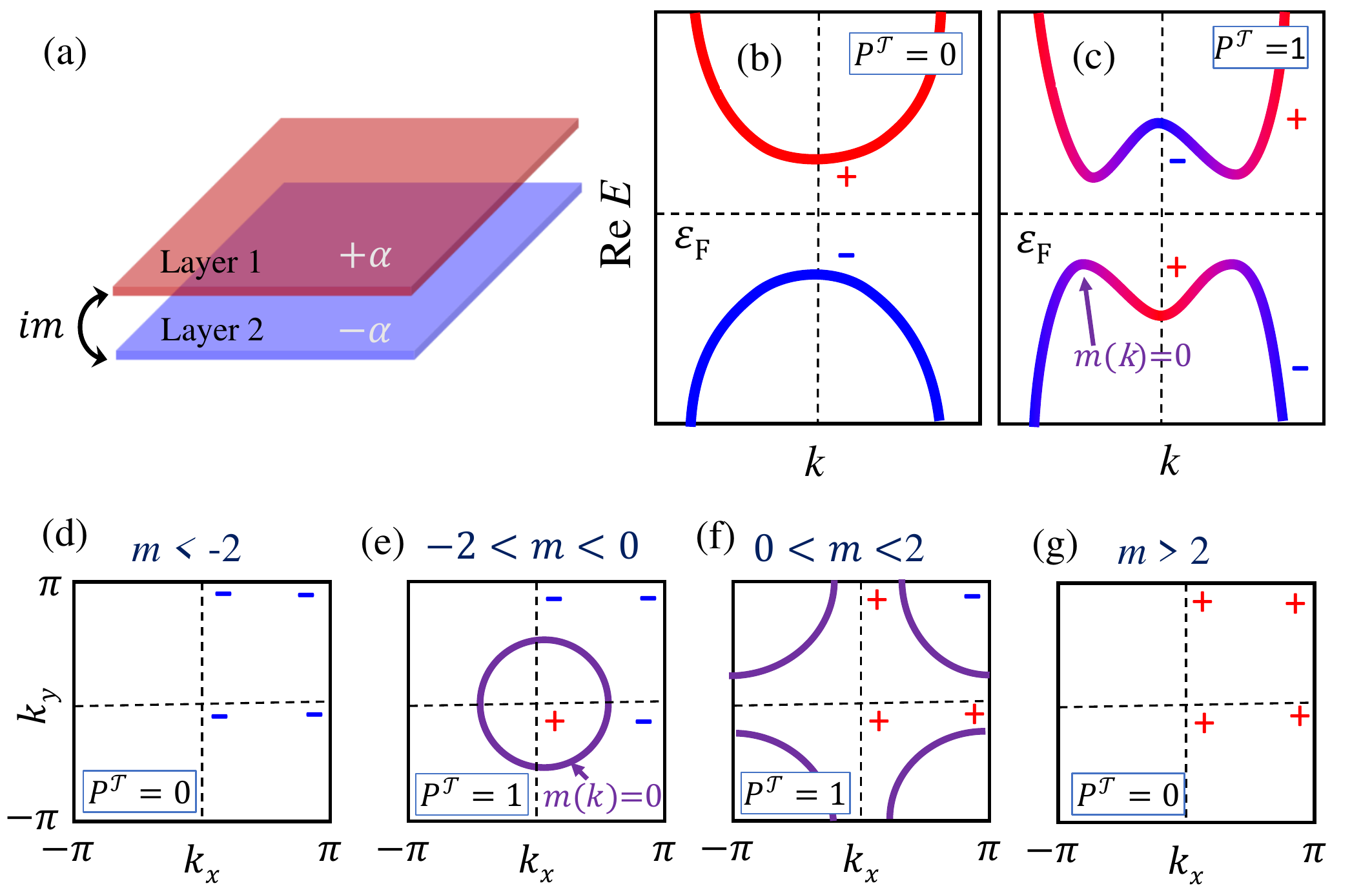} 
\caption{(a) Set of a Rashba bilayer with opposite Rashba spin-orbit coupling [$\pm\alpha({\bf k})$] and with NH complex hopping $im({\bf k})$ between the two layers. The model demonstrates a TR invariant 2D topological insulator (Eq.~\eqref{NHRashbabilayer}). (b) Schematic band diagram for a trivial insulator. (c) Band diagram of a non-trivial phase with red and blue color dictating opposite parity eigenvalues at a given ${\bf k}$-point. (d) \& (g) Parity profile of the valence band for the trivial phase when $|m|>2$ having the same parity at all TR invariant $k$-points. (e) \& (f) Parity profile of the two non-trivial phases, where the parity is inverted across the $m({\bf k})=0$ nodal line giving TR polarization $\mathcal{P}^{\mathcal{T}}=1$ while the TR invariant Chern number (if defined) is $C^{\mathcal{T}}=\pm 1$, respectively.}
\label{Fig:NHRashBabilayer}
\end{figure}

Here we take a multifaceted example, where the symmetry protected $\mathbb{Z}_2$ topological index can be identified through various methods proposed above. In Hermitian topological systems, one of the present author introduced the idea of helicity inversion as a driving principle for topological insulators in various dimensions.\cite{DasNC,QSHDW,SRayOL,SU3} We generalize this idea here where we engineer non-Hermiticity via adopting imaginary hopping between counter helical states. The setup is shown in Fig.~\ref{Fig:NHRashBabilayer}(a). We consider a bilayer of 2D electron gases with opposite Rashba-type spin-orbit coupling (SOC) in the adjacent layers such as $\pm\alpha({\bf k})$ where $\alpha({\bf k})=\alpha_{\rm R}(\sin k_y -i\sin k_x)$, where $\alpha_{\rm R}$ is the spin-orbit coupling (SOC) strength (real). The coupling between the two Rashba-layers are taken to the purely imaginary, and momentum-dependent as $im({\bf k})$ with $m({\bf k})=m + \cos k_x + \cos k_y$, where $m$ is constant and real. The Hamiltonian is expressed in a basis of two spins ($\uparrow,\downarrow$) and two layers (${\rm L_{1,2}}$) as   
\be
H({\bf k})=
\begin{pmatrix}
    0                    &    \alpha({\bf k}) & im({\bf k}) & 0\\
    \alpha^*({\bf k}) &    0   & 0 & im({\bf k}) \\
    im({\bf k})   &  0 & 0 & -\alpha({\bf k}) \\
    0 & im({\bf k}) & -\alpha^*({\bf k}) & 0
\end{pmatrix},
\label{NHRashbabilayer}
\ee
In the Hermitian case where $im({\bf k})$ is purely real, this Hamiltonian gives a strong TR invariant topological insulator for $|m|<2$.\cite{DasNC} To aid further discussion, we express the Hamiltonian in terms of $4\times 4$ Dirac matrices as
\be
H= {\bf d}\cdot{\bf \Gamma},
\label{NHRashbabilayer2}
\ee
where $\Gamma_{x,y,z}$ = ($\tau_z\otimes \sigma_x$,  $\tau_z\otimes \sigma_y$, $\tau_z\otimes \sigma_x$) with $\sigma_i$, and $\tau_i$ are the $2\times 2$ Pauli matrices in the spin and layer basis, respectively. Corresponding components of the ${\bf d}$-vector are $d_{x,y,z}$ = ($\alpha_{\rm R}\sin k_y$,  $\alpha_{\rm R}\sin k_x$,$im({\bf k})$). In the NH case, the Hamiltonian is symmetric under two TR operators $\mathcal{T}_{\pm}^2=\pm 1$, parity $\mathcal{P}$, as well as chiral $\mathcal{S}$ symmetries. The two TR symmetries are $\mathcal{T}_{\pm}=\mathcal{U}_{\pm}\mathcal{K}$, where $\mathcal{U}_-=i(I\otimes \sigma_y)$, and $\mathcal{U}_+=i(\tau_y\otimes\sigma_y)$, and $\mathcal{K}$ is the complex conjugation. Parity and chiral operators are $\mathcal{P}=\Gamma_z$, and $\mathcal{S}=i\tau_y\otimes \sigma_x$. In addition, the Hamiltonian is also pseudo-Hermitian with metric $\eta_{+}=\tau_y\otimes \sigma_y$. Under these operators the  NH Hamiltonian transforms as
\begin{subequations}
\bea
\label{NHExmTR}
\mathcal{U}_{\pm}H({\bf k})\mathcal{U}_{\pm}^{-1}&=&H^*(-{\bf k}),\\
\label{NHExmP}
\mathcal{P}H({\bf k})\mathcal{P}^{-1}&=&H(-{\bf k}),\\
\label{NHExmC}
\mathcal{S}H({\bf k})\mathcal{S}^{-1}&=&-H(-{\bf k}),\\
\label{NHExmeta}
\eta_+H^{\dag}({\bf k})\eta_+^{-1}&=&H({\bf k}).
\eea
\end{subequations}
The most striking part of the Hamiltonian is that we have $\{\eta_+,\mathcal{T}_+\}=0$ (Eq.~\eqref{PseudoHCond1}) which gives the Kramers degeneracy according to Sec.~\ref{Sec:VB1}. Hence we can define the TR polarization according to Eq.~\eqref{NHTRpolarization}. In fact, owing to the parity of this system, we can also employ Eq.~\eqref{NHTRparity}. Since parity operator is $\mathcal{P}=\Gamma_z$, the parity eigenvalue at a given ${\bf k}$ values is determined by the sign of $d_z=im({\bf k})$, and the TR invariant topological index is determined by
\be
(-1)^{P^{\mathcal{T}}} = \prod_{i=1}^I {\rm sgn}[m({{\bf k}_i})],
\label{NHExmPol}
\ee
where the four TR invariant ${\bf k}$-points are $(0,0)$, $(\pi,0)$, $(0,\pi)$, $(\pi,\pi)$. $m({\bf k})$ is even under parity, and thus possess a nodal ring in the 2D BZ for $|m|<2$, as shown in Fig.~\ref{Fig:NHRashBabilayer}(e-f). For $m<-2$, parity $p({\bf k}_i)=-1$ at all TR invariant ${\bf k}_i$-points, and thus according to Eq.~\eqref{NHExmPol}, we get $P^{\mathcal{T}}=0$. For $-2<m<0$, parity $p(0,0)=+1$, while all the TR invariant ${\bf k}$-points have $p=-1$, giving $P^{\mathcal{T}}=1$. Interestingly, for $0<m<2$, all the parity eigenvalues are inverted to $p(\pi,\pi)=-1$, and the rest are +1, yet we get $P^{\mathcal{T}}=1$. Finally, for $m>2$, we have all ${\bf k}_i$ have $p=+1$, and we get $P^{\mathcal{T}}=0$. Therefore, the topological invariant does not depend on individual parity, but odd-number of party inversions, and gives a $\mathbb{Z}_2$ symmetry.

Interestingly, the Hamiltonian in Eq.~\eqref{NHRashbabilayer} does not exhibit a finite any TR Chern number in this present form. However, with a suitable choice of basis, we can have the two Kramers state to be a good quantum number of the system as discussed in Sec.~\ref{Sec:VB4}.  We find that with a unitary transformation $U=I_{4\times 4} + i\tau_y\otimes \sigma_z$, we obtain a block diagonal Hamiltonian 
\be
UHU^{-1}=
\begin{pmatrix}
    h({\bf k})                    &    0\\
    0 &   h^*(-{\bf k}) 
\end{pmatrix},
\label{NHRashbabilayerBlockDiag}
\ee
where $h({\bf k}) = {\bf d}\cdot{\bf {\sigma}}$ with all $d_{\mu}$ components have the same expressions as before. $h({\bf k})$ breaks both TR and parity, but with its TR  conjugate $h^*(-{\bf k})$ on the other block, the full Hamiltonian $H$ becomes TR and parity invariant. Interestingly, $h({\bf k})$ now has the same form as Eq.~\eqref{GenNHH} with $d_z$ being complex. Therefore, we can compute the Chern number for each block and define the TR invariant Chern number. Following Eq.~\eqref{HChern}, we can split the BZ into two regions with $d_z<0$, and $d_z>0$ and define the Chern number for each states in terms of the winding number of the angle $\phi({\bf k})=\tan^{-1}(d_y/d_x)$. The Chern numbers for the upper/lower block are\cite{Bernevigbook} 
\be
C_{\pm} = \pm \frac{1}{2}\Big[{\rm sgn}[m_{\Gamma}] - {\rm sgn}[m_{\rm M}]\Big],
\ee
where $\Gamma=(0,0)$, and M = $(\pi,\pi)$. Clearly, for the same reason as before, we get $C_{\pm}=\pm 1$ in the region of $-2<m<0$, and $C_{\pm}=\mp 1$ for $0<m<2$, and $C_{\pm}=0$ otherwise. This gives the total Chern number $C_{\rm tot}=C_++C_-=0$ for all values of $m$, but their difference $C^\mathcal{T}=(C_+-C_-)/2=\pm 1$ in non-trivial topological region. Comparing $C^\mathcal{T}$  with the TR polarization $P^{\mathcal{T}}$ obtained above we find $P^{\mathcal{T}}\equiv|C^{\mathcal{T}}|$. Therefore, in the presence of Kramers pair, the TR invariant Chern number also obtains a $\mathbb{Z}_2$ invariant, while for TR breaking case Chern number can take any integer value (if quantized) and hence has the $\mathbb{Z}$ symmetry.\cite{NHTIclass}

\section{Higher dimensions}\label{Sec:VI}

\subsection{3D topological insulators}

As much work have been done in 1D and 2D NH systems, the literature on 3D systems is considerably less. In principle, the winding number and TR polarizations defined in 1D above can be easily generalized to three and higher odd dimensions. In 3D, the topological invariant belonging to the $\mathbb{Z}$ class is the 3D extension of the winding number (Eq.~\eqref{PseudoAntiHWinding1})\cite{NHTIclass}
\be
w_{\rm 3D} = \frac{1}{24\pi^2}\oint_{\mathcal{C}}\epsilon_{\mu\nu\rho}{\rm Tr}\big[(q^{-1}\partial_{\mu}q)(q^{-1}\partial_{\nu}q)(q^{-1}\partial_{\rho}q)\big]d^3k,
\label{NH3Dwinding}
\ee
where $q$- is the block off-diagonal matrix of the full $\mathcal{Q}-$matrix as in Eq.~\eqref{blockdiagonalQ} if the Hamiltonian is pseudo-anti-Hermitian, or simply the block off-diagonal matrix of the Chiral Hamiltonian as in Eq.~\eqref{blockdiagonalH}. As it was pointed out by Gong {\it et. al.}\cite{NHTIclass}, if any two components of the $q^{-1}\partial_{\mu}q$ term commute, the winding number vanishes. 

As done by Fu and Kane for the case of Hermitian TR invariant topological insulator,\cite{FuKanePRB} we shall be able to extend the calculations of the TR topological invariant  of Eq.~\eqref{NHTRpolarization} to the 3D case, and we may anticipate to obtain the TR polarization from the parity eigenvalues (Eq.~\eqref{NHTRparity}) at all $I=8$ TR invariant ${\bf k}$-points. 

Starting with a generic Hamiltonian written in terms of the gap vectors ${\bf d}({\bf k})$ in 3D as in Eq.~\eqref{NHRashbabilayer2}, we can easily generalize the formula for 3D winding number to five components ${\bf d}$-vector. In the case when the corresponding five $\Gamma$ matrices follow Clifford algebra, the eigenvalues are PH symmetric $E_{\pm}({\bf k})=\pm d$, where $d=\sqrt{\sum_{i=1}^5 d_i^2}$ with $d_i$ being complex in general. Clearly, EPs are located at the ${\bf k}$-points where all the gap terms simultaneously vanish. In analogy with the 1D and 2D cases discussed above, when discrete EPs are present in the BZ, they mark the critical point for topological phase transition. Across this critical point, we have a trivial to non-trivial topological phase transition. For $2\times 2$ Hamiltonians with three $d$-components, we were able to locate the EPs on the $({\rm Re}d_x, {\rm Re} d_y)$ plane with respect to the $d_z=0$ contour, as discussed in Sec.~\ref{Sec:IVA1}. For five components ${\bf d}$-vector this analysis becomes difficult. We will discuss one example of a 3D NH Hamiltonian in the discussion of the surface states in Sec.~\ref{Sec:VIIID}. 

\subsection{3D Topological semimetals}\label{Sec:VIA}

\begin{figure}[h]
\includegraphics[width=0.99\columnwidth]{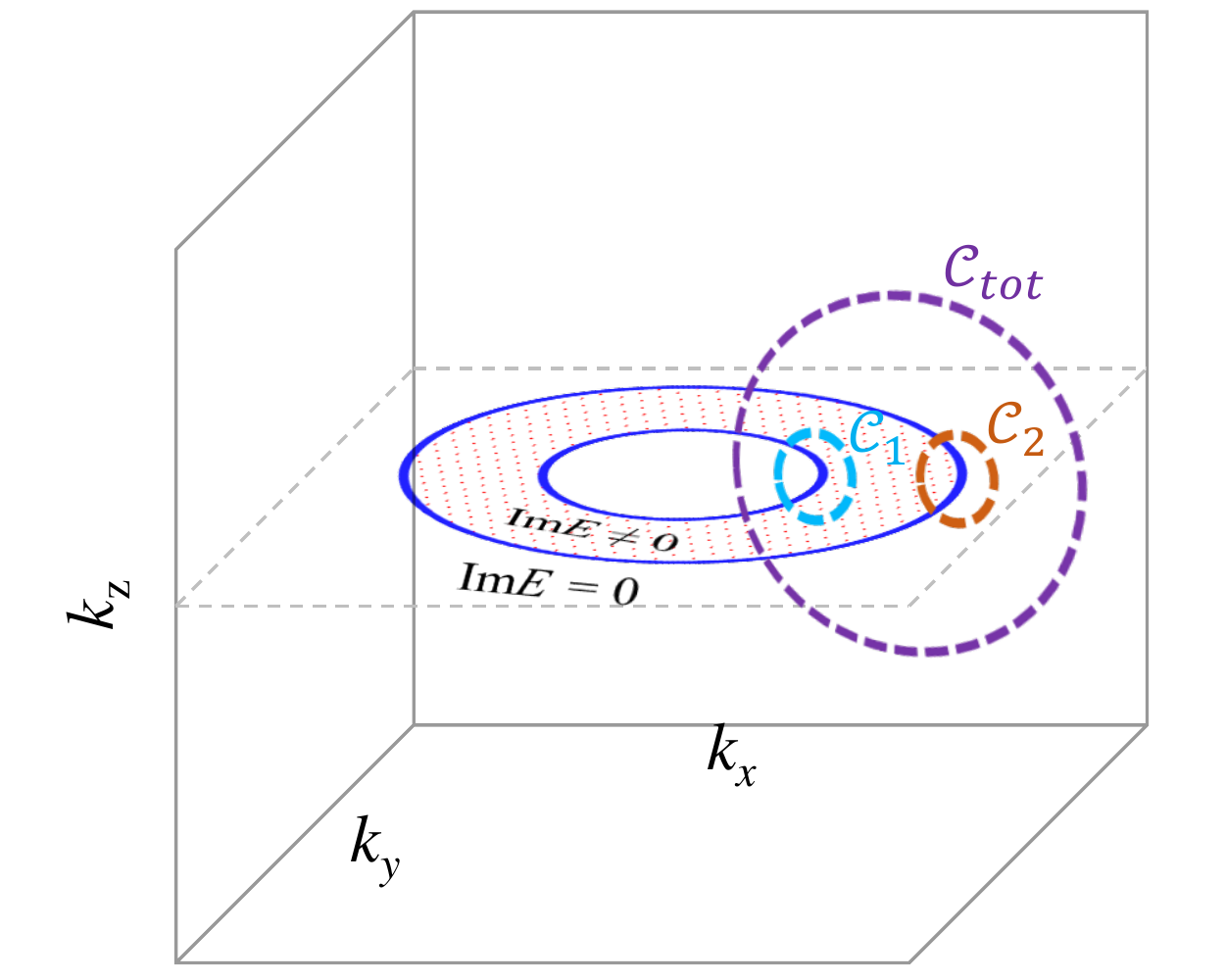} 
\caption{Plot of the exceptional ring (ER) for Eq.~\eqref{ER} on the $k_z=0$-plane. The region between the two concentric circles (blue) possess complex eigenvalues, while outside this ring energies are purely real. The three vertical closed contours $\mathcal{C}_i$ ($i$ = 1, 2, tot) give the loop that encloses inner, outer, and/or both exceptional nodal lines. The winding number for these three contours are $\pm 1/2$, and $\pm 1$, respectively.}
\label{Fig:ERing}
\end{figure}

In 3D, in addition to the discrete locations of EPs marking the topological phase boundary, EPs may form a continuous contour, namely exceptional nodal line (ENL),\cite{NHNodaLine1,NHNodaLine2,NHNodaLine3,NHNodaLine4}  or exceptional ring (ER),\cite{NHRing1,NHRing2Exp,NHRing3} or exceptional Hopf link (EHL) etc.\cite{NHLink1,NHHopf1,NHKnot1,NHKnot2} or even exceptional surface (ES).\cite{NHESurface,NHESurface2} The analogous cases in the Hermitian Hamiltonians are called Dirac nodal line,\cite{HNodalLine1,HNodalLine2,HNodalLine3,HNodalLine3,HNodalLine5} hopf link semimetals,\cite{HHopf1,HHopf2,HHopf3,HHopf4,HHopf5,HHopf6,HHopf7,HHopf7} where the band degeneracy occurs on a similar contour or ring etc. Let us assume the contour is formed on the $k_z=0$ - plane, and as one wraps around such a nodal line in the $k_x-k_z$ or $k_y-k_z$ (or equivalent) plane, we obtain a non-trivial winding number, see Fig.~\ref{Fig:ERing}. To describe such a state, we start with a $2\times2$ Hamiltonian ($H={\bf d}\cdot{\bf \sigma}$) which gives a NH topological nodal ring\cite{NHHopf1,NHNodaLine1}, with components:
\be
d_z = m + \cos k_x + \cos k_y + \cos k_z, ~~~ d_x=\sin k_z + i\delta, ~~~ d_y=0.
\label{ER}
\ee
$\delta$ term makes Hamiltonian NH. The $d_z=0$ contour gives a sphere, which turns into exceptional lines at the contours of $d_x=0$. Due to the specific form of $d_x$, such exceptional lines form on the constant $k_z$ contours. To establish a proof of principle, we take the long wavelength limit of the Hamiltonian, and focus on the $k_z=0$ plane, for simplicity. This yields $d_z \sim m+3-k^2$, and $d_x\sim i\delta$. The eigenvalues are  then $E_{\pm}(k) = \pm \sqrt{(m+3-k^2)^2 - \delta^2}$, with $k=\sqrt{k_x^2+k_y^2}$. This gives exceptional lines determined by the condition $(m+3-k^2)= \pm |\delta|$. From this condition we find the radii of two exceptional lines to be 
\be
k_{0,s} = \sqrt{m+3-s\delta},
\label{RLradii}
\ee
with $s=\pm 1$. It is understood that $|m|<3$ and $\delta<m$ for the exceptional lines to commence within the BZ. The energy eigenvalues in the region surrounded by the two exceptional lines are complex, but purely real outside. This is the reason, such a ring is called exceptional ring. Wang {\it et al.} \cite{NHNodaLine1} used both energy vorticity index $v_n$ as well as eigenstate winding number $w_n$ to demonstrate that as one wraps around each exceptional line, we obtain $v_n=\pm 1/2$ topological index (c.f. contours $\mathcal{C}_{1,2}$ in Fig.~\ref{Fig:ERing}), while the loop $\mathcal{C}_{\rm tot}$ that encloses the entire ring gives an integer winding number $\pm 1$, see Fig.~\ref{Fig:ERing}.  Yang {\it et al.}\cite{NHHopf1} categorized the real and complex energy regions as \PT~invariant and \PT broken regions, respectively. Similar situations may arise in NH superconductors as well.\cite{GhatakSCPRB} In the latter case, we can use the  Chern-Simons theory to obtain the total winding number (as in Eq.~\eqref{NH3Dwinding} above) which gives a $\mathbb{Z}_2$ symmetry. 



\section{Topological invariants in non-Hermitian Hamiltonians with real eigenvalues}\label{Sec:VII}
The NH Hamiltonian with purely real eigenvalues can be transformed into a Hermitian counterpart by a similarity transformation. Such systems are in general called crypto-Hermitian Hamiltonians.\cite{cryptoH0,cryptoH1,cryptoH2} There are usually two types of similarity transformations which guarantee real eigenspectrum: {\it anti-linear} pseudo-Hermitian\cite{PseudoHTheory,PseudoHReview} or \PT-symmetric\cite{Bender1998,BenderCopp, PToperator1,PToperator2,Bender2007} Hamiltonians. In both cases, the Hilbert space is redefined appropriately with respect to the corresponding similarity operator to obtain the corresponding theory of unitary. One of the key points of our interest here is that in both cases, the similarity operator is parameter dependent, which can yield new term in the Berry phase. Since here the eigen spectrum is real, one obtains purely real Berry phase. We discuss these two cases below.  

\subsection{Pseudo-Hermitian Hamiltonians}\label{Sec:VIIA}

For an {\it anti-linear} pseudo-Hermitian system, it is tempting to assume that the Berry phase formula (Eq.~\eqref{Eq:NHBerry}) simply modifies to a $\eta$-inner product form. However, crucial changes arise due to the fact that the {\it $\eta$ operator may be parameter dependent}, and hence contributes an additional term to the Berry phase. The key source of this modification is the modified time-evolution operator of a pseudo-Hermitian Hamiltonian itself. The time-dependent Schr\"odinder equation for an {\it anti-linear} pseudo-Hermitian case with metric $\eta_+$ is\cite{DaJiang1}
\be
i\hbar \frac{\partial}{\partial t}|\Psi\rangle = \left(H-\frac{i\hbar}{2}\eta_+^{-1}(t)\dot{\eta}_+(t)\right)|\Psi\rangle,
\label{TDSE}
\ee
where $|\Psi\rangle$ is the wavefunction of the system, and $\dot{\eta}_+$ is the time-derivative of the $\eta_+$-operator. 

Starting from Eq.~\eqref{TDSE}, the derivation of the Berry phase by Gong and Wang \cite{PTBerry,pseudoHBerry2,DaJiang1} gives
\bea
\gamma_n &=& i\oint_{\mathcal{C}} \Big[\langle \psi^{\rm L}_n({\bf k})|\eta_+({\bf k})|{\bf \nabla_k}\psi^{\rm R}_n({\bf k})\rangle \nonumber\\ 
&& \quad + \frac{1}{2}\langle \psi^{\rm L}_n({\bf k})|{\bf \nabla_{k}}\eta_+({\bf k})|\psi^{\rm R}_n({\bf k})\rangle\Big]\cdot d{\bf k}.
\label{berryPseudoH}
\eea
The first term resembles the complex Berry phase for the NH case in Eq.~\eqref{Eq:NHBerry}, while the second term is a consequence of the ${\bf k}$-dependence of the $\eta_+({\bf k})$-operator. The important consequence of the second term is that this term helps cancel the imaginary part of the first term on Eq.~\eqref{Eq:NHBerry} and hence the Berry phase comes purely real. This can be proved by the unitarity criterion of the eigenstates $\partial \langle \psi_n|\eta_+({\bf k})|\psi_n\rangle=0$, which yields $2{\rm Re}\langle \psi_n|\eta_+({\bf k})|\partial\psi_n\rangle = - \langle \psi_n|\partial\eta_+({\bf k})|\psi_n\rangle$ where we have used the fact that $\eta_+({\bf k})$ is Hermitian. Therefore, the second term indeed cancels the imaginary part of the first term in  Eq.~\eqref{berryPseudoH}, and one obtains a purely real Berry phase as
\bea
\label{berryPseudoH1}
\gamma_n &=& -{\rm Im}\oint_{\mathcal{C}} \Big[\langle \psi^{\rm L}_n({\bf k})|\eta_+({\bf k})|{\bf \nabla_k}\psi^{\rm R}_n({\bf k})\rangle\Big]\cdot d{\bf k}.
\label{berryPseudoH2}
\eea
This is a key result for the $\eta$-symmetric systems which ensures that the Berry phase is real. Generalization to 2D yields that the Berry curvature is also real and hence one obtains a purely real, and quantized Chern number. 

As an example, we take a NH SSH model\cite{SSHLieu} with $d_x-id_y=t_1+t_1'e^{-ik}$, and $d_x+id_y=t_2+t_2'e^{ik}$, where all parameters are real. This is a pseudo-Hermitian system under an anti-linear metric $\eta_{+}=\sigma_x\mathcal{K}$. Interestingly, according to Eq.~\eqref{PseudoH}, although this Hamiltonian is pseudo-Hermitian, however it fails to follow Eq.~\eqref{PseudoHcond2}, which we refer as pseudo-Hermiticity broken condition, and thus gives complex eigenvalues. However, as the parameters are tuned to follow the constraint $t'_1/t_1=t_2'/t_2 = c$ (where $c$ is a real number), the Hamiltonian becomes
\begin{eqnarray}
H({\bm k})=\left(
\begin{array}{cc} 
0 & t_1(1+ce^{-i k})\\ 
  t_2(1+c e^{i k}) & 0
\end{array} \right).
\label{pseudoHExm}
\end{eqnarray}
The eigenvalues $E_{\pm}(k) = \pm \sqrt{t_1t_2|1+ce^{ik}|}$ are real here (pseudo-Hermitian region). The corresponding eigenfunctions are $|\psi_{\pm}^{\rm R}\rangle=1/\sqrt{2}(\pm e^{i\phi(k)}~1)^{T}$, and $\langle\psi_{\pm}^{\rm L}|=1/\sqrt{2}(\pm e^{-i\phi(k)}~1)$, where $\phi(k)=\tan^{-1}(d_y/d_x)$ is now {\it real}. Here $d_x=t_+(1+c\cos{k})$, and $d_y=t_-(1+c\cos{k}) + t_+c\sin{k})$ where $t_{\pm}=(t_1\pm t_2)/2$. Both $d_x$ and $d_y$ are real since the eigenvalues are real. Solving Eq.~\eqref{berryPseudoH2}, we find the winding number becomes $w=1/2\int_{-\pi}^{\pi}\partial_k\phi(k) dk$. We note that we can split the total winding number into two components as discussed in Eq,~\eqref{EPwinding} with both EPs lying at the same point $({\rm Re}d_x,{\rm Re}d_y)=(0,0)$. In fact, a quicker answer can be obtained by the virtue of the chiral symmetry of the Hamiltonian  in Eq.~\eqref{pseudoHExm}. This helps us use Eq.~\eqref{ChiralWinding} where we identify $h_{1/2}=t_{1/2}(1+ce^{\pm ik})$. For each $h_{i}$ the winding number is 1/2 for $|c|<1$ (see Eq.~\eqref{ChiralWinding}) with the total winding number being 1. 

\subsection{\PT - symmetric Hamiltonians}\label{Sec:VIIB}
Next we discuss a \PT-symmetric NH Hamiltonian, according to Eq.~\eqref{PToperation}. \T~is an anti-linear operator, \P~is linear and Hermitian, and hence \PT~is also anti-linear (as the $\eta_+$ operator above, except here $H$ and \PT commute). Under such conditions, (\PT)$^2=\pm 1$ for spinless and spinful Hamiltonians, respectively. In the later case of (\PT)$^2= -1$, the Hamiltonian $H$, and \PT do not share the same eigenfunctions, despite they commute. In such a case, the \PT-symmetry may not in general guarantee real eigenvalues (for some exceptions in special cases, see Ref.~\cite{GhatakDasNewpaper}). In the other cases of (\PT)$^2= 1$, it is known that \PT symmetry is a necessary and sufficient condition to possess real eigenvalues.\cite{Bender1998,BenderCopp, PToperator1,PToperator2,Bender2007} In this case, the left and right eigenfunctions become \PT conjugate to each other, i.e., $|\psi_n^{\rm L}\rangle=\mathcal{PT}\psi_n^{\rm R}\rangle$. 

However, an important aspect of the \PT-symmetric system is that the \PT-inner product $\langle \psi_n|\mathcal{PT}|\psi_n\rangle$ is not always positive, definite. To overcome this problem, an intrinsic, linear symmetry, say $\mathcal{C}$, is introduced such that  $\langle \psi_n|\mathcal{CPT}|\psi_n\rangle=1$, which is called a $\mathcal{CPT}$ inner product.\cite{BenderCopp} Much like the $\eta_+$-operator above, $\mathcal{C}$-operator may not have any physical significance and is Hamiltonian dependent, with the property that $[H,\mathcal{C}]=0$, $[\mathcal{PT},\mathcal{C}]=0$, and hence $[H,\mathcal{CPT}]=0$. Most importantly, {\it $\mathcal{C}$-operator is parameter dependent}, and it helps in achieving real Berry phase. For \CPT-invariant Hamiltonians, the time-dependent Schr\"odinger equation is the same as Eq.~\eqref{TDSE} with $\eta_+\equiv \mathcal{C}$ (but not \CPT~since \PT~operator is not parameter dependent).\cite{GhatakDasNewpaper} The Berry phase formula also remains the same as in Eq.~\eqref{berryPseudoH}, with $\eta_+\equiv\mathcal{CPT}$ here. Following the same logic as in the pseudo-Hermitian case, we find that the Berry phase, Berry curvature, and Chern numbers are real here. 

We consider a \PT symmetric Hamiltonian with onsite gain and loss terms $i\lambda$ as
\begin{eqnarray}
H({\bm k})=\left(
\begin{array}{cc} 
i\lambda & t+t^{\prime}e^{-i k}\\ 
t+t^{\prime}e^{i k} & -i\lambda
\end{array} \right),
\label{PTHExm}
\end{eqnarray}
where $\lambda$, $t$ and $t^{\prime}$ are real. This NH system breaks both TR and parity, but is invariant under \PT. The TR symmetry is simply complex conjugation operator $\mathcal{T}=\mathcal{K}$, while $\mathcal{P}=\sigma_x$. Recasting the Hamiltonian in the usual ${\bf d}$-vector format we get $d_x= t+t'\cos k$, $d_y=t'\sin k$, both real, and $d_z=i\lambda$ is imaginary. Hence the eigenvalues are $E_{\pm}({\bf k}) = \pm \sqrt{d_x^2+d_y^2 -\lambda^2}$, setting the \PT invariant region to be $d_x^2+d_y^2>\lambda^2$. The corresponding eigenvectors are\cite{SSHJiang} 
\bea
|\psi_{\pm}^{\rm R}\rangle &=& \frac{1}{\sqrt{2E_{\pm}(E_{\pm}-i\lambda)}}( d_x-id_y ~~~E_{\pm}-i\lambda)^T,\\
\langle \psi_{\pm}^{\rm L}| &=& \frac{1}{\sqrt{2E_{\pm}(E_{\pm}-i\lambda)}}( d_x+id_y ~~~E_{\pm}-i\lambda).
\eea
The Hamiltonian is the same as Eq.~\eqref{su11}. Following now Eq.~\eqref{berryPseudoH2} with $\eta_+\equiv \mathcal{CPT}$, we find that the winding number formulation remains the same to be $w=\frac{1}{2\pi}\int_{-\pi}^{\pi}\partial \phi(k) dk$, where $\phi(k)=\tan^{-1}(d_y/d_x)$. Since $d_{x,y}$ are real, $w$ is also real and quantized. Evidently, the topological phase is still dictated by the same condition of $|t/t'|<1$ as in the Hermitian case, and $\lambda$ does not play a role here, except above a critical of it the system breaks the \PT-symmetry and the Berry phase becomes complex. However, within the \PT~ invariant region, since $\lambda$ does not play any role to the value of the Berry phase/winding number, the \PT-invariant topological phase is adiabatically connected to the corresponding Hermitian topological phase ($\lambda=0$). 

There have been many examples of \PT-symmetric topological phases including SSH model with a pair of gain and loss terms at the boundary\cite{SSHZhou,SSHLin} or inside the bulk\cite{SSHPTBulkimpurity}, Aubry-Andre-Harper lattice with a pair of gain and loss terms at impurity sites inside the bulk,\cite{NHAAHYuce,NHAAHTh} Kitaev model\cite{NHTSC,NHKitaevWang,NHKitaevKlett,NHQWireTh,NHKitaevLi,NHKitaevYuce}, photonic lattice\cite{NHTrimer}. In SSH chain, it is demonstrated that \PT-symmetric gain/loss terms inside the system increases the topological protection of the edge states, compared to having gain/loss terms at the edges.\cite{SSHZhou,SSHLin,SSHPTBulkimpurity} There have also been numerous experimental evidence of \PT-invariant topological systems in photonic lattices.\cite{NHExp2Photon,NHExp3Floquet,NHExp4QWalk,NHExp4QWalk2}

\section{Biorthogonal Bulk boundary correspondence}\label{Sec:VIII}

Topological invariants arise from the global features of the Hamiltonians and often rely on the periodic boundary conditions, especially when they are quantized. As one reduces the dimension of the system by one, and/or truncates the system to a finite size (but large enough to acquire the topological invariant inside the bulk),  it has the consequence of zero energy bound states at the boundary. For symmetry protected topological invariants, the zero-energy boundary states are robust to disorder and perturbations as long as the underlying symmetry is preserved; otherwise can be gapped out. There have been few conflicting examples regarding the robustness of such zero-energy, bound states in the NH topological phases, simply because (a) the energy is complex, and hence the bound states can decay in time and space, (b) in \PT-symmetric or pseudo-Hermitian systems with real energy eigenvalues in the bulk, the boundary states may or may not obey the same symmetry, and hence can dissipate. In other words, the zero-energy states at the boundary may or may not be normalizable. Several authors argued that under chiral symmetry the zero-energy boundary states remain topologically stable.\cite{NHHughes,NHTELee,Leykam,LFu,NHKramersEsaki,NHChernKawabata,SSHJiang,NHDomainWall,SSHYao,CommentEdge,NHEdgeYuce,NHBulkBoundary2}

To make generic comments on the topological protection of the boundary states, we can approach in two ways, in analogy with the Hermitian case. Zak showed that if we perform the integral of the Berry connection (Eq.~\eqref{HBerry}) in real space, the result which was a Berry phase in the momentum space, becomes the solution of a polarization at the boundary.\cite{Zak} This holds in all TR breaking topological insulators where charge polarization at the boundary is essentially linked to the bulk winding number or Chern number.\cite{Thoulesspump,Thoulesspump2} In TR invariant systems, the boundary polarization is called the TR polarization.\cite{KaneMele1,KaneMele2} For topological superconductors, the boundary states are the Majorana bound states.\cite{HMajoranaedge,ZhangRMP,HTSCReview} Secondly,  we noticed in all topological phases that the non-trivial topology is intrinsically linked to sign reversal of all gap terms inside the BZ. As we Fourier transform the gap terms to real space, we anticipate the gap terms to vanish at some position. The solution of Dirac fermions (near the degenerate points) in a domain wall is known to give a soliton like bound state solution, also known as Jackiw-Rebbi phase,\cite{JaciwRebbi} with half-integer fermion number (half-polarization).\cite{BellRajaraman} 
In this way, we can argue that the mechanism for the bulk topology is the same as the polarization solution at the boundary $-$ providing another perspective to the bulk-boundary correspondence. In NH topological insulators, especially in the cases without chiral symmetry, it is found that the boundary state may have an imaginary part which can render in a plane wave component of the edge states extending to the bulk - namely the `skin-effect' which is unique to NH topological systems. Furthermore, in the presence of balanced NH gain/loss potentials, the two edge states on both sides of the sample may become selectively attenuated and amplified, respectively, giving a new form of bulk-boundary correspondence. We examine these properties  below with the help of generic NH Hamiltonians, and finally take few representative examples.

\subsection{Zak phase}\label{Sec:VIIIA}
As argued in the context of Hermitian Hamiltonians,\cite{KaneChapteFranzabook,VanderbiltPolarization} heuristically if we assume $i\nabla_{\bf k}\equiv {\bf r}$, and substitute it in Eq.~\eqref{Eq:NHBerry}, we get Berry phase $\gamma_n\rightarrow \langle {\bf r}\rangle$, which is the dipole moment of charge = 1 particles. To check it more systematically, we focus on an 1D system of lattice constant $a$, and Fourier transform the Bloch states to the corresponding Wannier states as $|\psi_n^{\alpha}(k,x)\rangle =\sum_m e^{-ik(x-ma)} |\omega_n^{\alpha}(x-ma)\rangle$, where $|\omega_n^{\alpha}(x-ma)\rangle$ is a Wannier state at the $m^{\rm th}$ lattice site, $n$ is the band index, and $\alpha={\rm L/R}$. Following Zak\cite{Zak,NHZak}, Eq.~\eqref{Eq:NHBerry} can be written in the Wannier basis as
\be
\gamma_n^{\alpha\beta}=\frac{2\pi}{a}\sum_m\int_0^{a} (x-ma)\left\langle \omega_{n}^{\alpha}(x-ma) \big|\psi_{n}^{\beta}(x-ma)\right\rangle dx.
\label{NHZak}
\ee
To the best of our knowledge, it is not yet known whether the Wannier states become automatically biorthogonal even when the corresponding Bloch states are so. However, if we assume so, the biorthogonal condition reads as $\int_0^{a}\left\langle \omega_{n}^{\rm L}(x-ma) \big|\psi^{\rm R}_{n'}(x-m'a)\right\rangle dx = \delta_{nn'}\delta_{mm'}$. In such cases, the right hand side of Eq.~\eqref{NHZak} gives the biorthogonal polarization $P_n$,\cite{NHChernKunst} which leads to $P^{\alpha\beta}_n=\frac{e}{2\pi}\gamma_n^{\alpha\beta}$ ($e$ = electric charge). In various cases of NH Hamiltonians above, we found that the Berry phase is real and quantized. In these cases, the polarization at the edge will also be real and half-integer (since $\gamma^{\alpha\beta}=\pi$). Otherwise, for complex Berry phase, one obtains decaying  complex dipole moment, and the corresponding loss function arises from the imaginary part of the Berry phase.  

We recall that even in a biorthogonal system, each state is not individually orthonormal. This means, $P^{\alpha\alpha}_n$ is not always real even when $P^{\alpha\beta}_n$ is real. Here, $P^{\alpha\alpha}_n$ is time-dependent. Rudner and Levitov used the NH SSH model with balanced gain and loss as Eq.~\eqref{su11} to show that the time integral of $P^{\rm RR}_n$ is quantized.\cite{Levitov} More specifically, they have found that $\langle (\Delta m)_n\rangle=\frac{1}{e}\int_0^{\infty} P_n^{\rm RR}(t)dt=w_{\rm 1D}$, where $w_{\rm 1D}$ is the winding number of the same system given in Eq.~\eqref{PseudoAntiHWinding2}. $\langle (\Delta m)_n\rangle$ is the average displacement of the particle during a complete decay.\cite{Levitov} Interestingly, such a topological displacement is observed in a lattice of evanescently coupled optical waveguides with NH loss.\cite{NHExp1}

\subsection{Domain wall problem}\label{Sec:VIIIB}
Next we address the domain wall problem in the NH case. We start with a single-band case discussed in Sec.~\ref{Sec:IVB} with energy vorticity. The corresponding Hamiltonian presented in Eq.~\eqref{Eq:Singleband} is $H = t\cos k -\mu + i t\sin k -i\lambda$. We set $t=1$. To recast the solution in the Jackiw-Rebbi format, we expand the Hamiltonian near the EP in the long wavelength $k\rightarrow 0$ limit as $H\approx (1-\mu) + i(k-\lambda)$. In this limit, we can substitute $k\rightarrow -i\partial/\partial x$. For simplicity, we assume $m(x)$, $\lambda(x)$ are the inverse Fourier transformed components of the real part of $H$, i.e., $(1-\mu)$, and $\lambda$, respectively. So we solve the zero-energy Schr\"odinger equation 
\be
\frac{\partial \psi(x)}{\partial x} = [m(x)-i\lambda(x)]\psi(x).
\label{NHdomain}
\ee
The general solution is\cite{NHTELee}
\bea
\psi(x) &=& \psi(0)e^{\int^x (m(x')-i\lambda(x')) dx'},
\label{NHdomainpsi}
\eea
We notice that $m(x)$ contributes to the decaying part of the wavefunction away from the domain wall, while $\lambda$ gives the plane-wave like (unnormalization) solution. For the wavefunction to be normalizable on both sides of the domain wall, $m(x)$ must change sign across the domain wall, while the sign reversal of $\lambda$ does not affect the qualitative nature of the solution. Hence we assume $m(x)\sim mx$ where $x$ is measured from the domain wall boundary. $m(x)$ changes sign at $x=0$, and goes to $\mp m$ as $x\rightarrow \pm\infty$, and $\lambda(x)=\lambda$ $-$ a constant. This gives 
\be
\psi(x) \rightarrow \psi(0)e^{-mx^2}e^{-i\lambda x}.
\label{domainwallsolution}
\ee
In a Hermitian case ($\lambda=0$), we obtain only a Gaussian solution which gives normalizable soliton mode at $x=0$.\cite{KaneRMP,ZhangRMP} However, we obtain here an additional plane wave like component $e^{-i\lambda x}$ with wavevector $\lambda$, which is damped by $m$. Of course, as long as $1/\lambda$ is small compared to $a$, i.e. the wavelength of solution is small enough compared to the lattice constant $a$ (here $a=1$), we obtain a fairly localized zero-energy solution at the boundary. Interestingly, $|1/\lambda|<1$ is also the limit where energy vorticity is finite in the bulk (Sec.~\ref{Sec:IVB}).

Next we consider a generic 2D NH model for Chern insulator as Eq.~\eqref{RiceMele1}. We take the long-wavelength limit so that $d_x\sim k_x$, $d_y\sim k_y+i\delta$, ($t=1$) and $d_z\sim m+2-i\lambda$. In addition, we assume a periodic boundary condition along the $y$-direction, and an open boundary condition along the $x$-direction. So we can write the wavefunction as $\psi_{k_y}(x,y)=e^{ik_y y}\psi_{k_y}(x)$. Similarly, we assume $m(x)$, $\lambda(x)$, and $\delta(x)$ are the inverse Fourier transformed components of $m({\bf k})$, $\lambda$, and $\delta$, respectively. The corresponding Schr\"odinger equation is\cite{Leykam,LFu}
\bea
&&\left[-i\sigma_x\frac{\partial}{\partial x} + \big(k_y + i\delta(x)\big)\sigma_y \right.\nonumber\\
&&~~~ \left. + \big(m(x)+i\lambda(x)\big)\sigma_z\right]\psi_{k_y}(x) = E(k_y)\psi_{k_y}(x).
\label{NHdomainChern}
\eea
The solution of the above equation is a step function with respect to the domain wall position, $x=0$, 
\bea
\psi_{k_y}(x,y) = \left(
\begin{array}{cc} 
\psi_1\\ 
\psi_2
\end{array} \right)e^{ik_y y}\left[e^{\kappa_+ x}\theta(-x) + e^{\kappa_- x}\theta(x)\right].
\label{NHdomainChern}
\eea
where $\kappa_{\pm}=\pm m_{\pm} + \delta_{\pm} \pm i s_{\pm}\lambda_{\pm}$, where $m_{\pm}$, $\delta_{\pm}$, and $\lambda_{\pm}$ are the values of the corresponding quantities at $x\rightarrow \pm\infty$, and $ s_{\pm}={\rm sgn}[m_{\pm}]$. The eigenvalues are $E_{k_y}=s_- tk_y$. As in the single-band case above, without loosing generality, we assume that only the Dirac mass changes sign across the domain wall $m_{\pm}=\pm m$, while the NH potentials do not, i.e., $\delta_{\pm}=\delta$, and $\lambda_{\pm}=\lambda$. Then we obtain the solution as 
\bea
\psi_{k_y}(x,y) = \left(
\begin{array}{cc} 
\psi_1\\ 
\psi_2
\end{array} \right)e^{ik_y y}e^{-mx^2-\delta x}e^{i\lambda x}.
\label{NHdomainChern2}
\eea
Therefore, the solution is a plane wave along the periodic lattice ($y$-direction), but a mixed state along along the $x$-direction as in Eq.~\eqref{domainwallsolution}. Eq.~\eqref{NHdomainChern2} reveals an important message that the NH potential {\it $\delta$ does not contribute to the unnormalizable part of the wavefunction}, and with its presence the wavefunction is fully localized at the edge. On the other hand, the onsite NH potential $\lambda$ solely contributes to the last term in Eq.~ \eqref{NHdomainChern2}. This gives a delocalization component, damped by both Dirac mass $m$ and NH hopping $\delta$ terms. {\it If we set $\lambda=0$, the Hamiltonian has the Chiral symmetry and the edge states are fully localized.} We discuss the nature of the topological protection of the edge states along the $y$-direction in Sec.~\ref{Sec:VIIIC0} below.

\begin{figure}[t]
\includegraphics[width=0.8\columnwidth]{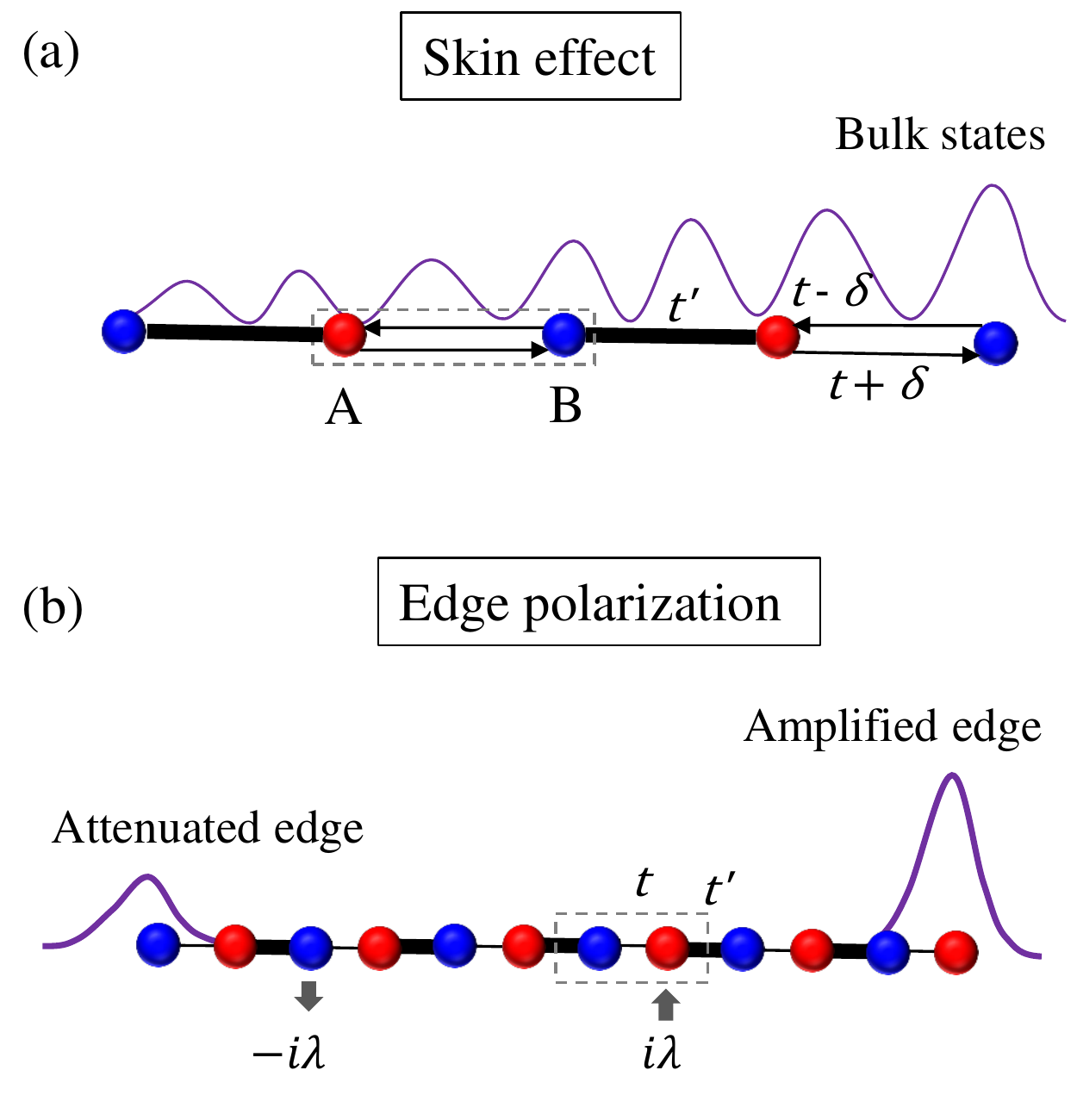} 
\caption{(a) A schematic demonstration of the NH `skin effect'. In this case, often owing to imbalanced tunneling amplitudes $t\pm \delta$ along right and left moving bloch states between the two sublattices `A' and `B', the bulk states gradually shift towards one side of the lattice. We take a SSH type chain for this demonstration,\cite{SSHYao,NHEPTI2} however the effect occurs in higher dimensions as well.\cite{NHChernYao,NHHopf1,NHNodaLine1}  (b) In a topological bulk state (we take the SSH chain as an example),  when balanced gain and loss terms ($\pm i\lambda$) are introduced (either randomly or periodically), it is found that the edge states on different sides behave oppositely: in one of them the probability (or polarization) is amplified while on the other side the probability is simultaneously attenuated.\cite{NHTrimer,SSHZhou,NHSSHFloquet,NHSSHFloquet2,NHAAHYuce,NHAAHYuce2,NHAAHTh}
} 
\label{Fig:Skineffect}
\end{figure}

\subsection{Skin-effect}\label{Sec:VIIIE}
For NH topological Hamiltonians there can arise interesting anisotropic localization effects of both bulk and edge states, depending on the details by which non-Hermiticity is achieved. We first discuss the so-called {\it skin-effect} where the {\it bulk} states can localize to one of the edge.\cite{SSHYao,NHEPTI2,NHChernYao,NHHopf1,NHNodaLine1} We demonstrate this effect in terms of the SSH model in Eq.~\eqref{SSH1}. In a physical picture, this Hamiltonian stems from a bipartite lattice with two sublattices (say, `A' and `B') per unit cell, as shown in Fig.~\ref{Fig:Skineffect}(a). The inter-sublattice tunneling amplitude between the two adjacent unit cell is $t^{\prime}$. And the tunneling amplitude between the two sublattices within a unit cell is $t \pm \delta$ for right and left moving states, respectively. Recall that the non-trivial topological region is defined here by $|t-t'|<\delta<|t+t'|$. Therefore, in the topological region, a left moving Bloch state with energy dispersion $(t-\delta)e^{-ik}$ has a higher propensity to hop to the next unit cell with tunneling amplitude $t'>|t-\delta|$, than being reflected with dispersion $(t+\delta)e^{ik}$. The tunneling probability is thus $T^2=|(t+\delta)/(t-\delta)|>1$. This means, in every hopping along the left-hand direction, the Block wave's amplitude increases by $T$ to $Te^{-ik}\rightarrow e^{-i k+\ln{T}}$. For a Hermitian case when $\delta=0$, $T=1$, and hence the Bloch waves remain fully plane-wave like. However, since $T>1$ and is real in the non-trivial topological phase, the amplitude of the Bloch wave in traversing to the left side keeps on increasing, and the bulk states become localized at the left edge of the sample. This is the NH `skin-effect' deduced in Refs.~\cite{SSHYao,NHEPTI2,NHChernYao,NHHopf1,NHNodaLine1}. The existence of NH {\it skin effect} is shown via numerical simulation in 1D SSH model,\cite{SSHYao,NHEPTI2} 2D Chern insulator,\cite{NHChernYao} 3D topological nodal semimetals\cite{NHHopf1,NHNodaLine1} and other NH topological systems.\cite{SkinRonny}  In fact, one can cast the problem in a non-Bloch wave formalism by defining a complex wavevector $k'=k+i\ln T$, and obtain distinct topological invariants in the bulk band structure.\cite{SSHYao} Equivalently, one can treat the asymmetric coupling amplitudes $\delta$ in a SSH chain as an imaginary gauge field, which then induces a NH Aharonov-Bohm effect under periodic boundary condition, and a `skin-effect' in finite lattice.\cite{JinSong}

A related effect arises in the presence of balanced gain and loss onsite terms, which breaks chiral symmetry. With particular examples in 1D topological insulators, it was shown that when balanced impurity potentials ($\pm i\lambda$) are introduced in each unit cell or at random impurity sites (with the total onsite imaginary potential being zero), two different edge states on two different sides of the 1D lattice become simultaneously amplified and attenuated at the same rate.\cite{NHTrimer,SSHZhou,NHSSHFloquet,NHSSHFloquet2,NHAAHYuce,NHAAHYuce2,NHAAHTh} This phenomena is shown schematically in Fig.~\ref{Fig:Skineffect}(b). This is analogous to a current in which particles are being transferred from one side to the other, but without any applied voltage. However, if the system preserves the \PT-symmetry, the edge states become protected, as shown for the similar SSH model with balanced gain and loss at edge,\cite{SSHZhou,SSHLin} or inside the bulk\cite{SSHPTBulkimpurity} and also for the  Aubry-Andre-Harper lattice.\cite{NHAAHYuce,NHAAHTh} This case is discussed further in Sec.~\ref{Sec:IX}.

\subsection{Protection from backscattering}\label{Sec:VIIIC0}
The topological protection of boundary states to disorder is yet an open question in NH topological insulators. For Hermitian Hamiltonians, the edge states are protected from backscattering as long as disorder potentials respect the underlying symmetry. For NH cases, one has to be careful in generalization this concept, since the violation of this protection can come from various sources. For example, across a degenerate point in Hermitian Hamiltonians, the two corresponding eigenstates are orthogonal. If these two states turn out to the right-hand and left-hand moving states across a disorder site (which is often the case for chiral symmetric, or TR symmetric edge states), backscattering is naturally protected here, as long as disorder potential commutes with the Hamiltonian. For NH Hamiltonians, in general, the right-hand and left hand states across EPs are {\it linearly dependent}. Hence, across an impurity site $-$ irrespective of whether the impurity potential is Hermitian or NH $-$ the expectation value of the impurity potential with respect to the right- and left-hand states is, in general, not guaranteed to vanish. We caution that quantitative discussions of topological protection is not available in the literature for generic cases. Few specific cases can be discussed qualitatively under three categories: (a) Bulk Hamiltonian is an Hermitian topological insulator, while the impurity potential is NH, (b) The bulk Hamiltonian is NH, and the impurity potential is Hermitian, (b) Both bulk Hamiltonian and the impurity potentials are NH. 

(a) The first case is analogous to the case studied by Hatano-Nelson,\cite{HatanoNelson} and others\cite{LonghiSRep,LonghiPRB} in 1D. They showed that in 1D systems, the NH impurity potential renders delocalization of the states as opposed to Anderson localization for Hermitian impurity potentials. In what follows, the edge states for a Hermitian Hamiltonian should remain delocalized under the NH Hamiltonians.

(b) Consider that there is no EP in the energy spectrum. Then the Hermitian disorder potential  leads to topological protection in the following specific case. For NH topological insulators, we found in Eq.~\eqref{NHdomainChern2} that the edge states are fully delocalized along the edge with a plane wave like solution $\psi_{k_y}(y)\sim \exp(ik_y y)$. Let us assume that the right and left hand eigenstates $\psi_{k_y}^{\rm R}$, $\psi_{-k_y}^{\rm L}$ follow the biorthogonal condition $\langle \psi_{-k_y}^{\rm L}|\psi_{k_y}^{\rm R}\rangle=0$ under some underlying  symmetry. Now if the Hermitian impurity potential $V$ commutes with the Hamiltonian $[H,V]=0$, then the matrix-element $\langle \psi_{-k_y}^{\rm L}|V|\psi_{k_y}^{\rm R}\rangle$ is guaranteed to vanish. This means the edge states remain topologically protected from {\it backscattering}. We remind the reader that we are not considering the peculiarity of the edge state along the direction of the bulk (i.e., along the $x$-direction in Eq.~\eqref{NHdomainChern2}). 

(c) The final case of both Hamiltonian and impurity potential being NH can be qualitatively understood from the above two discussions. Again for the specific case of delocalized edge states  $\psi_{k_y}(y)\sim \exp(ik_y y)$ with biorthogonal condition, under a NH disorder potential, say, $iV$, the edge states can remain delocalized if the Hatano-Nelson criterion is satisfied.\cite{HatanoNelson} Hence they remain topologically protected under the symmetry condition as in (b). Any modification to the states along the direction of the bulk as in (b) with NH or Hermitian disorder is not yet considered in the literature.

\begin{figure}[t]
\includegraphics[width=0.9\columnwidth]{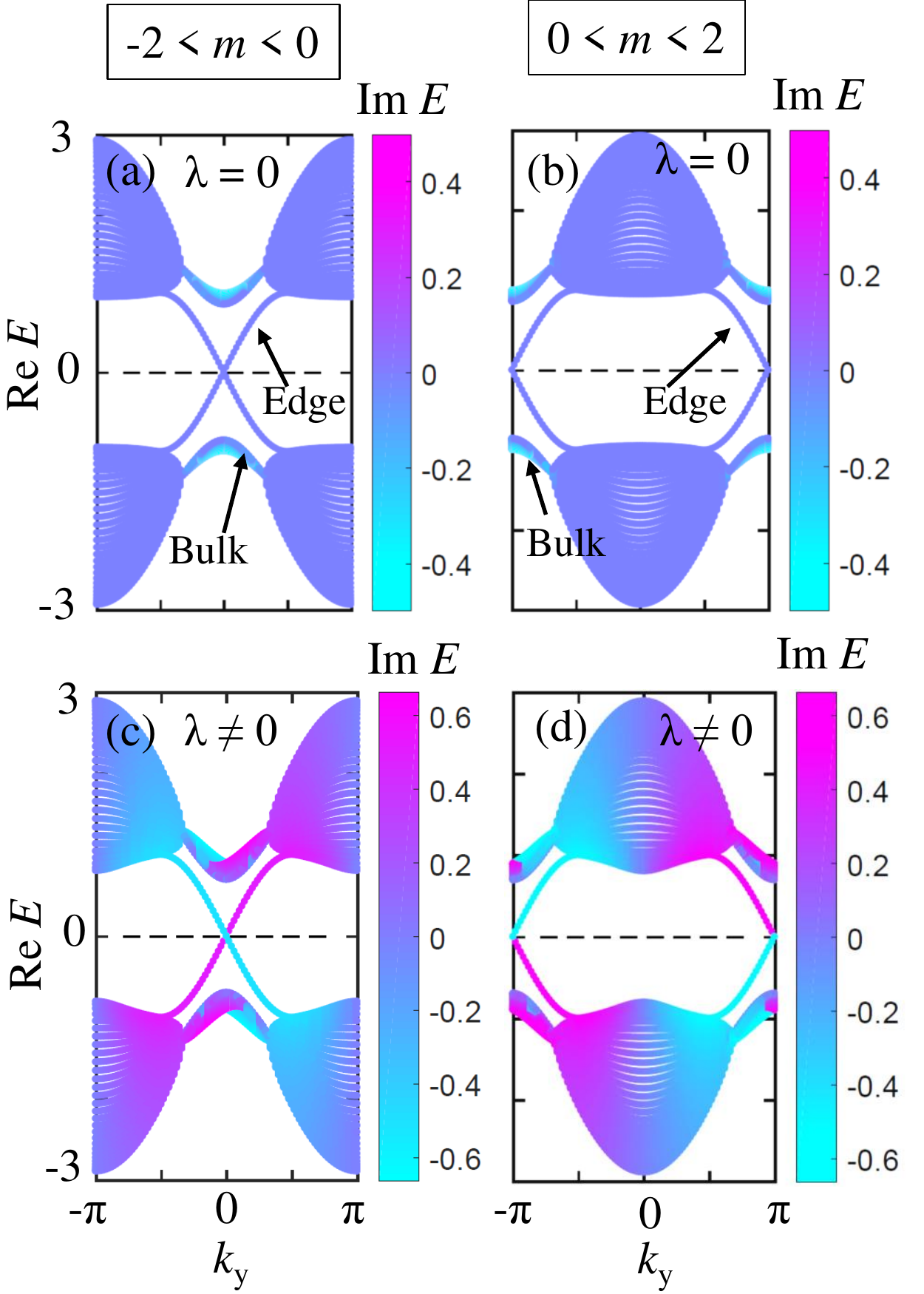} 
\caption{Edge state profile of a 2D Chern insulator (Eq.~\eqref{NH2DEdge}) with (upper panel) and without (lower panel) chiral symmetry. Right and left panels are for Chern number $C=\pm 1$, respectively. Here we assumed a periodic boundary condition along the $y$-direction and an open boundary condition along the $x$-axis. Cyan to magenta colormap depicts the value of $\pm {\rm Im} E({\bf k})$. We observe two counter-propagating edge states, each of which lies on two different edges of the lattice.\cite{NHChernKawabata} We notice that when the chiral symmetry is present ($\lambda=0$), the edge state energy is purely real (upper panel), and as the chiral symmetry is broken with $\lambda\ne 0$, the edge state energies become complex, and thus dissipates into the bulk (see Eq.~\eqref{NHdomainChern2}).
} 
\label{Fig:2DEdge}
\end{figure}

\subsection{Finite lattice models}\label{Sec:VIIIC}

We supplement the above domain wall solutions obtained in the long wavelength ($k\rightarrow 0$) limit to a full lattice model. We consider the same 2D Chern insulator (Eq.~\eqref{RiceMele1}) with periodic and open boundary conditions along the $y$, and $x$-directions, respectively. The Hamiltonian in the real space then reads as\cite{NHChernKawabata,NHTransferMatrix}
\bea
H&=&\sum_m\sum_{k_x}\Big[\psi_{m+1,k_y}^{\dag}\frac{1}{2}\big(\sigma_x + \sigma_z\big)\psi_{m,k_y} + {\rm h.c.}\nonumber\\
&&+\psi_{m,k_y}^{\dag}\big(m+\cos k_y + i\lambda)\sigma_z + (\sin k_y- i\delta)\sigma_y\big) \psi_{m,k_y}\Big],
\label{NH2DEdge}
\eea
where $\psi_{m,k_y}$ is the annihilation operator at the $m^{\rm th}$-site. We have also included the chiral symmetry breaking imaginary gain and loss onsite terms $\pm i\lambda$, in addition to the NH inter-species hopping $i\delta$. We solve the above equation in a 30$\times $30 lattice. The energy levels for the two topological phases of $C=\pm1$ are shown in Fig.~\ref{Fig:2DEdge}. We find two edge states: one from the right and other one from the left side of the lattice. The edge states have linear dispersions near the Fermi level, as predicted in the continuum solution in Eq.~\eqref{NHdomainChern}. They adiabatically connect to the bulk quantum well states at higher energy, in consistent with the bulk-boundary correspondence.\cite{KaneRMP,ZhangRMP,DasRMP} For $-2<m<0$ region, the $d_z=0$ nodal line encircles the ${\bf k}=0$ point (in Fig.~\ref{Fig:2DEdge}), and correspondingly the edge state passes through $E=0$ at this ${\bf k}=0$-point. In the $0<m<2$ region, the gapless points shift to the $k_y=\pm\pi$ points as the center of the $d_z=0$ contour shifts to this ${\bf k}$-point.

One of the notable aspects of the edge solution is the presence and absence of the imaginary part of the edge state dispersions without and with the chiral symmetry breaking potential $i\lambda$, respectively. They are shown on the top and bottom panels of Fig.~\ref{Fig:2DEdge}, respectively. On the top panel, we find that the edge states are purely real in both regions due to the presence of the chiral symmetry. As we turn on $i\lambda$, we notice that the edge states are no longer real, rather gain imaginary components. As also obtained in the domain wall problem in Eq.~\eqref{NHdomainChern2}, the chiral symmetry breaking perturbation is responsible to the plane-wave like solution of the wavefunction, thereby makes the edge states propagate to the bulk (`skin-effect'). However, as long as the chiral symmetry is present, the edge states are robust and localized to the boundary. Another interesting property of the chiral symmetry breaking gain/loss term is that as long as the gain and loss terms are balanced, the two edge states separately become amplified and attenuated. This can be useful for many applications.

\begin{figure*}[ht] 
\includegraphics[width=1.9\columnwidth]{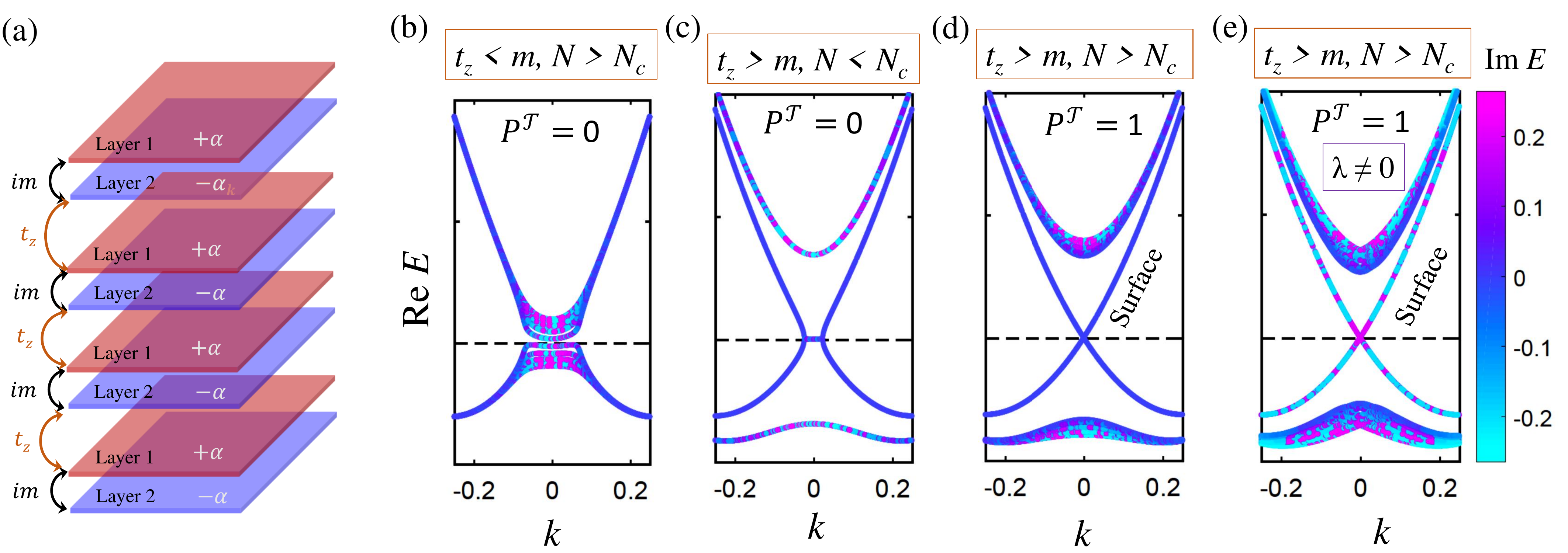} 
\caption{2D surface states of a 3D TR invariant topological insulator (Eq.~\eqref{3DHam}). (a) We construct a heterostructure of the complex 2D QSH insulator discussed in Eq.~\eqref{NHRashbabilayer}, with nearest neighbor inter-Rashba bilayer coupling $t_z$ (real). We observe a topological phase transition above the critical value of the number of such bilayers $N_c$, as well as for $t_z>m$. (b) For $t_z<m$, the bulk bands do not undergo any band inversion and thus we obtain trivial quantum well states with complex energy. (c) For $t_z>m$, a band inversion occurs, but for a thin film with number of layers $N<N_c$, the two surface states across the bulk can hybridize and create an exceptional surface state. (d) For both  $t_z>m$, and $N>N_c$ we obtain robust, gapless surface state with purely real energies. (e) Finally we add a similar chiral breaking term $\pm i\lambda$ with gain and loss to two Rashba layers in each Rashba-bilayer. Here we observe that although gapless surface states exist, but their energies are now complex, and hence they dissipate. }
\label{Fig:3DSurface}
\end{figure*}

\subsection{Surface state in 3D topological insulators}\label{Sec:VIIID}

As a demonstration to the existence of robust gapless surface states in 3D topological insulators, we construct a 3D model by stacking the 2D TR invariant Rashba-bilayer system (Eq.~\eqref{NHRashbabilayer}) along the $z$-direction, as shown in Fig.~\ref{Fig:3DSurface}. The inter Rashba bilayer hopping ($t_z$) is assumed to be different  than the intra-Rashba-bilayer dispersion $m({\bf k})$. The Hamiltonian for such a finite size heterostructure is\cite{DasNC}
\begin{eqnarray}
H_{\rm 3D}({\bm k})=\left(
\begin{array}{ccccc} 
H_{\rm 2D} & T({\bf k}) &  0  & 0 & \dots\\
T^{\dag}({\bf k})  & H_{\rm 2D} & T({\bf k}) & 0 & \dots\\
0 & T^{\dag}({\bf k}) & H_{\rm 2D} & T({\bf k}) & \dots\\
0 & 0 & T^{\dag}({\bf k}) & H_{\rm 2D} & \dots\\
\vdots & \vdots & \vdots & \vdots &\ddots
\end{array} \right),
\label{3DHam}
\end{eqnarray}
where $T({\bf k}) = t_zI_{4\times 4}$ with $t_z$ being real. The NH part still arises from the imaginary Dirac mass $im({\bf k})$ as in the 2D case in Eq.~\eqref{NHRashbabilayer}. We start with the parameter set which gives a 2D TR invariant topological insulator for an isolated Rashba bilayer. Above a critical number of such Rashba bilayers (say $N_c$), and for $t_z>m$, we find a transition from weak to strong topological transition, marked by the appearance of 2D surface states on the $(k_x,k_y)$ plane. Across the phase transition as a function of both $t_z$ and number of bilayers $N$, we observe the presence of trivial surface states at zero energy, which represent exceptional surface states where both real and imaginary parts are simultaneously zero. This is unique to NH topological insulators. Above the topological phase transition, we find that the surface states are purely real, and are adiabatically connected to the bulk bands. Furthermore, as a chiral symmetry breaking potential $i\lambda\Gamma_z$ is added to the onsite potential in each Rashba-bilayer, we find that the surface states gain imaginary term and hence dissipate. This conclusion is robust to all dimensions we study here.

\section{Other non-Hermitian topological systems}\label{Sec:IX}

\subsection{Photonics}\label{Sec:IXA}

\begin{figure}[h]
\includegraphics[width=0.95\columnwidth]{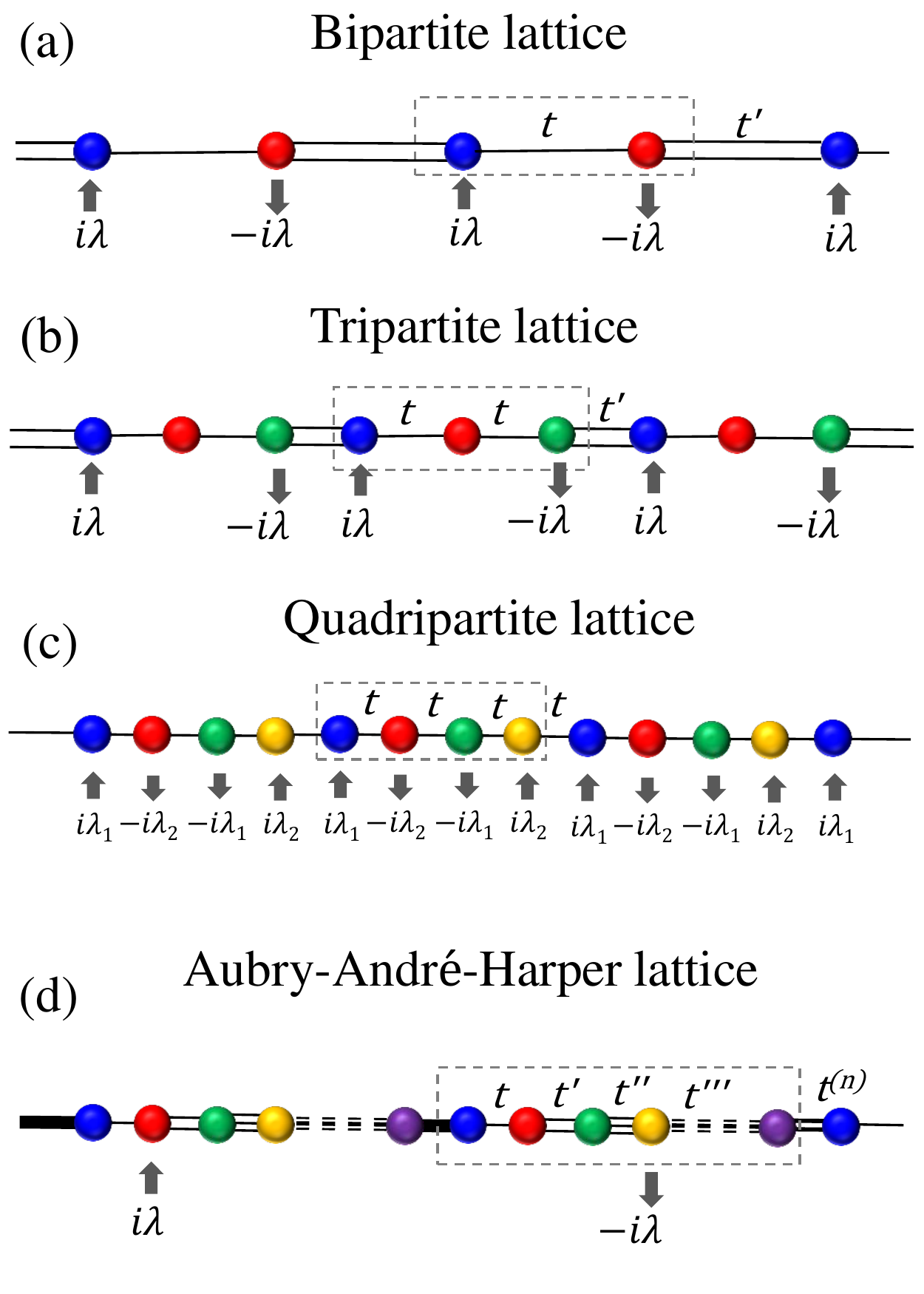} 
\caption{(a) A typical bipartite lattice in 1D - the SSH model, with balanced gain and loss terms ($\pm i\lambda$) in every unit cell. $t$ and $t^{\prime}$ are the two nearest-neighbor hopping terms. (b) A tripartite lattice with balanced gain/loss terms, according to Eq.~\eqref{NHtrimerH}. In this case, one of the sites (red dot) does not have any onsite potential. (c) A quadripartite lattice with different onsite potentials $\pm i\lambda_{1,2}$, as given in Eq.~\eqref{NHquarterH}. (d) Finally, a AAH lattice, as used in Eq.~\eqref{mcabjs4} with $N$ number of lattice sites. The nearest-neighbor hopping terms are periodically modulated as $t^{(n)}=t(1+\gamma\cos{(2\pi\beta n + \Phi)})$, where $n$ is lattice site, and $t$, $\gamma$, $\beta$, and $\Phi$ are real parameters. Unlike the other three cases discussed above, in this AAH model, balanced gain/loss terms are only added at two random impurity sites.
} 
\label{Fig:Othermodels}
\end{figure}

NH quantum systems are readily obtained in photonic setups. So, the general expectation is that the above-discussed topological phases can be easily realized/engineer in photonic systems. However, the essential bottleneck is how closely the electromagnetic wave equation is related to  the Sch\"odinger equation of quantum theory? This question has being addressed rigorously in the literature for decades. At least in the Heisenberg representation, the essential matrix format of the eigenvalue problem does represent all the salient  properties of the Hilbert space. Therefore, the topological invariants obtained in the condensed matter setup are well applicable in the photonic systems. There have been several theoretical studies\cite{PhotonicTI,PhotonicTI2,NHESurface2,NHTrimer,NHQuarter} and, most importantly, experimental manifestations\cite{NHExp1,NHExp2Photon,NHExp3Floquet,NHExp4QWalk,NHExp4QWalk12,NHExp5FermiArc,NHExp6,NHExp7,NHExp8Laser,NHExp9Photon,NHExp10Photon,NHExp11,NHExp12} of the topological phases in optically controlled systems.

Non-Hermiticity is rather easily achieved in photonic systems, where complex potential is obtained through complex refractive index in the electromagnetic wave equations. The \PT~invariance of the complex refractive index $n(x)$ renders \PT~invariance to the system.\cite{Ruter,guo,sch,bit,ram}. A system of coupled multi-channel optical waveguides, arranged in a periodic way, can produce a pseudo-Hermitian photonic system when the alternative channels receive some amplification via optical pumping. With such setups, the phase transition associated with \PT-symmetry and pseudo-Hermiticity breaking can be easily tuned. 

Following Ref.~\cite{NHTrimer}, we give the example of a periodic lattice of three coupled waveguides per unit cell. A Hermitian version of such tripartite SU(3) topological insulator was studied by us\cite{SU3} and others\cite{HTrimer2}. Here we consider that such a trimer lattice is subjected to balanced gain and loss terms $\pm i\lambda$, separated by a passive waveguide in between. In other words, in this tripartite lattice, the three sublattices, say, A, B, and C possess complex onsite potential as ($i\lambda,0,-i\lambda$) with $\lambda$ being real. In addition, we assume the intracell nearest neighbor sublattice hopping is $t$, while the corresponding intercell hopping is $t'$, with $t$, $t'$ being real. The corresponding setup is depicted in Fig.~\ref{Fig:Othermodels}(b). Within the coupled mode theory, the single-mode coupled waveguide lattice  can be modeled by a tight-binding model (in analogy with the non-interacting Hamiltonian we consider in the condensed matter systems).\cite{NHTrimer} The three coupled waveguide equation of motion can now be expressed in terms of the three-components spinor to obtain the Hamiltonian
\be
H_{k}=\left(
\begin{array}{ccc}
i\lambda & t & t'e^{ik} \\
t & 0 & t \\
t'e^{-ik} & t & -i\lambda
\end{array}
\right).
\label{NHtrimerH}
\end{equation}
The above Hamiltonian is a generalization to the bipartite SSH model with balanced gain and loss terms as studied in Eq.~\eqref{SSHwithoutChiral} above. The Hamiltonian in Eq.~\eqref{NHtrimerH} is \PT-symmetric, with the \PT-symmetry breaking occurring at $(\lambda
^{2}-2t^{2}-(t')^{2})^{3}/3^{3}+(2t^{2}t'\cos k)^{2}/2^{2}=0$. Interestingly, the system is topologically non-trivial in the parameter range of $|t/t'|<1$ with winding number obtained from Eq.~\eqref{Eq:NHBerry} to be $1$, 0 and $1$ for the three bands, and all zero otherwise. As in the case of bipartite SSH model in Eq.~\eqref{SSHwithoutChiral}, the gain/loss terms $i\lambda$ do not directly enter into the winding number formula, except it mobilizes the EP across the adiabatic loop, and hence play an indirect role in the topological phase transition. 

The solution of the above Hamiltonian with an open boundary condition gives edge states with complex energy (since the system does not have chiral symmetry) with gain and loss terms on the right and left hand edges, respectively:
\be
E_{\pm}^{\alpha}=\frac{1}{2}\left(-s i\lambda \pm \sqrt{4t^{2}-\lambda^{2}}\right),
\label{NHtrigredge}
\ee
where $\alpha={\rm R/L}$ (right and left edge energies) with corresponding $s=\pm$. Clearly, the right hand edge gives a lossy site where the corresponding polarization (Zak phase) is reduced, while the left hand side gives an amplified polarization where the polarization is amplified in each adiabatic cycle in the bulk with winding number $1$. The two edge states are symmetric about the energy zero. Above a gain (loss) threshold, zero energy edge modes appear, and they are amplified and damped when propagating in opposite directions. Jin\cite{NHTrimer} also pointed out several potential applications of this unique topological edge states. 

In another example, Takata and Natomi\cite{NHQuarter} considered a {\it pseudo-Hermitian} (but not \PT-symmetric) quadripartite optical lattice. Here intracell and intercell hopping amplitudes are assumed to be the same, i.e., $t=t'$, but the balanced gain/loss onsite terms are assumed to be sublattice selective as $(i\lambda_1,-i\lambda_2,-i\lambda_1, i\lambda_2)$ in the four-nearest neighbor sites. The schematics of such a lattice in shown in Fig.~\ref{Fig:Othermodels}(c). The corresponding Hamiltonian reads as
\be
{H}(k) = \left( \begin{array}{cccc}
i \lambda_1 & t & 0 & t e^{-i k} \\
t & -i \lambda_2 & t & 0 \\
0 & t & -i \lambda_1 & t \\
t e^{i k} & 0 & t & i \lambda_2 \\
\end{array} \right), 
\label{NHquarterH}
\end{equation}
The eigenfrequency detuning $\omega (k)$ is given by,
\begin{equation}
\omega(k) = \pm \frac{1}{\sqrt{2}} \sqrt{A \pm \sqrt{ A^2 - B^2 - 16 t^4 \sin^2 \frac{k}{2}}}.
\label{NHquarterE}
\end{equation}
where $A = 4 t^2 - \lambda_1 ^2 - \lambda_2^2$ and $B = 2 \lambda_1\lambda_2$. This means the eigenvalues are real in the parameter space of $A>0$, and $A^2-B^2-16t^4\sin^2(k/2)>0$ region. Since this \PT-broken Hamiltonian possess real eigenvalues, there must exists a corresponding {\it anti-linear} metric $\eta_+\mathcal{K}$, which is $\eta_+=\cos (k/2)\sigma_x\otimes I_{2\times 2} + \sin (k/2)\sigma_y\otimes I_{2\times 2}$. Furthermore, owing to the PH symmetry of the eigenvalues, we can also find an pseudo-anti-Hermitian metric $\eta_-=I_{2\times 2}\otimes \sigma_z$. Therefore, the topological properties of this Hamiltonian can be easily obtained by defining topologically equivalent  off-block diagonal spectral Hamiltonian $\mathcal{Q}$ (Eq.~\eqref{blockdiagonalQ}) and the corresponding winding number can be obtained from Eq.~\eqref{PseudoAntiHWinding2}. The Hamiltonian is topologically non-trivial with winding number $w=1$ for $\lambda_1>0$, and $\lambda_2>2$ for $t=1$, and trivial otherwise.\cite{NHQuarter} It is interesting to note that here the topological phase transition is driven entirely by the the NH perturbation terms $\lambda_i$, and hence with dynamical tuning of this parameter one can generate dynamical topological quantum phase transition in the same setup.\cite{NHDQTI,NHRiceMeleDQ,NHDQPT,NHDQTIQuench,NHDynamicalClass}

\subsection{Topology with gain and loss at impurity sites}\label{Sec:IXB}
In Eqs.~\eqref{NHtrimerH}, \eqref{NHquarterH} we discussed 1D photonic chains with balanced loss and gain in each unit cell, which results in balanced lossy and amplified polarizations at the right and left edges in the topological phase.\cite{NHTrimer} Here we consider a counter-example in which we take a {\it trivial} bulk band, and introduce gain and loss terms $\pm i\lambda$ at two different {\it impurity} sites, or at the two ends, but not in every unit cell. Such models are studied in the literature with SSH lattice\cite{SSHZhou,NHSSHFloquet,NHSSHFloquet2}, or Aubry-Andr\'e-Herper (AAH) lattice,\cite{NHAAHYuce,NHAAHYuce2,NHAAHTh,NHAAHChen} or Kitaev chain of $p$-wave superconductor\cite{NHKitaevWang,NHKitaevKlett} in 1D. We consider an 1D off-diagonal AAH lattice with modulated tight binding hoppings, which is subjected to gain and loss perturbations at two different impurity sites\cite{NHAAHYuce}:
\begin{eqnarray}\label{mcabjs4}
H&=&-t\sum_{n=1}^{N-1}\left(1+\gamma \cos{(2\pi \beta n+\Phi)}\right)  c^{\dagger}_{n} c_{n+1}+h.c.\nonumber\\
&&+i\lambda c^{\dagger}_j c_j- i\lambda c^{\dagger}_{N-j+1} c_{N-j+1}),
\end{eqnarray}
where $t$ is fixed (real) tunneling amplitude, $c^{\dagger}_n$ and $c_n$ denote the creation and annihilation operators of fermionic particles on site $n$, respectively.  The constant $\gamma$ parameter is the strength of the modulation, and $\beta$ controls the periodicity of the modulation with phase $\Phi$. Here the particles are injected on the $j$-th site, and removed from the $(N-j+1)$-th site, where $N$ is the number of lattice sites.  

The system preserves $\mathcal{PT}$-symmetry when the modulation of the parameter is restricted between $1+\lambda \cos(\Phi)$ and $1-\lambda \cos(\Phi)$ at fixed $\Phi$. In the \PT-unbroken phase, topological phase transition is characterized by the existence of zero-energy edge modes. When $N$ is odd, zero energy edge modes exist for all values of $\Phi$. For even $N$, a non-trivial topological phase appears in the region of $\pi/2<\Phi<3\pi/2$.

\subsection{Other systems}
Weyl fermions have been recently observed in Hermitian condensed matter systems.\cite{WeylRMP,DasRMP,WeylRao,WeylDas} In the NH systems, the existences of Weyl fermions across discrete points, or at exceptional rings are also presented in the literature.\cite{NHWeylTh,NHWeylTh2, WeylColdAtomTh,NHRing1} Experimental evidence of Weyl exceptional point is also presented in a bipartite optical waveguide.\cite{NHRing2Exp} 
 Dynamical quantum phase transition from trivial to non-trivial topological phases is demonstrated in some photonic systems\cite{NHDQTI, NHRiceMeleDQ,NHDQPT} with sudden quench\cite{NHDQTIQuench,NHDynamicalClass}. There have also been several works on Floquet topological insulators under periodic time-evolutions,\cite{NHSSHFloquet,NHSSHFloquet2,NHFloquet} with experimental supports on similar systems.\cite{NHExp3Floquet} In recent years, there are proposals for higher order topological phases, in which the boundary states arise in two lesser dimensions.\cite{NHHOTI1,NHHOTI2,NHHOTI3,NHHOTI4} Topological phases are also obtained in NH quantum XY model with complex magnetic field.\cite{NHXYSong} 
Finally, NH topological superconductivity has also been studied in 1D Kitaev model.\cite{NHTSC,NHESurface,NHKitaevWang,NHKitaevKlett,NHKitaevMandal,NHKitaevMcdonald,NHQWireTh,NHKitaevLi,NHKitaevYuce} Beyond the usual fermionic and photonic lattices, topological phonon, polaritons\cite{NHPhonon} and plasmons\cite{NHPlasmons,NHPlasmons2} are discussed recently in NH systems. NH topological phases are also proposed in electromagnetic \cite{NHLCRHO,NHGraphene,NHMaxwell}, as well as in mechanical systems\cite{NHMechanical,NHOpMech1,NHOpMech2}.

\section{Discussion and conclusions}

The band degeneracy, either the Dirac/Weyl point in Hermitian systems or the EP in the NH systems, lies at the heart of topological phases in both systems. However, its characteristics and manifestations are very different in the two systems. For the Hermitian case, the two degenerate states are linearly independent, and hence scattering between these two states, by those impurities and perturbations which commute with the Hamiltonian, is prohibited. On the contrary, at EP the eigenstates become linearly dependent, and hence the Hilbert space becomes ill defined. Experimentally such EPs are already observed.\cite{EPExp1,EPExp2,Ruter,guo} However, the good news is that in the case of topological insulators in NH settings, the adiabatic loop for the Berry phase calculation does not have to pass through an EP, but only it needs to enclose an EP. In other words, the system does not have to physically possess an EP, but only needs to lie in the parameter range where the gap terms enclose a possible EP, i.e., all the gap terms vanish but not at the same point. This is a crucial condition for the NH Hamiltonians to possess topological invariants. 

After some initial controversy, the field has more or less settled now with the consensus that topological phases can exist in NH systems with modified bulk-boundary correspondence. It is shown with numerous examples that as long as chiral symmetry is present, topological edge states are localized with zero energy without dissipations. Without chiral symmetry, topological edge states can exhibit interesting properties. For example, with balanced gain/loss perturbations in the bulk, the edge states on both sides of the lattice selectively become lossy (say, on the right hand edge) and amplified (say, on the left hand edge). This means, when polarization, or particle density decays in one side, in another side it increases by equal amount. This phenomenon is somewhat analogous to the electric/thermal conductivity from one side to another, but with the exception that here it is obtained through NH bulk topological invariant without applied voltage or heat bath in a translationally invariant system. This property opens up the possibility of devising an optical diode. Furthermore, one obtains a `skin effect' in NH systems where not only the edge states, but also the bulk states move to the edge and dissipates. This feature gives the opportunity to study new surface phenomena without taming the bulk. The tuning facility between half-integer to integer topological number may be relevant for many applications such as half-twisted vector beam production.\cite{VectorBeam} 

As much progress is made in terms of observations of the NH Hamiltonians and its topological phases in photonic systems, such an evidence is not yet present in condensed matter setup. With the advent of pump-probe mechanism, where excited states can now be studied easily, exploration of NH Hamiltonians and its quantum and topological phases should be taken up in the future studies.

The notion of \PT-symmetry and non-Hermiticity have now been considered in many other experimental setups like in optomechanics,\cite{opto1,opto2,opto3,opto4} photonics \cite{photo1, photo2,photo4} and plasmonics,\cite{plas1, plas2} metamaterials,\cite{mm1,mm2,mm3} photonic topological insulators\cite{tpi1,tpi2} etc. \PT-symmetry breaking leads to many interesting phenomena such as a loss induced optical transparency,\cite{otp} nonreciprocal wave propagation in power oscillations,\cite{powosc1,powosc2} \PT-solitons,\cite{PTs1,PTs2,PTs3,PTs4} double refraction.\cite{powosc1,dob2}. Some more advanced optical experiments in \PT-symmetric high-power laser systems and laser oscillators have also been done.\cite{PTlaser1,PTlaser2,PTlaser3,PTlaser4,PTlaser5,PTlaser6,PTlaser7,PTlaser8,PTlaser9} `Anti-Laser', which is the TR partner to a laser emission\cite{antl1,antl2,antl3,antl34,antl345,antl4} can be used in engineering Kramers pairs, and TR invariant $\mathbb{Z}_2$ topological invariants.\cite{epTI1,epTI2,NHTELee,Leykam}

\section{Appendix}

\subsection{Useful identities for complex Berry phases}\label{AppendixA}\nonumber 

Using the biorthogonal condition $\langle{\psi}_n^{\rm L}|{\psi}_n^{\rm R}\rangle=1$,  we get $\langle{\psi}_n^{\rm L}|\partial{\psi}_n^{\rm R}\rangle=-\langle \partial{\psi}_n^{\rm L}|{\psi}_n^{\rm R}\rangle = - (\langle \partial{\psi}_n^{\rm R}|{\psi}_n^{\rm L}\rangle)^*$. Substituting this identity in Eq.~\eqref{Eq:NHBerry}, we find $\gamma^{\rm LR}_{n} = (\gamma^{\rm RL}_{n})^*$. Therefore, the difference between these two Berry phases $\gamma_{-}=(\gamma_n^{\rm LR}+\gamma_n^{\rm RL})/2$ is a real number. 

The time-dependent Schr\"odinger equation for each left and right eigenstates are $i\partial_t |{\psi}_n^{\rm R}\rangle = H|{\psi}_n^{\rm R}\rangle$, and  $i\partial_t |{\psi}_n^{\rm L}\rangle = H^{\dag}|{\psi}_n^{\rm L}\rangle$ (these two equations ensure that the bi-norm is preserved, i.e., $\partial_t \langle {\psi}_n^{\rm L}|{\psi}_n^{\rm R}\rangle=0$ ). Using these two Schr\"odinger equations, we can obtain the time-evolution of the individual norms to be as 
\begin{subequations}
\bea
\partial_t \langle {\psi}_n^{\rm R}|{\psi}_n^{\rm R}\rangle &=& i\langle {\psi}_n^{\rm R}|\delta H|{\psi}_n^{\rm R}\rangle,\\
\partial_t \langle {\psi}_n^{\rm L}|{\psi}_n^{\rm L}\rangle &=& -i\langle {\psi}_n^{\rm L}|\delta H|{\psi}_n^{\rm L}\rangle,
\label{eq1:AppendixA}
\eea 
\end{subequations}
where $\delta H = H^{\dag}-H$ is the NH part of the Hamiltonian.  Since $\delta H $ is anti-Hermitian, its eigenvalues are purely imaginary, and hence in right hand side of Eq~\eqref{eq1:AppendixA} is real. Furthermore, $\partial_t \langle {\psi}_n^{\alpha}|{\psi}_n^{\alpha}\rangle = 2{\rm Re}[\langle  {\psi}_n^{\alpha}|\partial_t{\psi}_n^{\alpha}\rangle]$. Therefore, the imaginary part of corresponding Berry phases 
\bea
{\rm Im}\left[{\gamma}_n^{\alpha\alpha}\right] &=& \oint_{\mathcal{C}}{\rm Re}[\langle  {\psi}_n^{\alpha}|\partial_t{\psi}_n^{\alpha}\rangle]\nonumber\\
& =& \pm \frac{1}{2}\oint_{\mathcal{C}} \partial_t\langle{\psi}_n^{\alpha}|\delta H|{\psi}_n^{\alpha}\rangle dt,\nonumber\\
& =& \pm \frac{1}{2}\langle {\psi}_n^{\alpha}|\delta H|{\psi}_n^{\alpha}\rangle,
\eea
where $\pm$ signs are for $\alpha={\rm L/R}$ states. Therefore, the imaginary part of the Berry phase comes solely from the NH part ($\delta H$) of the Hamiltonian, and as $\delta H\rightarrow 0$, the Berry phase formula coincide with that of the hermitian one and becomes real. 

\subsection{Appendix to pseudo-Hermitian eigenspectrum}\label{AppendixPseudoH}
Combination of Eq.~\eqref{Lefteigen} and Eq.~\eqref{PseudoH} leads to 
\bea
H^{\dag}(\eta|\psi_n^{\rm R} \rangle)  &=& E_n (\eta|\psi_n^{\rm R} \rangle). 
\label{AppPseudoH1}
\eea
where $E_n$, $E_n^*$ are the corresponding eigenvalues. The eigenstates are linearly independent (except at the EPs) and thus can be bi-orthogonalized as $|{\psi}_n^{\rm R}\rangle=|\psi_n^{\rm R}\rangle/\sqrt{\langle \psi_n^{\rm L}|\psi_n^{\rm R} \rangle}$, and $|{\psi}_n^{\rm L}\rangle=|\psi_n^{\rm L}\rangle/\sqrt{\langle \psi_n^{\rm L}|\psi_n^{\rm R} \rangle}$, which then gives $\langle {\psi}_n^{\rm L}|{\psi}_m^{\rm R}\rangle=\delta_{nm}$, and $\sum_n |{\psi}_m^{\rm R}\rangle \langle{\psi}_n^{\rm L}|=1$ (in the literature, the left and right, biorthogonal eigenstates are often denoted by $|\psi_n\rangle\rangle$ and $|\psi_n\rangle$, respectively).


\begin{acknowledgments} 
We acknowledge Jasper van Wezel, Franco Nori, Cem Y\"uce, Daniel Leykam, and  Zhong Wang for useful comments on our preprint.  AG acknowledges the financial support from Science and Engineering Research Board (SERB), Department of Science \& Technology (DST), Govt. of India for the National Post Doctoral Fellowship. TD acknowledges the financial support from the Infosys Science foundation through the Young Investigator Award.
\end{acknowledgments}

\end{document}